\documentclass[a4paper,UKenglish,cleveref,autoref,thm-restate,nolineno]{socg-lipics-v2021}

\hideLIPIcs  

\usepackage{mathtools,
            nicematrix,
            tikz-cd,
            booktabs,
            longtable,
            diagbox}
\usepackage[linesnumbered,algoruled,lined,noend]{algorithm2e}
\DeclareMathAlphabet{\mathbbol}{U}{bbold}{m}{n} 
\usepackage{placeins} 

\newcommand{\B}{\mathcal B}
\newcommand{\F}{\mathbb F}
\newcommand{\PP}{\mathcal P}
\newcommand{\QQ}{\mathcal Q}
\newcommand{\Q}{\mathbb Q}
\newcommand{\kk}{\mathbbol k}
\newcommand{\R}{\mathbb R}
\newcommand{\Z}{\mathbb Z}
\DeclareMathOperator{\rank}{rank}
\DeclareMathOperator{\Span}{span}
\newcommand{\barcode}{\mathrm{Barcode}}
\newcommand{\Matrix}{\mathrm{Matrix}}
\DeclareMathOperator{\im}{im}
\DeclareMathOperator{\col}{col}
\DeclareMathOperator{\Tor}{Tor}
\DeclareMathOperator{\Coker}{coker}
\DeclareMathOperator{\GL}{GL}
\DeclareMathOperator{\low}{low}
\DeclareMathOperator{\piv}{piv}
\DeclareMathOperator{\Red}{Red}

\newcommand{\firstcol}[1]{\makebox[2em][c]{$#1$}} 

\title{Computing Integral Persistence Efficiently}

\author{Tamal K. Dey}{Department of Computer Science, Purdue University, West Lafayette, Indiana, USA} {tamaldey@purdue.edu}{https://orcid.org/0000-0001-5160-9738}{Supported by NSF funds DMS-2301360 and CCF-2437030.}
\author {Rapha\"{e}l Tinarrage}{Institute of Science and Technology Austria, Klosterneuburg, Austria} {raphael.tinarrage@ist.ac.at}{https://orcid.org/0000-0002-1404-1095}{}

\authorrunning{T.~Dey, R.~Tinarrage}
\Copyright{Tamal K. Dey and Rapha\"{e}l Tinarrage}



\supplementdetails{Code}{https://github.com/raphaeltinarrage/integral-persistence}

\acknowledgements{
    This work grew out of discussions during the Foundations of Computational Geometry and Topology program at the Institute for Computational and Experimental Research in Mathematics (ICERM) in Providence, Rhode Island.
    The program was supported by the National Science Foundation under Grant No.\ DMS-2424556.
}

\EventEditors{John Q. Open and Joan R. Access}
\EventNoEds{2}
\EventLongTitle{42nd Conference on Very Important Topics (CVIT 2016)}
\EventShortTitle{CVIT 2016}
\EventAcronym{CVIT}
\EventYear{2016}
\EventDate{December 24--27, 2016}
\EventLocation{Little Whinging, United Kingdom}
\EventLogo{}
\SeriesVolume{42}
\ArticleNo{23}

\AtBeginDocument{%
  \hypersetup{pdfsubject={Preprint}}%
} 

\begin{document}

\maketitle

\begin{abstract}
    While persistence modules with field coefficients are well understood, working with integral coefficients remains a challenge.
    In particular, torsion obstructs the existence of an integral persistence barcode.
    We investigate this problem from two perspectives.
    First, from an algorithmic viewpoint, we present an algorithm that computes the barcodes of a simplicial filtration over all fields at once in cubic time for any fixed dimension.
    This is the first algorithm with this time complexity.
    Our algorithm starts from a reduction with rational coefficients and corrects it at all exceptional primes via efficient updates akin to the vineyard algorithm.
    As a consequence, we obtain a nearly cubic algorithm for deciding whether the barcode is field-independent.
    In experiments, we showcase the advantages of our approach over established algorithms.
    Second, from a theoretical viewpoint, we distinguish three sources of torsion and identify their effects on barcodes.
    Under a mild assumption, we show that the barcode of a simplex-wise filtration over any field can be obtained from its barcode over the rationals through three successive transformations: swapping certain death times (map torsion), extending certain bars to the right (tensor torsion), and extending certain bars to the left (Tor torsion).
\end{abstract}


\setcounter{tocdepth}{1}
\begingroup
    \fontsize{10}{11}\selectfont
    \setlength{\parskip}{4.25pt}
    \makeatletter
        \renewcommand*\l@section{\@dottedtocline{1}{0em}{2em}}
        \renewcommand*\@dotsep{3}
    \makeatother
    \renewcommand{\contentsname}{{\large Contents}}
    \makeatletter
        \section*{\contentsname}
        \vspace{-0.5em}
        \@starttoc{toc}
    \makeatother
\endgroup
\newpage


\section{Introduction}

\subsection{Persistence modules and integral coefficients}
\label{subsec:intro_persistence}

\subparagraph{Persistence theory.}

In persistent homology, the standard input to many algorithms is a \emph{simplicial filtration}, that is, a  nested sequence of simplicial complexes.
We denote it
\begin{equation*}
    K^\bullet\colon\quad
    K^0 \subset K^1 \subset \cdots \subset K^T
\end{equation*}
where $K^0 = \emptyset$.
Given a field $\kk$---such as a finite field $\F_p$ with $p$ prime---and an integer $d\geq 0$, the simplicial filtration above yields a sequence of homology groups and linear maps, denoted%
\begin{equation}
    \label{eq:persistence_module_field_coefficient}
    H_d(K^\bullet;\kk)\colon\quad
    H_d(K^0;\kk) \to H_d(K^1;\kk) \to \cdots \to H_d(K^T;\kk).
\end{equation}
This structure is called a \emph{persistence module}.
Associated with it is a \emph{persistence barcode}, encoded as a finite multiset of half-open intervals $[b_i, d_i)$ where $b_i, d_i \in \{0,\dots,T,+\infty\}$ and $b_i<d_i$.
In the category of finitely indexed and pointwise finite-dimensional persistence modules, the barcode is a complete invariant up to isomorphism \cite{ZomorodianCarlsson2005,CrawleyBoevey2015,ChazalDeSilvaGlisseOudot2016}.
We denote it by
\begin{equation}
    \label{eq:persistence_barcode}
    \barcode (H_d(K^\bullet;\kk)) = \{\!\{ [b_i, d_i) \mid 1 \leq i \leq r \}\!\}.
\end{equation}
The classical \emph{persistence algorithm with sequential reduction} computes this barcode~\cite{EdelsbrunnerLetscherZomorodian2002}.~Its worst-case complexity is $\Theta(N^3)$ field operations where $N$ is the number of simplices in~$K^\bullet$~\cite{Morozov2005}.

\subparagraph{Field choice problem.}

A limitation of the framework above is that the field $\kk$ must be fixed in advance, and different choices of $\kk$ may lead to different barcodes.
This phenomenon is governed by \emph{torsion}, either in the homology groups themselves or in the maps, as we describe in \Cref{sec:torsion_on_barcodes}.
More precisely, Obayashi and Yoshiwaki proved in \cite{ObayashiYoshiwaki2023} that $\barcode (H_d(K^\bullet;\kk))$ does not depend on $\kk$ provided the \emph{integral homology} and \emph{integral relative homology}~groups
\begin{equation}
    \label{eq:three_homology_groups}
    H_d(K^t;\Z),~~
    H_{d-1}(K^t;\Z),
    ~\text{ and }~ 
    H_d(K^t,K^s;\Z)     
\end{equation}
are free for all $0\leq s \leq t \leq T$.
They also provide an algorithm to test this condition for \emph{simplex-wise filtrations}, i.e., when simplices are added one by one.
It consists of running the classical persistence algorithm over $\Z$ on the boundary matrix, requiring $O(N^3)$ operations.

\subparagraph{Integral decomposition.}

Working directly with integral coefficients offers another perspective on the problem.
Similar to \Cref{eq:persistence_module_field_coefficient}, integral homology yields a diagram~of~$\Z$-modules%
\begin{equation}
    \label{eq:persistence_module_integer_coefficient}
    H_d(K^\bullet;\Z)\colon\;
    H_d(K^0;\Z) \to H_d(K^1;\Z) \to \cdots \to H_d(K^T;\Z).
\end{equation}
Unlike persistence modules over a field, such \emph{integral persistence modules} need not decompose into interval modules, hence need not admit a barcode.
Nevertheless, Luo and Henselman-Petrusek proved in \cite{LuoHenselman-Petrusek2025,LuoHenselman-Petrusek2026} that a decomposition exists whenever the relative homology~groups
\[
    H_d(K^t,K^s;\Z), 
    \qquad 0\leq s \leq t \leq T,
\]
are free.
They also proposed an algorithm to decide field independence for \emph{arbitrary filtrations}.
However, it requires reducing $T^2$ matrices to Smith normal form, resulting in large complexity.

\subparagraph{Contribution.}

Unlike the work above, we do not assume field independence.
First, we give a \emph{cubic} algorithm to compute $\barcode (H_d(K^\bullet;\kk))$ over all fields $\kk$.
We deduce a \emph{near-cubic} algorithm to decide field independence for \emph{arbitrary} filtrations (\Cref{th:field_independence_problem}).
Second, we identify a hierarchy of torsion phenomena and describe how each affects the barcode (\Cref{th:birth_death_profiles}).
We do so by isolating the three terms in \Cref{eq:three_homology_groups} via a sequence~of~modules%
\[
    H_d(K^\bullet;\kk)
    \hookleftarrow
    H_d(K^\bullet;\Z)\otimes\kk 
    \twoheadrightarrow 
    H_d^\mathrm{free}(K^\bullet;\Z)\otimes\kk.
\]

\subsection{Multi-field algorithms}
\label{subsec:intro_algorithms}

\subparagraph{Brute-force algorithm.}

In the absence of a suitable invariant for integral modules, an alternative is to work with all fields at once.
As recalled in \Cref{subsec:reduction_exceptional_primes}, there are only finitely many \emph{exceptional primes} that yield barcodes different from the $\Q$-barcode.
We denote them%
\begin{equation}
    \label{eq:definition_exceptional_primes}
    \PP 
    = \PP(K^\bullet)
    = \big\{p \text{ prime} \mid \exists d\geq0,\; \barcode\big(H_d(K^\bullet;\F_p)\big) \neq \barcode\big(H_d(K^\bullet;\Q)\big) \big\}.    
\end{equation}
Moreover, $\PP$ can be identified via the persistence algorithm over $\Q$ (\Cref{lem:generic_reduction_identify_primes}).
By \emph{brute-force algorithm}, we mean running the algorithm independently for $\Q$ and each $\F_p$, $p\in\PP$.~It~has
\[
    O\bigl(N^3\times \bigl(1+|\PP|\bigr)\bigr)
\]
worst-case complexity.
Moreover, we exhibit in \Cref{prop:number_exceptional_primes_newman} a family of filtrations whose number of exceptional primes reaches $\Omega(N/\log N)$ and for which the algorithm has~complexity%
\[
    \Omega\bigl(N^4 / \log N\bigr).
\]

\subparagraph{Modular reconstruction.}

In \cite{BoissonnatMaria2014,Boissonnat_2019}, Boissonnat and Maria proposed the \emph{modular reconstruction} algorithm. 
Given a simplex-wise filtration $K^\bullet$ and a list of $L$ fields $\F_{p_1},\dots,\F_{p_L}$, it computes their barcodes simultaneously. 
While the classical persistence algorithm reduces the boundary matrix over a specified field $\F_{p_i}$, their algorithm works in the ring $\Z/(p_1\cdots p_L)\Z$.
The Chinese Remainder Theorem then enables simultaneous reductions.
Its complexity is
\[
    O\bigl(\bigl[L\times(\QQ_L-\QQ_*)+\QQ_L^3\bigr]\times A_L\bigr),
\]
where $A_L$ is the cost of arithmetic operations in $\Z/(p_1\cdots p_L)\Z$, 
$\QQ_L$ is the number of distinct bars $[i,j)$ appearing in the barcode of $H_d(K^\bullet;\kk)$ for some $d\geq0$ and $\kk$ in the list,
and $\QQ_*$ is the number of bars for a fixed reference field.
For certain $L$, this approach can outperform brute force.
However, when all exceptional primes are included, $\QQ_L$ can be as large as $\Omega(N^{3/2}/(\log N)^{1/2})$ (\Cref{prop:number_exceptional_bars}).
In this case, the cubic term in the stated complexity~is%
\[
    \QQ_L^3 = 
    \Omega\bigl( N^{9/2}/(\log N)^{3/2} \bigr).
\]

\subparagraph{Omni-field persistence.}

As part of his library \texttt{Dionysus2} \cite{morozov_dionysus2}, Morozov implemented \emph{omni-field persistence}, an algorithm for computing the barcode of $H_d(K^\bullet;\kk)$ over \emph{all} fields $\F_p$.
The algorithm runs a reduction over $\Q$ and marks a column as exceptional when a prime-specific behavior is encountered.
This occurs when the column has an exceptional pivot modulo $p$, or when the column requires exceptional previous columns to be reduced.
Its complexity is thus%
\[
    O\biggl(N\times \sum_{j=0}^N \biggl(N + \sum_{p\in\PP_j} n_{j,p} \biggr) \biggr),
\]
where $\PP_j$ is the set of primes for which the $j$th column receives an exceptional reduction, and $n_{j,p}$ is the number of column operations during this reduction.
\Cref{prop:number_exceptional_columns_omnifield} exhibits filtrations where exceptional columns are recruited many times, yielding a complexity of%
\[
    \Omega\big(N^4/\log N \big).
\]

\subparagraph{Our approach.}

We give an algorithm (\Cref{alg:main_algorithm}) that computes the barcode of $H_d(K^\bullet;\Q)$, $\PP$, and the barcodes of $H_d(K^\bullet;\F_p)$ for all $p\in\PP$ in \emph{cubic time}. 
Its complexity is~(\Cref{th:main_algorithm})
\[
    O\big( N^3\log(\ell+1) \big),
    \qquad
    \ell = \dim K^T.
\]
The algorithm avoids redundant computations by reusing the rational reduction wherever possible. We show how to perform efficient updates, in the spirit of the vineyard algorithm~\cite{CohenSteiner2006}.

\subsection{Organization of the paper}



\begin{description}
    \item[\Cref{sec:algorithmic_result}]
        We present our main algorithm
        (\Cref{alg:main_algorithm,th:main_algorithm}).
        We also explain its application to the field-independence problem (\Cref{alg:field_independence_problem,th:field_independence_problem}).

    \item[\Cref{sec:numerical_experiments}]
        We benchmark our implementation against existing methods on the Linial--Meshulam process (\Cref{subsec:benchmark}).
        We then investigate field dependence in random Rips filtrations (\Cref{subsec:torsion_random}), with a brief discussion of our results on homological torsion.

    \item[\Cref{sec:torsion_on_barcodes}]
        We present our full theoretical analysis of torsion (\Cref{th:birth_death_profiles}).
        In \Cref{subsec:three_sources_of_torsion}, we distinguish Tor, tensor, and map torsion.
        In \Cref{subsec:three_persistence_modules}, we isolate their contributions via the field-coefficient, tensor, and free persistence modules (\Cref{eq:persistence_module_field_coefficient,eq:persistence_module_tensor,eq:persistence_module_free}).

    \item[\Cref{sec:computation_presentation_matrices}]
        We develop the algorithmic counterpart of this analysis: computing presentation matrices for these modules.
        We treat the tensor module in \Cref{subsec:presentation_matrix_tensor} (\Cref{alg:presentation_matrix_tensor_module,th:presentation_matrix_tensor_module}) and the free module in \Cref{subsec:presentation_matrix_free} (\Cref{alg:presentation_matrix_free_module,th:presentation_matrix_free_module}). 

    \item[\Cref{sec:filtrations_with_large_torsion}]
        We give complexity results for the methods mentioned in \Cref{subsec:intro_algorithms}: brute force (\Cref{prop:number_exceptional_primes_newman}), modular reconstruction (\Cref{prop:number_exceptional_bars}), and omni-field (\Cref{prop:number_exceptional_columns_omnifield}).

    \item[Appendix] 
        We collect all proofs (\Cref{sec:proofs}), all algorithm descriptions (\Cref{sec:algorithms}), and a list of the notation used throughout the paper (\Cref{sec:notation}).
\end{description}

\section{Computing persistence barcodes over all fields simultaneously}
\label{sec:algorithmic_result}

In this section, we develop an algorithm (\Cref{alg:main_algorithm}) to compute the barcodes of $H_d(K^\bullet;\F_p)$ over all fields $\F_p$ simultaneously.
The main novelty is \emph{priority-based reduction}, introduced in \Cref{subsec:reduction_fixed_field}, which can be efficiently updated.
We apply it to our problem in \Cref{subsec:reduction_exceptional_primes}.

\subsection{Reduction algorithm over a fixed field}
\label{subsec:reduction_fixed_field}

Throughout this subsection, we consider an $n\times m$ matrix $D$ with coefficients in a field $\kk$, such as a boundary or a presentation matrix (see \Cref{sec:computation_presentation_matrices}).
We first recall the classical persistence algorithm in \Cref{subsubsec:classical_reduction}.
Then, in preparation for our main algorithm, we introduce in \Cref{subsubsec:row_priority} a variant in which the rows are assigned a prescribed \emph{priority}.
In \Cref{subsubsec:efficient_updates_priority}, we develop an algorithm for updating a reduction when the priority changes.

\subsubsection{Classical reduction algorithm}
\label{subsubsec:classical_reduction}

\subparagraph{Matrix reduction and pivots.}

Let $D$ have shape $n\times m$.
A \emph{reduction} of $D$ is a matrix 
\[
    R = DV
\]
where $V$ is an invertible upper-triangular $m\times m$ matrix with diagonal entries $1_\kk$, and
\begin{equation*}
    \forall j,j' \in \{1,\dots,m\},\quad j\neq j' \implies \low(R,j) \neq \low(R,j')    
\end{equation*}
where $\low(R,j)$ is the \emph{pivot index}, defined as the index of the lowest nonzero entry of the $j$th column of $R$, and is undefined if the column is zero.
We call its entry the \emph{pivot} and denote
\[
    \piv(R,j) = R[\low(R,j),j].
\]
For convenience, when the $j$th column is zero, we use the following symbol and default~value:%
\[
    \low(R,j) = \bot_j, 
    \qquad 
    \piv(R,j) = 1_\kk.
\]

By the Pairing Uniqueness Lemma \cite{CohenSteiner2006}, the index $\low(R,j)$ does not depend on the specific transformation $V$ but only on $D$.
The same observation holds for $\piv(R,j)$ \cite[Theorem 5]{Hang2021Umatch}.

Finally, when $D$ is the boundary matrix of a filtration $(K^t)_{t=0}^N$, the pivot indices $\low(R,j)$ determine the persistence barcode.
See \Cref{eq:barcode_from_pivots} in \Cref{subsubsec:full_algorithm} for the precise formula.

\subparagraph{Augmented matrix.}

To describe our algorithm, it will be convenient to stack on top of $D$ the identity matrix $I_m$.
This \emph{augmented matrix} has shape $(n+m)\times m$ and is denoted
\begin{equation}
    \label{eq:augmented_matrix}
    \widetilde{D} = 
    \begin{pmatrix}
        I_m \\\hline
        D
    \end{pmatrix}.
\end{equation}
Note that if $R = DV$ is a reduction of $D$, then the following matrix is a reduction of $\widetilde{D}$:
\[
    \widetilde{R} = 
    \widetilde{D}V = 
    \begin{pmatrix}
        V \\\hline
        R
    \end{pmatrix}.
\]

This notation has two advantages.
First, for visualization, the transformation $V$ can be read directly from the upper part of $\widetilde{R}$.
Second, by definition of $V$, the columns of $\widetilde{R}$ are always nonzero.
This eliminates the need to handle zero columns of $R$ separately.

The first $m$ rows of $\widetilde{R}$ are indexed by $\bot_1,\dots,\bot_m$ and the last $n$ rows by $1,\dots,n$.
Thus, the index of the lowest nonzero entry of its $j$th column is a symbol $\bot_i$ or an integer $i$, depending on whether it is zero or not in $R$.
We recover the notation previously introduced:%
\[
    \forall j\in\{1,\dots,m\},
    \quad
    \low(\widetilde{R},j)=\low(R,j)
    \quad\text{and}\quad
    \piv(\widetilde{R},j)=\piv(R,j).
\]

\begin{example}
    We show below a reduction of a $3\times4$ matrix $D$ over the rational coefficients $\kk=\Q$.
    Its pivot indices are respectively $(3,1,\bot_3,2)$, with corresponding entries boxed in $\widetilde{R}$.
    \begin{equation}
        \label{eq:example_reduction}
        \widetilde{D} = 
        \begin{pNiceArray}{cccc}
            1 & 0 & 0 & 0 \\
            0 & 1 & 0 & 0 \\
            0 & 0 & 1 & 0 \\
            0 & 0 & 0 & 1 \\\hline
            1 & 0 & 1 & 0 \\
            2 & 1 & 1 & 4 \\
            2 & 1 & 1 & 1 
        \end{pNiceArray}
        \qquad\longrightarrow\qquad
        \widetilde{R} = 
        \begin{pNiceArray}{cccc}[first-col]
            \firstcol{\bot_1}   & 1 & \mathllap{-}\tfrac12 & \mathllap{-}1 & \mathllap{-}\tfrac12 \\
            \firstcol{\bot_2}   & 0 & 1 & 1 & 0 \\
            \firstcol{\bot_3}   & 0 & 0 & \boxed{1} & 0 \\
            \firstcol{\bot_4}   & 0 & 0 & 0 & 1 \\\hline
            \firstcol{1}        & 1 & \mathclap{\boxed{-\tfrac12}} & 0 & \mathllap{-}\tfrac12 \\
            \firstcol{2}        & 2 & 0 & 0 & \boxed{3} \\
            \firstcol{3}        & \boxed{2} & 0 & 0 & 0 
        \end{pNiceArray}.
    \end{equation}
\end{example}

\subparagraph{Classical algorithm.}

The \emph{persistence algorithm with sequential reduction} \cite{EdelsbrunnerLetscherZomorodian2002} reduces $\widetilde{D}$ to $\widetilde{R}$ by Gaussian elimination.
Start with $\widetilde{R}\gets\widetilde{D}$.
The algorithm sweeps its columns $\col_1,\col_2,\dots$ from left to right.
At the $j$th stage, it identifies the lowest nonzero entry of $\col_j$, say of index $i$.
If $i = \low(\widetilde{R},j')$ for some $j'<j$---that is, if this pivot index is already \emph{owned} by another column---then the algorithm zeroes out the entry $\widetilde{R}[i,j]$ via the column operation
\[
    \col_j \gets \col_j - \frac{\widetilde{R}[i,j]}{\widetilde{R}[i,j']}\col_{j'}.
\]
This process is repeated until the lowest nonzero index of $\col_j$ is not owned, in which case it is saved as its pivot.
The output matrix $\widetilde{R}$ is a reduction of $\widetilde{D}$ that we denote
\[
    \Red(\widetilde{D}) = \widetilde{R}.
\]
Note that, in this article, we apply the algorithm to the augmented matrix $\widetilde{D}$, which has full column rank.
Thus, a column cannot be zeroed out; instead, it can acquire a pivot $\bot_j$.

\begin{example}
    We show below the first stages of the reduction of the matrix $\widetilde{D}$ in \Cref{eq:example_reduction}.
    Once a column is reduced, its pivot is boxed, and the next conflicting entry is shown in gray.
    \begin{equation*}
        \begin{pNiceArray}{cccc}
            1 & 0 & 0 & 0 \\
            0 & 1 & 0 & 0 \\
            0 & 0 & 1 & 0 \\
            0 & 0 & 0 & 1 \\
            \hline
            1 & 0 & 1 & 0 \\
            2 & 1 & 1 & 4 \\
            \boxed{2} & \fcolorbox{black}{gray!15}{1} & 1 & 1
        \end{pNiceArray}
        \to 
        \begin{pNiceArray}{cccc}
            1 & \mathllap{-}\tfrac12 & 0 & 0 \\
            0 & 1 & 0 & 0 \\
            0 & 0 & 1 & 0 \\
            0 & 0 & 0 & 1 \\
            \hline
            1 & \mathclap{\boxed{-\tfrac12}} & 1 & 0 \\
            2 & 0 & 1 & 4 \\
            \boxed{2} & 0 & \fcolorbox{black}{gray!15}{1} & 1 
        \end{pNiceArray}
        \to 
        \begin{pNiceArray}{cccc}
            1 & \mathllap{-}\tfrac12 & \mathllap{-}\tfrac12 & 0 \\
            0 & 1 & 0 & 0 \\
            0 & 0 & 1 & 0 \\
            0 & 0 & 0 & 1 \\
            \hline
            1 & \mathclap{\boxed{-\tfrac12}} & \fcolorbox{black}{gray!15}{$-\tfrac12$} & 0 \\
            2 & 0 & 0 & 4 \\
            \boxed{2} & 0 & 0 & 1 
        \end{pNiceArray}
        \to 
        \begin{pNiceArray}{cccc}
            1 & \mathllap{-}\tfrac12 & \mathllap{-}1 & 0 \\
            0 & 1 & 1 & 0 \\
            0 & 0 & \boxed{1} & 0 \\
            0 & 0 & 0 & 1 \\
            \hline
            1 & \mathclap{\boxed{-\tfrac12}} & 0 & 0 \\
            2 & 0 & 0 & 4 \\
            \boxed{2} & 0 & 0 & \fcolorbox{black}{gray!15}{1} 
        \end{pNiceArray}
        \to 
    \end{equation*}
\end{example}

\subsubsection{Row priority and exhaustive reduction}
\label{subsubsec:row_priority}

\subparagraph{Row priority.}

Consider a matrix $\widetilde{D}$ over $\kk$, as in \Cref{eq:augmented_matrix}, and index its rows by
\[
    \{\bot_1,\dots,\bot_m,1,\dots,n\}.
\]
Suppose we apply the persistence algorithm to $\widetilde{D}$, and denote by $\widetilde{R}$ the working matrix.
When a column $\col_j$ is being processed, its pivot $\piv(\widetilde{R},j)$ is identified as its \emph{lowest} nonzero entry.
Thus, its entries are inspected from bottom to top, according to the order
\begin{equation}
    \label{eq:canonical_order}
    n \succ \cdots \succ 1 \succ \bot_m \succ \cdots \succ \bot_1.
\end{equation}
We now generalize this convention by allowing an arbitrary ordering of the rows.

\begin{definition}
    \label{def:row_priority}
    A \emph{row priority} is an order on $\{\bot_1,\dots,\bot_m,1,\dots,n\}$ encoded as~a~permutation
    \[
        \tau = (\tau_1,\dots,\tau_{n+m}),
        \quad\text{ meaning }\quad
        \tau_1 \succ \dots \succ \tau_{n+m}.
    \]    
\end{definition}

For a nonzero column $j$ of a matrix $\widetilde{R}$, we define its \emph{$\tau$-pivot index} by
\[
    \low(\widetilde{R},j,\tau) = \tau_{i^*},
    \quad
    i^* = \min\bigl\{ i \in \{1,\dots,m+n\} \mid \widetilde{R}[\tau_i,j]\neq 0\bigr\}.
\]
In other words, $\tau_{i^*}$ is the highest row index, with respect to the order $\tau$, whose corresponding entry in the column $\widetilde{R}[:,j]$ is nonzero.
This entry is called the $j$th \emph{$\tau$-pivot} and is denoted
\[
    \piv(\widetilde{R},j,\tau) = 
    \widetilde{R}\bigl[\low(\widetilde{R},j,\tau),\,j\bigr].
\]

We call \emph{canonical order} the order of \Cref{eq:canonical_order} and denote it by $\tau_\mathrm{can}$.
With respect to this order, the notions of $\tau$-pivot index and $\tau$-pivot coincide with their classical counterparts:
\[
    \low(\widetilde{R},j,\tau_\mathrm{can}) = \low(\widetilde{R},j),
    \qquad
    \piv(\widetilde{R},j,\tau_\mathrm{can}) = \piv(\widetilde{R},j).
\]

\subparagraph{Reduction with row priority.}

We extend the notion of reduction to an arbitrary row priority.

\begin{definition}
    \label{def:reduction_row_priority}
    Consider a matrix $D\in\kk^{n\times m}$ and its augmentation $\widetilde{D}=\left(\begin{smallmatrix}I_m\\D\end{smallmatrix}\right)\in\kk^{(n+m)\times m}$.
    We call \emph{$\tau$-reduction} of $\widetilde{D}$ a matrix $\widetilde{R}\in\kk^{(n+m)\times m}$ that can be written as 
    \[
        \widetilde{R} = \widetilde{D}V 
    \]
    for an upper triangular matrix $V\in\kk^{m\times m}$ with diagonal entries $1_\kk$, and such that
    \[
        \forall j,j' \in \{1,\dots,m\},\quad j\neq j' \implies \low\bigl(\widetilde{R},j,\tau\bigr) \neq \low\bigl(\widetilde{R},j',\tau\bigr).
    \]
\end{definition}

The persistence algorithm extends naturally to a row priority $\tau$: at each step, the pivot is chosen according to $\tau$ rather than the canonical order.
See \Cref{ex:pivot_sequence_total} for an illustration.
We call this variant the \emph{persistence algorithm with priority $\tau$}.
Its output is~denoted
\[
    \Red\bigl(\widetilde{D},\tau\bigr) = \widetilde{R}.
\]
It is a $\tau$-reduction of $\widetilde{D}$ in the sense of \Cref{def:reduction_row_priority}.
Note that for the canonical order $\tau=\tau_\mathrm{can}$ defined in \Cref{eq:canonical_order}, the algorithm recovers the classical reduction:
\[
    \Red\bigl(\widetilde{D},\tau_\mathrm{can}\bigr) = \Red\bigl(\widetilde{D}\bigr).
\]

The procedure is formalized in \Cref{alg:persistence_algorithm_priority}.
Equivalently, it can be described as
\textbf{(1)}~permuting the rows of $\widetilde{D}$ into increasing $\tau$-order; 
\textbf{(2)}~applying the classical persistence algorithm; and 
\textbf{(3)}~restoring the original row order.
As a consequence, the properties of the classical algorithm carry over to this setting.
We directly deduce the correctness of the algorithm.

\begin{lemma}
    \label{lem:persistence_algorithm_priority}
    For an augmented matrix $\widetilde{D}$ of size $(n+m)\times m$ over $\kk$ and a permutation $\tau$, \Cref{alg:persistence_algorithm_priority} outputs a $\tau$-reduction $\widetilde{R}$ of $\widetilde{D}$.
    It performs $O(m^2)$ column operations.
\end{lemma}

\begin{example}
    \label{ex:pivot_sequence_total}
    Let $\widetilde{D}$ be the matrix in \Cref{eq:example_reduction}.
    Consider the permutation
    \[
        \tau = (1,3,\bot_3,2,\bot_4,\bot_2,\bot_1).
    \]
    \Cref{alg:persistence_algorithm_priority} yields the following $\tau$-reduction of $\widetilde{D}$.
    It has pivot indices $(1,3,\bot_3,2)$.
    \begin{equation*}
        \widetilde{D} = 
        \begin{pNiceArray}{cccc}
            1 & 0 & 0 & 0 \\
            0 & 1 & 0 & 0 \\
            0 & 0 & 1 & 0 \\
            0 & 0 & 0 & 1 \\\hline
            1 & 0 & 1 & 0 \\
            2 & 1 & 1 & 4 \\
            2 & 1 & 1 & 1 
        \end{pNiceArray}
        \qquad\longrightarrow\qquad
        \Red\bigl(\widetilde{D},\tau\bigr) = 
        \begin{pNiceArray}{cccc}[first-col]
            \firstcol{\bot_1}   & 1 & 0 & \mathllap{-}1 & 0 \\
            \firstcol{\bot_2}   & 0 & 1 & 1 & \mathllap{-}1 \\
            \firstcol{\bot_3}   & 0 & 0 & \boxed{1} & 0 \\
            \firstcol{\bot_4}   & 0 & 0 & 0 & 1 \\\hline
            \firstcol{1}        & \boxed{1} & 0 & 0 & 0 \\
            \firstcol{2}        & 2 & 1 & 0 & \boxed{3} \\
            \firstcol{3}        & 2 & \boxed{1} & 0 & 0 
        \end{pNiceArray}.
    \end{equation*}
    The fact that the pivot indices of $\Red\bigl(\widetilde{D},\tau\bigr)$ coincide with the prefix of $\tau$ is a general property of permutations arising from an $\F_p$-reduction of $\widetilde{D}$; see \Cref{lem:pivot_sequence_total} in \Cref{subsubsec:repairing_generic_reduction} for details.
\end{example}

\subparagraph{Exhaustive reduction.}

A useful priority is obtained from the pivot sequence of the canonical reduction itself.
More precisely, denote the canonical reduction of $\widetilde{D}$ and its pivot indices by
\begin{equation}
    \label{eq:canonical_reduction}
    \widetilde{R}_\mathrm{can}
    =
    \Red\bigl(\widetilde{D},\tau_\mathrm{can}\bigr)
    =
    \Red\bigl(\widetilde{D}\bigr),
    \qquad
    p_j = \low(\widetilde{R}_\mathrm{can},j),
    \qquad j\in\{1,\dots,m\}.
\end{equation}
The indices $p_1,\dots,p_m$ form a subset of $\{\bot_1,\dots,\bot_m,1,\dots,n\}$.
Let us write the remaining indices as $q_1,\dots,q_n$ in arbitrary order.
We define the associated \emph{pivot priority} by
\begin{equation}
    \label{eq:permutation_piv}
    \tau_\mathrm{piv}
    =
    (p_1,\dots,p_m,q_1,\dots,q_n),
    \quad\text{ meaning }\quad
    p_1 \succ \dots \succ p_m \succ q_1 \succ \dots \succ q_n.
\end{equation}

Thus, the pivot indices have higher priority than all other rows.
Applying the persistence algorithm to $\widetilde{D}$ with this priority (\Cref{alg:persistence_algorithm_priority}) yields its \emph{exhaustive reduction}, denoted by%
\begin{equation}
    \label{eq:exhaustive_reduction}
    \widetilde{R}_\mathrm{piv}
    =
    \Red\bigl(\widetilde{D},\tau_\mathrm{piv}\bigr).
\end{equation}
Its main properties are summarized in the following lemma.

\begin{lemma}[Proof p.~\pageref{proof:exhaustive_reduction}]
    \label{lem:exhaustive_reduction}
    Let $\widetilde{R}_\mathrm{can}$ and $\widetilde{R}_\mathrm{piv}$ be the canonical and exhaustive reductions of $\widetilde{D}$ as in \Cref{eq:canonical_reduction,eq:exhaustive_reduction}.
    They have the same pivot indices and pivot entries:
    \[
        \forall j\in\{1,\dots,m\},
        \quad
        \low(\widetilde{R}_\mathrm{piv},j,\tau_\mathrm{piv})
        =
        \low(\widetilde{R}_\mathrm{can},j)
        \quad
        \text{and}
        \quad
        \piv(\widetilde{R}_\mathrm{piv},j,\tau_\mathrm{piv})
        =
        \piv(\widetilde{R}_\mathrm{can},j).
    \]
    Moreover, every pivot row is zero in all subsequent columns: for every $1\leq j<j'\leq m$,
    \[
        \widetilde{R}_\mathrm{piv}[p_j,j']=0
        \qquad\text{where}\qquad
        p_j = \low(\widetilde{R}_\mathrm{can},j).
    \]
\end{lemma}

Thus, exhaustive reduction preserves the canonical pivots and additionally zeros out all entries in each pivot row to the right of its owning column.
It can be computed in two passes: 
\textbf{(1)} perform a usual reduction on $\widetilde{D}$
and \textbf{(2)} for all columns, clear the entries at earlier~pivots.%

\begin{example}
    As an illustration, consider the matrix $\widetilde{D}$ of \Cref{eq:example_reduction} over $\Q$.
    Its pivot index sequence is $(3,1,\bot_3,2)$.
    We append the remaining indices in canonical order and form%
    \[
        \tau_\mathrm{piv}
        =
        (3,1,\bot_3,2,\bot_4,\bot_2,\bot_1).
    \]
    The canonical and exhaustive reductions of $\widetilde{D}$ are as follows (only the last column differs):
    \begin{equation}
        \label{eq:example_exhaustive_reduction}
        \begin{gathered}
            \widetilde{D} \\[-1pt]
            \begin{pNiceArray}{cccc}
                1 & 0 & 0 & 0 \\
                0 & 1 & 0 & 0 \\
                0 & 0 & 1 & 0 \\
                0 & 0 & 0 & 1 \\\hline
                1 & 0 & 1 & 0 \\
                2 & 1 & 1 & 4 \\
                2 & 1 & 1 & 1
            \end{pNiceArray}
        \end{gathered}
        \quad\longrightarrow\quad
        \begin{gathered}
            ~~~~~\widetilde{R}_\mathrm{can} = \Red\bigl(\widetilde{D},\tau_\mathrm{can}\bigr) \\[-1pt]
            \begin{pNiceArray}{cccc}[first-col]
                \firstcol{\bot_1} & 1 & \mathllap{-}\tfrac12 & -1 & \mathllap{-}\tfrac12 \\
                \firstcol{\bot_2} & 0 & 1 & 1 & 0 \\
                \firstcol{\bot_3} & 0 & 0 & \boxed{1} & 0 \\
                \firstcol{\bot_4} & 0 & 0 & 0 & 1 \\\hline
                \firstcol{1} & 1 & \mathclap{\boxed{-\tfrac12}} & 0 & \mathllap{-}\tfrac12 \\
                \firstcol{2} & 2 & 0 & 0 & \boxed{3} \\
                \firstcol{3} & \boxed{2} & 0 & 0 & 0
            \end{pNiceArray}
        \end{gathered}
        \quad\longrightarrow\quad
        \begin{gathered}
            ~~~~~\widetilde{R}_\mathrm{piv} = \Red\bigl(\widetilde{D},\tau_\mathrm{piv}\bigr) \\[-1pt]
            \begin{pNiceArray}{cccc}[first-col]
                \firstcol{\bot_1} & 1 & \mathllap{-}\tfrac12 & \mathllap{-}1 & 0 \\
                \firstcol{\bot_2} & 0 & 1 & 1 & \mathllap{-}1 \\
                \firstcol{\bot_3} & 0 & 0 & \boxed{1} & 0 \\
                \firstcol{\bot_4} & 0 & 0 & 0 & 1 \\\hline
                \firstcol{1} & 1 & \mathclap{\boxed{-\tfrac12}} & 0 & 0 \\
                \firstcol{2} & 2 & 0 & 0 & \boxed{3} \\
                \firstcol{3} & \boxed{2} & 0 & 0 & 0
            \end{pNiceArray}
        \end{gathered}.
    \end{equation}
\end{example}

\subsubsection{Efficient updates of priority-reduced matrices}
\label{subsubsec:efficient_updates_priority}

\subparagraph{Update under adjacent transposition.}

Consider an augmented matrix $\widetilde{D}$ over $\kk$ and suppose that we have computed its $\tau$-reduction for a permutation $\tau$ of $\{\bot_1,\dots,\bot_m,1,\dots,n\}$.
Denote%
\[
    \widetilde{R} = \Red\bigl(\widetilde{D},\tau\bigr),
    \qquad
    \tau = (\tau_1,\dots,\tau_{n+m}).
\]
In addition, consider another permutation $\tau'$ that differs from $\tau$ by an \emph{adjacent transposition}, i.e., the exchange of two adjacent values.
Thus, there is an index $i\in\{1,\dots,n+m-1\}$ with
\[
    \tau' =
    (\tau_1,\dots,\tau_{i-1},\;
    \underset{\scalebox{1.25}[-1.25]{$~~~\curvearrowleft$}}
    {\tau_{i+1},\; \tau_i},\;
    \tau_{i+2},\dots,\tau_{n+m}).
\]
In other words, we exchange the priority of two rows: $\tau_i\succ\tau_{i+1}$ for $\tau$ but $\tau_{i+1}\succ\tau_i$ for $\tau'$.

In \Cref{alg:transposition_priority}, we explain how to deduce a $\tau'$-reduction $\widetilde{R}'$ of $\widetilde{D}$ from $\widetilde{R}$.
Only a few cases occur; they are solved by the updates below.
Crucially, each case requires at most one column operation.
This procedure is closely related to the vineyard algorithm \cite{CohenSteiner2006}; see~\Cref{rem:vineyard_algorithm}.%

\newcommand{\swapschematic}[6]{%
    \begin{pNiceArray}{ccccc}[first-row,first-col]
        & & #1 & & #2 & \\
            & & \vdots & & \vdots & \\
        \firstcol{\tau_{i+1}}
            & \cdots & #3 & \cdots & #4 & \cdots \\
        \firstcol{\tau_i}
            & \cdots & #5 & \cdots & #6 & \cdots \\
            & & 0 & & 0 &
    \end{pNiceArray}%
}

\begin{enumerate}
    \item 
    Suppose that $\tau_i$ and $\tau_{i+1}$ are $\tau$-owned in $\widetilde{R}$, i.e., that there exist indices $a,b$ such that
    \[
        \tau_i = \low\bigl(\widetilde{R},a,\tau\bigr),
        \qquad
        \tau_{i+1} = \low\bigl(\widetilde{R},b,\tau\bigr).
    \]
    In this case, define
    \[
        \alpha = \widetilde{R}[\tau_i,a]\neq0,    
        \qquad
        \beta = \widetilde{R}[\tau_{i+1},b]\neq0,
        \qquad
        x = \widetilde{R}[\tau_{i+1},a].
    \]
    \begin{enumerate}
        \item 
            \label{it:update_transposition_1a}
            If $x=0$, do nothing.
            Keep the matrix $\widetilde{R}' = \widetilde{R}$ and pivots $                \low\bigl(\widetilde{R}',\bullet,\tau'\bigr) = \low\bigl(\widetilde{R},\bullet,\tau\bigr)$.
            \[
                \swapschematic
                    {a}{b}
                    {0}{\boxed{\beta}}
                    {\boxed{\alpha}}{0}
                \quad\xrightarrow{~\text{Case~\ref{it:update_transposition_1a}}~}\quad
                \swapschematic
                    {a}{b}
                    {0}{\boxed{\beta}}
                    {\boxed{\alpha}}{0}.
                \]
        \item 
        \label{it:update_transposition_1b}
            If $x\neq 0$ and $a<b$, 
            perform the following column operation and swap the pivots:
            \begin{gather*}
                \col_b\gets\col_b-\frac{\beta}{x}\col_a,
                ~~~
                \low\bigl(\widetilde{R}',a,\tau'\bigr) = \low\bigl(\widetilde{R},b,\tau\bigr),
                ~~
                \low\bigl(\widetilde{R}',b,\tau'\bigr) = \low\bigl(\widetilde{R},a,\tau\bigr).%
                \\[1ex]
                \swapschematic
                    {a}{b}
                    {x}{\boxed{\beta}}
                    {\boxed{\alpha}}{0}
                \quad\xrightarrow{~\text{Case~\ref{it:update_transposition_1b}}~}\quad
                \swapschematic
                    {a}{b}
                    {\boxed{x}}{0}
                    {\alpha}{\boxed{-\frac{\alpha\beta}{x}}}.
            \end{gather*}
        \item 
            \label{it:update_transposition_1c}
            If $x\neq 0$ and $a>b$, 
            perform the following column operation and keep the pivots:
            \begin{gather*}
                \col_a\gets\col_a-\frac{x}{\beta}\col_b,
                \qquad
                \low\bigl(\widetilde{R}',\bullet,\tau'\bigr) = \low\bigl(\widetilde{R},\bullet,\tau\bigr).%
                \\[1ex]
                \swapschematic
                    {b}{a}
                    {\boxed{\beta}}{x}
                    {0}{\boxed{\alpha}}
                \quad\xrightarrow{~\text{Case~\ref{it:update_transposition_1c}}~}\quad
                \swapschematic
                    {b}{a}
                    {\boxed{\beta}}{0}
                    {0}{\boxed{\alpha}}.
            \end{gather*}
    \end{enumerate}
    \item
        If at least one of $\tau_i$ and $\tau_{i+1}$ is not $\tau$-owned,
        no column operation is needed, and $\widetilde{R}'=\widetilde{R}$.
        \begin{enumerate}
            \item 
            \label{it:update_transposition_2a}
            Suppose that $\tau_i$ is owned by $\col_a$, $\tau_{i+1}$ is unowned, and $x\neq 0$. 
            Then one pivot changes:
            \begin{gather*}
                \low\bigl(\widetilde{R}',a,\tau'\bigr)=\tau_{i+1}.
                \\[1ex]
                \begin{pNiceArray}{ccc}[first-row,first-col]
                    & \cdots & a & \cdots \\
                    &        & \vdots & \\
                    \firstcol{\tau_i}     & \cdots & \boxed{\alpha} & \cdots \\
                    \firstcol{\tau_{i+1}} & \cdots & x              & \cdots \\
                    &        & 0 & 
                \end{pNiceArray}
                \quad\xrightarrow{~\text{Case~\ref{it:update_transposition_2a}}~}\quad
                \begin{pNiceArray}{ccc}[first-row,first-col]
                    & \cdots & a & \cdots \\
                    &        & \vdots & \\
                    \firstcol{\tau_i}     & \cdots & \alpha        & \cdots \\
                    \firstcol{\tau_{i+1}} & \cdots & \boxed{x}     & \cdots \\
                    &        & 0 &
                \end{pNiceArray}.
            \end{gather*}
            \item
            \label{it:update_transposition_2b}
            In all other cases, the pivots are unchanged: $\low\bigl(\widetilde{R}',\bullet,\tau'\bigr) = \low\bigl(\widetilde{R},\bullet,\tau\bigr)$.
            For instance,
            \[
            \begin{pNiceArray}{ccc}[first-row,first-col]
                & \cdots & b & \cdots \\
                &        & \vdots & \\
                \firstcol{\tau_i}     & \cdots & 0              & \cdots \\
                \firstcol{\tau_{i+1}} & \cdots & \boxed{\beta}  & \cdots \\
                &        & 0 &
            \end{pNiceArray}
            \quad\xrightarrow{~\text{Case~\ref{it:update_transposition_2b}}~}\quad
            \begin{pNiceArray}{ccc}[first-row,first-col]
                & \cdots & b & \cdots \\
                &        & \vdots & \\
                \firstcol{\tau_i}     & \cdots & 0              & \cdots \\
                \firstcol{\tau_{i+1}} & \cdots & \boxed{\beta}  & \cdots \\
                &        & 0 &
            \end{pNiceArray}.
        \]
        \end{enumerate}
\end{enumerate}

\begin{example}
    Consider the matrices $\widetilde{D},\widetilde{R}_\mathrm{can},\widetilde{R}_\mathrm{piv}$ of \Cref{eq:example_exhaustive_reduction}.
    Note that we have
    \begin{alignat*}{2}
        \widetilde{R}_\mathrm{can}
        &= \Red\bigl(\widetilde{D},\tau\bigr),
        \qquad & \tau
        &= (3,2,1,\bot_4,\bot_3,\bot_2,\bot_1),\\
        \widetilde{R}_\mathrm{piv}
        &= \Red\bigl(\widetilde{D},\tau'\bigr),
        \qquad & \tau'
        &= (3,1,2,\bot_4,\bot_3,\bot_2,\bot_1).
    \end{alignat*}
    The permutations differ by an adjacent transposition, and Case~\ref{it:update_transposition_1c} takes $\widetilde{R}_\mathrm{can}$ to $\widetilde{R}_\mathrm{piv}$.
\end{example}

\subparagraph{Kendall distance.}

We now consider arbitrary permutations $\tau,\tau'$ of $\{\bot_1,\dots,\bot_m,1,\dots,n\}$.
The \emph{Kendall tau distance}, introduced in \cite{Kendall1938}, is equal to the number of discordant pairs $\{i,j\}$:%
\begin{equation}
    \label{eq:kendall_distance_discordant_pairs}
    K(\tau,\tau') 
    =
    \bigl|\bigl\{ \{p,q\} \subset \{\bot_1,\dots,\bot_m,1,\dots,n\} 
    \mid 
    p\neq q,\;
    p\prec_\tau q \iff p \succ_{\tau'} q
    \bigr\}\bigr|.
\end{equation}
Equivalently, $K(\tau,\tau')$ is equal to the minimum number of adjacent transpositions required to turn $\tau$ into $\tau'$.
We will use this quantity to measure the distance between permutations.
We point out that $K(\tau,\tau')$ is at most $(n+m)(n+m-1)/2$, attained when $\tau'$ is the reverse of $\tau$.%

\subparagraph{Full update.}

We finally consider a matrix $\widetilde{D}$, a $\tau$-reduction $\widetilde{R}$, and another permutation $\tau'$.
In \Cref{alg:full_correction_priority}, we propose a procedure to transform $\widetilde{R}$ into a $\tau'$-reduction~$\widetilde{R}'$.
It works by scanning the indices $i=1,\dots,n+m$ in order.
Whenever $\tau_i\neq\tau_i'$, perform the following~steps:

\begin{enumerate}
    \item
    Identify the index $j>i$ such that $\tau_j=\tau_i'$.
    \item
    In the current permutation $\tau$, move $\tau_j$ to the $i$th position by successive adjacent transpositions.
    At each transposition, apply \Cref{alg:transposition_priority} to update the matrix $\widetilde{R}$.
\end{enumerate}

The number of transpositions processed is equal to the Kendall distance.
We deduce:%

\begin{lemma}[Proof p.~\pageref{proof:full_correction_priority}]
    \label{lem:full_correction_priority}
    Given a $\tau$-reduction $\widetilde{R}$ of $\widetilde{D}$ and another row priority $\tau'$, \Cref{alg:full_correction_priority} outputs a $\tau'$-reduction $\widetilde{R}'$ of $\widetilde{D}$.
    It performs at most $K(\tau,\tau')$ column operations.
\end{lemma}

Note that a simpler alternative to \Cref{alg:full_correction_priority} is to compute a $\tau'$-reduction from scratch via \Cref{alg:persistence_algorithm_priority}.
However, this may require significantly more operations, as shown in~\Cref{rem:reduction_update_vs_from_scratch}.%
    
\subsection{Reduction algorithm at all exceptional primes}
\label{subsec:reduction_exceptional_primes}

From now on, we suppose that $\widetilde{D}$ has coefficients in $\Z$.
Given a field $\kk \in \{\Q\}\cup\{\F_p \mid p \text{ prime}\}$, we denote by $\widetilde{D}^\kk$ the matrix obtained by interpreting the entries of $\widetilde{D}$ in $\kk$.
Moreover, we denote by $\widetilde{R}^\kk$ a reduction of $\widetilde{D}^\kk$, for instance obtained by the persistence algorithm over $\kk$.

In \Cref{subsubsec:exceptional_primes}, we study how the field $\kk$ affects the pivot sequence of $\widetilde{R}^\kk$ and identify the exceptional primes.
In \Cref{subsubsec:repairing_generic_reduction}, we show how to recover these pivots using the priority updates of \Cref{subsubsec:efficient_updates_priority}.
These ideas are combined into our main algorithm in \Cref{subsubsec:full_algorithm}.%

\subsubsection{Exceptional primes and generic pivots}
\label{subsubsec:exceptional_primes}

\subparagraph{$\kk$-pivots.}

Let $\widetilde{R}^\kk$ be a reduction of $\widetilde{D}^\kk$.
Here, we only consider the usual notion of reduction (\Cref{subsubsec:classical_reduction}); priority-based reductions (as in \Cref{subsubsec:row_priority}) will be employed in \Cref{subsubsec:repairing_generic_reduction} and with $\Q$ coefficients only.
We are interested in the \emph{pivot index sequence} of $\widetilde{R}^\kk$, denoted~by%
\[
    \low(\kk,\bullet)
    =
    \bigl( \low(\kk,j) \bigr)_{j=1}^m,
    \qquad
    \low(\kk,j) = \low\bigl(\widetilde{R}^\kk,j\bigr).
\]
It is an $m$-tuple of distinct elements of $\{\bot_1,\dots,\bot_m,1,\dots,n\}$.
For certain primes $p$, $\low(\F_p,\bullet)$ may disagree with the \emph{generic sequence} $\low(\Q,\bullet)$.
We call them \emph{exceptional primes} and~denote%
\begin{equation}
    \label{eq_exceptional_primes}
    \PP 
    = \PP(D)
    = \{p \text{ prime} \mid \low(\F_p,\bullet) \neq \low(\Q,\bullet)\}.
\end{equation}
In this section, we describe a procedure to identify these primes, using only the reduction~$\widetilde{R}^\Q$.%

\begin{example}
    Consider the matrix $\widetilde{D}\in\Z^{(3+4)\times4}$ of \Cref{eq:example_reduction}.
    Its generic sequence is $\low(\Q,\bullet)=(3,1,\bot_3,2)$.
    Moreover, its exceptional primes are $\PP = \{2,3\}$; they correspond to $\low(\F_2,\bullet)=(1,3,\bot_3,2)$ and $\low(\F_3,\bullet)=(3,1,\bot_3,\bot_4)$, as the following calculations show.
    \[
        \begin{gathered}
            \widetilde{D}^{\F_2} \\[-2pt]
            \begin{pNiceArray}{cccc}
                1 & 0 & 0 & 0 \\
                0 & 1 & 0 & 0 \\
                0 & 0 & 1 & 0 \\
                0 & 0 & 0 & 1 \\
                \hline
                1 & 0 & 1 & 0 \\
                0 & 1 & 1 & 0 \\
                0 & 1 & 1 & 1 
            \end{pNiceArray}
        \end{gathered}
        \;\to\;
        \begin{gathered}
            ~~~\Red\bigl(\widetilde{D}^{\F_2}\bigr) \\[-2pt]
            \begin{pNiceArray}{cccc}[first-col]
                \firstcol{\bot_1}   & 1 & 0 & 1 & 0 \\
                \firstcol{\bot_2}   & 0 & 1 & 1 & 1 \\
                \firstcol{\bot_3}   & 0 & 0 & \mathclap{\boxed{1}} & 0 \\
                \firstcol{\bot_4}   & 0 & 0 & 0 & 1 \\\hline
                \firstcol{1}        & \mathclap{\boxed{1}} & 0 & 0 & 0 \\
                \firstcol{2}        & 0 & 1 & 0 & \mathclap{\boxed{1}} \\
                \firstcol{3}        & 0 & \mathclap{\boxed{1}} & 0 & 0 
            \end{pNiceArray},
        \end{gathered}
        \qquad\qquad
        \begin{gathered}
            \widetilde{D}^{\F_3} \\[-2pt]
            \begin{pNiceArray}{cccc}
                1 & 0 & 0 & 0 \\
                0 & 1 & 0 & 0 \\
                0 & 0 & 1 & 0 \\
                0 & 0 & 0 & 1 \\
                \hline
                1 & 0 & 1 & 0 \\
                2 & 1 & 1 & 1 \\
                2 & 1 & 1 & 1 
            \end{pNiceArray}
        \end{gathered}
        \;\to\;
        \begin{gathered}
            ~~~\Red\bigl(\widetilde{D}^{\F_3}\bigr) \\[-2pt] 
            \begin{pNiceArray}{cccc}[first-col]
                \firstcol{\bot_1}   & 1 & 1 & 2 & 0 \\
                \firstcol{\bot_2}   & 0 & 1 & 1 & 2 \\
                \firstcol{\bot_3}   & 0 & 0 & \mathclap{\boxed{1}} & 0 \\
                \firstcol{\bot_4}   & 0 & 0 & 0 & \mathclap{\boxed{1}} \\
                \hline
                \firstcol{1}        & 1 & \mathclap{\boxed{1}} & 0 & 0 \\
                \firstcol{2}        & 2 & 0 & 0 & 0 \\
                \firstcol{3}        & \mathclap{~\boxed{2}} & 0 & 0 & 0 
            \end{pNiceArray}.
        \end{gathered}
    \]
\end{example}

\subparagraph{Principal minors.}

For every $j\in\{1,\dots,m\}$, let the $j$th \emph{principal minor} of $\widetilde{D}$ be the product%
\begin{equation}
    \label{eq:principal_minor}
    \Delta_j = \piv(\Q,1)\cdots\piv(\Q,j).
\end{equation}
Since the pivots correspond to entries of $\widetilde{R}$, it can be written as a determinant (\Cref{lem:pivot_rank_profile}):
\begin{equation}
    \label{eq:principal_minor_det}
    \Delta_j = \det \widetilde{D}\big[\big(\low(\Q,1),\dots,\low(\Q,j)\big), 1\colon j\big].
\end{equation}
Thus, unlike the pivots, the $\Delta_j$ are integers.
Moreover, they determine the exceptional~primes.

\begin{lemma}[Proof p.~\pageref{proof:generic_reduction_identify_primes}]
    \label{lem:generic_reduction_identify_primes}
    For every prime $p$ and index $j\in\{1,\dots,m\}$, we have
    \[
        \bigl\{ \low(\F_p,1),\dots,\low(\F_p,j) \bigr\} = 
        \bigl\{ \low(\Q,1),\dots,\low(\Q,j) \bigr\}
        \iff
        p \text{ does not divide } \Delta_j.
    \]
    In particular, the exceptional primes are those dividing at least one of the principal minors:
    \[
        \PP = \{p \text{ prime} \mid p \text{ divides } \Delta_j \text{ for some } j\in\{1,\dots,m\}\}.
    \]    
\end{lemma}

\subparagraph{Quantitative bounds for boundary matrices.}

We deduce a bound on the cardinality of $\PP$.
It involves the shape $n\times m$ of the initial matrix $D$, its sup norm, and its rational rank,~denoted%
\[
    \|D\|_\infty = \max_{\substack{1\leq i \leq n \\ 1\leq j \leq m}} |D[i,j]|,
    \qquad
    r = \rank_\Q (D) \leq \min(n,m).
\]
In the case where $D$ is the boundary matrix of a simplicial filtration with $N$ simplices, its shape satisfies $n=m=N$, its norm is $\|D\|_\infty=1$, and its rational rank is at most~$(N-1)/2$.%

\begin{lemma}[Proof p.~\pageref{proof:number_exceptional_primes}]
    \label{lem:number_exceptional_primes}
    The exceptional primes of an arbitrary matrix $D\in\Z^{n\times m}$ satisfy
    \[
        |\PP|\leq r^2 \bigg( 1 + \frac{\log \|D\|_\infty }{\log(r+1)} \bigg).
    \]
    In particular, if $D$ is the boundary matrix of a simplicial filtration $K^\bullet$ with $N$ simplices, then 
    \[
        |\PP|\leq \left( \frac{N-1}{2} \right)^2.
    \]
\end{lemma}

\begin{remark}    
    In \Cref{sec:filtrations_with_large_torsion}, we build filtrations with many exceptional primes.
    In particular, \Cref{prop:number_exceptional_primes_newman} gives a family of 2-dimensional filtrations $K^\bullet_N$ such that $|\PP| = \Omega( N/\log N )$.
\end{remark}

\subsubsection{Repairing the generic reduction over an exceptional field}
\label{subsubsec:repairing_generic_reduction}

Fix a prime $p$.
We introduce our main routine: how to compute, from an exhaustive rational reduction $\widetilde{R}^\Q$ and its pivot sequence $\low(\Q,\bullet)$, the sequence $\low(\F_p,\bullet)$.
The idea is to simulate the $\F_p$-pivots in $\widetilde{R}^\Q$ via the notion of priority-reduction introduced in \Cref{subsubsec:row_priority}.

\subparagraph{Pivot priorities.}

We complete $\low(\Q,\bullet)$ and $\low(\F_p,\bullet)$ into full row priorities (\Cref{def:row_priority}), as was done in \Cref{eq:permutation_piv}, but we choose here convenient orders for the non-pivot~indices.%

\begin{definition}
    \label{def:permutation_piv_p}
    The \emph{$\Q$-pivot priority} is the following permutation of $\{\bot_1,\dots,\bot_m,1,\dots,n\}$:
    \[
        \tau_\mathrm{piv}^{\Q}
        =
        (\low(\Q,1),\dots,\low(\Q,m),\,q_1,\dots,q_n)
    \]
    where the non-pivot indices $q_1,\dots,q_n$ are in canonical order.
    Similarly, the \emph{$\F_p$-pivot priority}~is%
    \begin{equation*}
        \tau_\mathrm{piv}^{\F_p}
        =
        (\low(\F_p,1),\dots,\low(\F_p,m),\,q'_1,\dots,q'_n),
    \end{equation*}
    where the non-pivot indices $q'_1,\dots,q'_n$ are listed in their order in $\tau_\mathrm{piv}^{\Q}$.
\end{definition}

Now, consider a reduction of the $\Q$-coefficient matrix $\widetilde{D}^\Q$ with respect to this~priority:%
\[
    \widetilde{R}^\Q_{p\text{-}\mathrm{piv}} = \Red\left(\widetilde{D}^\Q,\tau_\mathrm{piv}^{\F_p}\right).
\]
Interestingly, the pivot indices of $\widetilde{R}^\Q_{p\text{-}\mathrm{piv}}$ under priority $\tau_\mathrm{piv}^{\F_p}$ are the first $m$ entries of $\tau_\mathrm{piv}^{\F_p}$:

\begin{lemma}[Proof p.~\pageref{proof:pivot_sequence_total}]
    \label{lem:pivot_sequence_total}
    For every $j\in\{1,\dots,m\}$,
    $
    \low\!\left(\widetilde{R}^\Q_{p\text{-}\mathrm{piv}},j,\tau_\mathrm{piv}^{\F_p}\right)
    = \low(\F_p,j).
    $
\end{lemma}

In other words, reducing $\widetilde{D}^\Q$ over $\Q$ with priority $\tau_\mathrm{piv}^{\F_p}$ recovers the canonical pivot sequence of the original reduction $\widetilde{R}^{\F_p}$ of $\widetilde{D}^{\F_p}$ over $\F_p$.
This phenomenon is illustrated in~\Cref{ex:pivot_sequence_total}.%

\subparagraph{Iterative correction routine.}

The $\F_p$-order $\tau_\mathrm{piv}^{\F_p}$ is not known a priori.
We unravel it from $\widetilde{R}^\Q$ column by column.
First, we intertwine $\tau_\mathrm{piv}^{\Q\vphantom{\F_p}}$ and $\tau_\mathrm{piv}^{\F_p}$ as follows.
We start from $\tau_0=\tau_\mathrm{piv}^{\Q\vphantom{\F_p}}$.
Then, for every $j\in\{1,\dots,m\}$, we obtain $\tau_j$ from $\tau_{j-1}$ by moving $\low(\F_p,j)$ to position $j$:
\begin{equation}
    \label{eq:intermediate_priority}
    \tau_j = \biggl( \low(\F_p,1),\, \dots,\, \low(\F_p,j),\, \underbrace{~~~~~~\dots_{\vphantom{p}}~~~~~~}_{\substack{
            \text{remaining entries of }\\
            \tau_\mathrm{piv}^{\Q}\text{ in original order}
        }}
    \biggr).
\end{equation}
Our routine iteratively builds matrices $\widetilde{R}_0^\Q,\widetilde{R}_1^\Q,\dots,\widetilde{R}_m^\Q$ with coefficients in $\Q$ that satisfy:%

\begingroup
\renewcommand{\theenumi}{H\arabic{enumi}}
\renewcommand{\labelenumi}{(\theenumi)}
\begin{enumerate}
    \item\label{hyp:1} 
        $\widetilde{R}^\Q_j$ is a $\tau_j$-reduction of $\widetilde{D}^\Q$.
    \item\label{hyp:2}
        The first $j$ values of $\low(\widetilde{R}^\Q_j,\bullet,\tau_j)$ and $\low(\F_p,\bullet)$ coincide.
    \item\label{hyp:3} 
        The entries of $\widetilde{R}^\Q_j[\colon\!,1\colon\!j]$ do not contain a denominator divisible by $p$.
\end{enumerate}
\endgroup

By \ref{hyp:1}, the input $\widetilde{R}^\Q_0$ must be an \emph{exhaustive} canonical reduction, as in \Cref{lem:exhaustive_reduction}.
Moreover, by \ref{hyp:2}, the output $\widetilde{R}^\Q_m$ has the required $\F_p$-pivots, as in \Cref{lem:pivot_sequence_total}.
Last, \ref{hyp:3} implies that the first $j$ columns of $\widetilde{R}^\Q_j$ can be reached by safe mod-$p$ operations (no division by $p$ is involved).

The important observation is that the next $\F_p$-pivot can be read from $\widetilde{R}^\Q_j$ taken modulo~$p$.%

\begin{lemma}[Proof p.~\pageref{proof:pivot_sequence_prefix}]
    \label{lem:pivot_sequence_prefix}
    If \ref{hyp:1}--\ref{hyp:3} hold, the denominators of entries in $\widetilde{R}^\Q_j[\colon\!,1\colon\!(j+1)]$ are not divisible by $p$, and the lowest nonzero entry of $\widetilde{R}^\Q_j[:,j+1] \mod{p}$ has index $\low(\F_p,j+1)$.
\end{lemma}

Thus, from $\widetilde{R}^\Q_j$, we can identify $\low(\F_p,j+1)$, and therefore $\tau_{j+1}$.
Since $\tau_{j+1}$ differs from $\tau_j$ by a few adjacent transpositions, $\widetilde{R}_{j+1}$ can be obtained efficiently from $\widetilde{R}_j$ via \Cref{alg:full_correction_priority} of \Cref{subsubsec:efficient_updates_priority}; see \Cref{lem:full_correction_priority}.
The resulting procedure is presented in \Cref{alg:full_correction_modp}.

\begin{lemma}[Proof p.~\pageref{proof:full_correction_modp}]
    \label{lem:full_correction_modp}
    From an exhaustive reduction $\widetilde{R}^\Q$ of $\widetilde{D}$, \Cref{alg:full_correction_modp} outputs a $\tau_\mathrm{piv}^{\F_p}$-reduction whose pivots are $\low(\F_p,\bullet)$.
    It performs at most $K\bigl(\tau_\mathrm{piv}^{\Q\vphantom{\F_p}},\tau_\mathrm{piv}^{\F_p}\bigr)$ column operations.
\end{lemma}

\begin{example}
    We show the first step of \Cref{alg:full_correction_modp} for $p=2$, starting from the exhaustive reduction $\widetilde{R}^\Q=\widetilde{R}_\mathrm{piv}$ of $\widetilde{D}$ in \Cref{eq:example_exhaustive_reduction}.
    The pivots are boxed for the current order $\tau_i$.
    \begin{equation*}
        \begin{gathered}
            ~~~~\widetilde{R}_0^\Q=\widetilde{R}_{\vphantom{0}}^\Q \\[-2pt]
            \begin{pNiceArray}{cccc}[first-col]
                \firstcol{\bot_1}   & 1 & \mathllap{-}\tfrac12 & \mathllap{-}1 & 0 \\
                \firstcol{\bot_2}   & 0 & 1 & 1 & \mathllap{-}1 \\
                \firstcol{\bot_3}   & 0 & 0 & \boxed{1} & 0 \\
                \firstcol{\bot_4}   & 0 & 0 & 0 & 1 \\\hline
                \firstcol{1}   & 1 & \mathclap{\boxed{-\tfrac12}} & 0 & 0 \\
                \firstcol{2}   & 2 & 0 & 0 & \boxed{3} \\
                \firstcol{3}   & \boxed{2} & 0 & 0 & 0
            \end{pNiceArray}
            \\[2pt]
            ~~~~~\tau_0 \!=\! (3,1,\bot_3,2,\dots)
        \end{gathered}
       \quad\xrightarrow[]{~\col_2\;\gets\;\col_2+\frac12\col_1~}\quad
        \begin{gathered}
            ~~~~\widetilde{R}_1^\Q \\[-2pt]
            \begin{pNiceArray}{cccc}[first-col]
                \firstcol{\bot_1}   & 1 & 0 & \mathllap{-}1 & 0 \\
                \firstcol{\bot_2}   & 0 & 1 & 1 & \mathllap{-}1 \\
                \firstcol{\bot_3}   & 0 & 0 & \boxed{1} & 0 \\
                \firstcol{\bot_4}   & 0 & 0 & 0 & 1 \\\hline
                \firstcol{1}   & \boxed{1} & 0 & 0 & 0 \\
                \firstcol{2}   & 2 & 1 & 0 & \boxed{3} \\
                \firstcol{3}   & 2 & \boxed{1} & 0 & 0
            \end{pNiceArray}.
            \\[2pt]
            ~~~~~\tau_1 \!=\! (1,3,\bot_3,2,\dots)
        \end{gathered}
     \end{equation*}
\end{example}

\subsubsection{Complete algorithm and the field-independence problem}
\label{subsubsec:full_algorithm}

\subparagraph{Description of the algorithm.}

Given an integral $n\times m$ matrix $D$ and its augmentation $\widetilde{D}$, we aim to compute the pivot sequences $\low(\F_p,\bullet)$ for all primes $p$.
It suffices to store $\low(\Q,\bullet)$ and the sequences at exceptional primes $p\in\PP$.
The algorithm works in four steps.

\begin{enumerate}
    \item 
        \label{it:main_algorithm_step1}
        Compute a reduction $\widetilde{R}_\mathrm{can}^\Q$ over $\Q$ with the usual algorithm; save $\piv(\Q,\bullet)$ and $\low(\Q,\bullet)$.
    \item
        \label{it:main_algorithm_step2}
        Identify the exceptional primes as divisors of the minors $\Delta_i=\piv(\Q,1)\cdots\piv(\Q,i)$.
    \item 
        \label{it:main_algorithm_step3}
        Deduce an exhaustive reduction $\widetilde{R}_\mathrm{piv}^\Q$ over $\Q$ by priority-based reduction (\Cref{alg:persistence_algorithm_priority}).
    \item
        \label{it:main_algorithm_step4}
        For each exceptional prime $p$:\\
        \hspace*{1.5em}compute the sequence $\low(\F_p,\bullet)$ from $\widetilde{R}_\mathrm{piv}^\Q$ with \Cref{alg:full_correction_modp}.
\end{enumerate}

\noindent
The full procedure is written under \Cref{alg:main_algorithm}.
See \Cref{rem:implementation_details} for implementation details.%

\subparagraph{Amortized analysis.}

Our algorithm calls \Cref{alg:full_correction_modp} at each exceptional prime.
There are at most quadratically many such primes (\Cref{lem:number_exceptional_primes}), and each call requires at most $K(\tau_\mathrm{piv}^{\Q\vphantom{\F_p}},\tau_\mathrm{piv}^{\F_p})$ column operations (\Cref{lem:full_correction_modp}), which is quadratic in the worst case.
This yields a quartic upper bound on the total number of column operations.
It can be substantially~improved.

To control the number of operations, we consider the maximal norm of the columns of $D$:%
\[
    \|D\|_{2,\infty} = \max_{\substack{1\leq j \leq m}} \|D[\colon,j]\|.
\]
Note that if $D\in\Z^{N\times N}$ is the boundary matrix of a simplicial filtration $(K^t)_{t=0}^N$, then each of its columns has at most $\ell+1$ nonzero entries, where $\ell=\dim K^N$.
Moreover, its entries are $0$, $-1$, or $+1$, by definition of the integral boundary map.
We deduce that $\|D\|_{2,\infty}=\sqrt{\ell+1}$.%

\begin{lemma}[Proof p.~\pageref{proof:full_correction_amortized}]
    \label{lem:full_correction_amortized}
    For $D\in\Z^{n\times m}$, the sum of exceptional Kendall distances~satisfies%
    \[
        \sum_{p\in\PP} K\bigl(\tau_\mathrm{piv}^{\Q\vphantom{\F_p}},\tau_\mathrm{piv}^{\F_p}\bigr)
        \leq 2(m+n)^2 \log \|D\|_{2,\infty}.
    \]    
\end{lemma}

We deduce a bound on the complexity of \Cref{alg:main_algorithm}.
By \emph{arithmetic operation}, we mean any operation $+$, $-$, $\times$, $\div$ in $\Q$ or $\F_p$, computing gcd, factorizing integers, or testing~divisibility.%

\begin{theorem}[Proof p.~\pageref{proof:main_algorithm}]
    \label{th:main_algorithm}
    \Cref{alg:main_algorithm} takes as an input a matrix $D\in\Z^{n\times m}$ and outputs its exceptional primes $\PP$ and the pivot indices $\low(\Q,\bullet)$ and $\low(\F_p,\bullet)$ for all $p\in\PP$.
    It uses%
    \[
        O\big((m+n)^3 \max(1,\log\|D\|_{2,\infty}) \big)
    \]
    arithmetic operations.
    If $D$ is the boundary matrix of a simplex-wise filtration $(K^t)_{t=0}^N$, we~get%
    \[
        O\big( N^3\log(\ell+1) \big),
        \qquad
        \ell=\dim K^N.
    \]
\end{theorem}

\subparagraph{Barcodes.}

Let $D$ be the boundary matrix of $(K^t)_{t=0}^N$, $\widetilde D$ its augmentation, and~$\widetilde{R} = \left(\begin{smallmatrix}V \\ R \end{smallmatrix}\right)$ a reduction.
A simplex $\sigma_i$ creates a homology class when the $i$th column of the unaugmented reduction $R$ is zero.
That is, when $\low(\F_p,i) = \bot_i$.
The corresponding bar $[i,d_i)$ dies at a finite time or persists indefinitely, depending on whether another column has pivot index $i$:
\[
    d_i=
    \begin{cases}
    j & \text{if } \low(\F_p,j)=i\text{ for some }j\in\{1,\dots,N\},\\
    +\infty & \text{otherwise}.
    \end{cases}
\]
The persistence barcode is then given by
\begin{equation}
    \label{eq:barcode_from_pivots}
    \barcode\big(H_d(K^\bullet;\F_p)\big)
    =
    \bigl\{\!\bigl\{
        [i,d_i)
        \mid i\in\{1,\dots,N\},\;
        \dim(\sigma_i)=d,\;
        \low(\F_p,i)=\bot_i
    \bigr\}\!\bigr\}.
\end{equation}

For a filtration $(K^t)_{t=0}^T$ that is not simplex-wise, we first refine it into a simplex-wise filtration $(\overline{K^t})_{t=0}^N$, where $N$ is the number of simplices, by arbitrarily breaking ties between simplices with the same filtration value. 
The original barcode is then recovered from \Cref{eq:barcode_from_pivots} by restoring the original filtration values and discarding empty intervals.

\subparagraph{Field independence.}

Consider an arbitrary simplicial filtration $K^\bullet$.
The field-independence problem, mentioned in \Cref{subsec:intro_persistence}, asks whether $\barcode (H_d(K^\bullet;\kk))$ is independent of the field $\kk$.
The considerations above suggest a solution, which we give in \Cref{alg:field_independence_problem}:

\begin{enumerate}
    \item 
        Refine the filtration, compute the barcodes, and restore the initial filtration times.
    \item
        Check whether all exceptional $\F_p$-barcodes coincide with the $\Q$-barcode.
\end{enumerate}

\noindent
Since testing the equality of two barcodes of size $N$ requires $O(N\log N)$ time, we get:

\begin{theorem}[Proof p.~\pageref{proof:field_independence_problem}]
    \label{th:field_independence_problem}
    \Cref{alg:field_independence_problem} takes as an input a simplicial filtration $(K^t)_{t=0}^T$, potentially not simplex-wise, and decides if the $\barcode(H_d(K^\bullet;\F_p))$, $p$ prime, all coincide.
    Its complexity, in terms of the number of simplices $N$ and the dimension $\ell=\dim K^T$, is%
    \[
        O\big(N^3 \log (N) \log(\ell+1) \big).
    \]
\end{theorem}

\section{Numerical experiments}
\label{sec:numerical_experiments}

We benchmark our implementation in \Cref{subsec:benchmark} and study random complexes in~\Cref{subsec:torsion_random}.%

\subsection{Benchmark on the Linial--Meshulam process}
\label{subsec:benchmark}

\subparagraph{Dataset.}

As in \cite{Boissonnat_2019,ObayashiYoshiwaki2023}, we benchmark our algorithm on the 2-dimensional \emph{Linial--Meshulam process}.
We start from the complete $1$-skeleton on the vertex set $\{1,\dots,n\}$ and add triangles $\sigma_1,\sigma_2,\dots$ in uniformly random order.
Instead of adding all $\binom{n}{3}$ triangles, we take the first 
\[
    \left\lfloor \frac{2c}{n}\binom{n}{3} \right\rfloor,
    \quad c = 2.7538.
\]
Thus, the full filtration contains $N =n + \binom{n}{2} + \lfloor \frac{2c}{n} \binom{n}{3} \rfloor$ simplices.
We mention that $\lfloor \frac{c}{n}\binom{n}{3} \rfloor$ (without a 2) is the cutoff near which numerical experiments observed a \emph{torsion burst}:~the brief appearance of a torsion subgroup of large order in $H_1(K^t;\Z)$.
Thus, we expect the filtration to show many exceptional primes $\PP$.
We refer to \cite{LinialPeled2016,KahleLutzNewmanParsons2020} for background on this~phenomenon.%

The algorithms described below expect simplex-wise filtrations.
Thus, instead of starting from the full 1-skeleton, we insert vertices and edges successively, in lexicographic order.
This refinement may introduce exceptional primes that are not intrinsic to the original~filtration.%

\subparagraph{Our implementation.}

We implemented \Cref{alg:main_algorithm} in \texttt{C++}.
The code follows the pseudocode with two additional optimizations:
\textbf{(1)} in the augmented matrix $\widetilde{R} = \widetilde{D}V = \left(\begin{smallmatrix} V \\ R \end{smallmatrix}\right)$ (\Cref{eq:augmented_matrix}), the upper block $V$ is represented lazily; and
\textbf{(2)} each column of the lower block $R$ is stored sparsely, using integer numerators and a common column denominator.

\subparagraph{Competitors.}

We compare our method with the three approaches described in \Cref{subsec:intro_algorithms}.
We use the \texttt{C++} implementation of \texttt{GUDHI} \cite{gudhi:PersistentCohomology} for brute force and modular reconstruction, and \texttt{Dionysus2} \cite{morozov_dionysus2} for omni-field persistence.
For modular reconstruction, we patched the \texttt{GUDHI} implementation to accept a list of primes rather than an interval.
To facilitate comparison, we decompose the running time of each method into four phases, summarized in~\Cref{tab:benchmark_phases}.

\begin{table}[htbp]
    \centering
    \small
    \setlength{\tabcolsep}{3pt}
    \renewcommand{\arraystretch}{1.3}
    \begin{tabular}{
            >{\raggedright\arraybackslash}p{0.175\linewidth}
            >{\raggedright\arraybackslash}p{0.18\linewidth}
            >{\raggedright\arraybackslash}p{0.245\linewidth}
            >{\raggedright\arraybackslash}p{0.165\linewidth}
            >{\raggedright\arraybackslash}p{0.14\linewidth}
        }
        \toprule
        {%
            \ifdefined\socgtabular
              \let\tabular\socgtabular
              \let\endtabular\endsocgtabular
            \fi
            \diagbox[
                innerwidth=\linewidth,
                height=2.5\baselineskip
            ]{Phase}{Method}%
        }
        & Brute force & Modular reconstruction & Omni-field & Our method \\
        \midrule
        Prime discovery
        & Steps \ref{it:main_algorithm_step1}--\ref{it:main_algorithm_step2}
        & Steps \ref{it:main_algorithm_step1}--\ref{it:main_algorithm_step2}
        & ---
        & Steps \ref{it:main_algorithm_step1}--\ref{it:main_algorithm_step2}
        \\
        Initialization
        & Simplex tree
        & Simplex tree and \newline partial inverse masks
        & Filtration
        & Step \ref{it:main_algorithm_step3}
        \\
        Main phase
        & Reduction over each $\F_p$, $p\in\PP$
        & Reduction
        & Reduction
        & Step \ref{it:main_algorithm_step4}
        \\
        Barcode output
        & ---
        & Decode pivots
        & Decode columns
        & Decode pivots
        \\
        \bottomrule
    \end{tabular}
    \caption{
        To measure running times, we divide each method into four phases.
        For the first two, which lack a native procedure for detecting exceptional primes, we use Steps \ref{it:main_algorithm_step1}--\ref{it:main_algorithm_step2} of our~\Cref{alg:main_algorithm}.%
    }
    \label{tab:benchmark_phases}
\end{table}

\subparagraph{First experiment.}

\begin{figure}[htbp]
    \centering
    \includegraphics[width=1\linewidth]{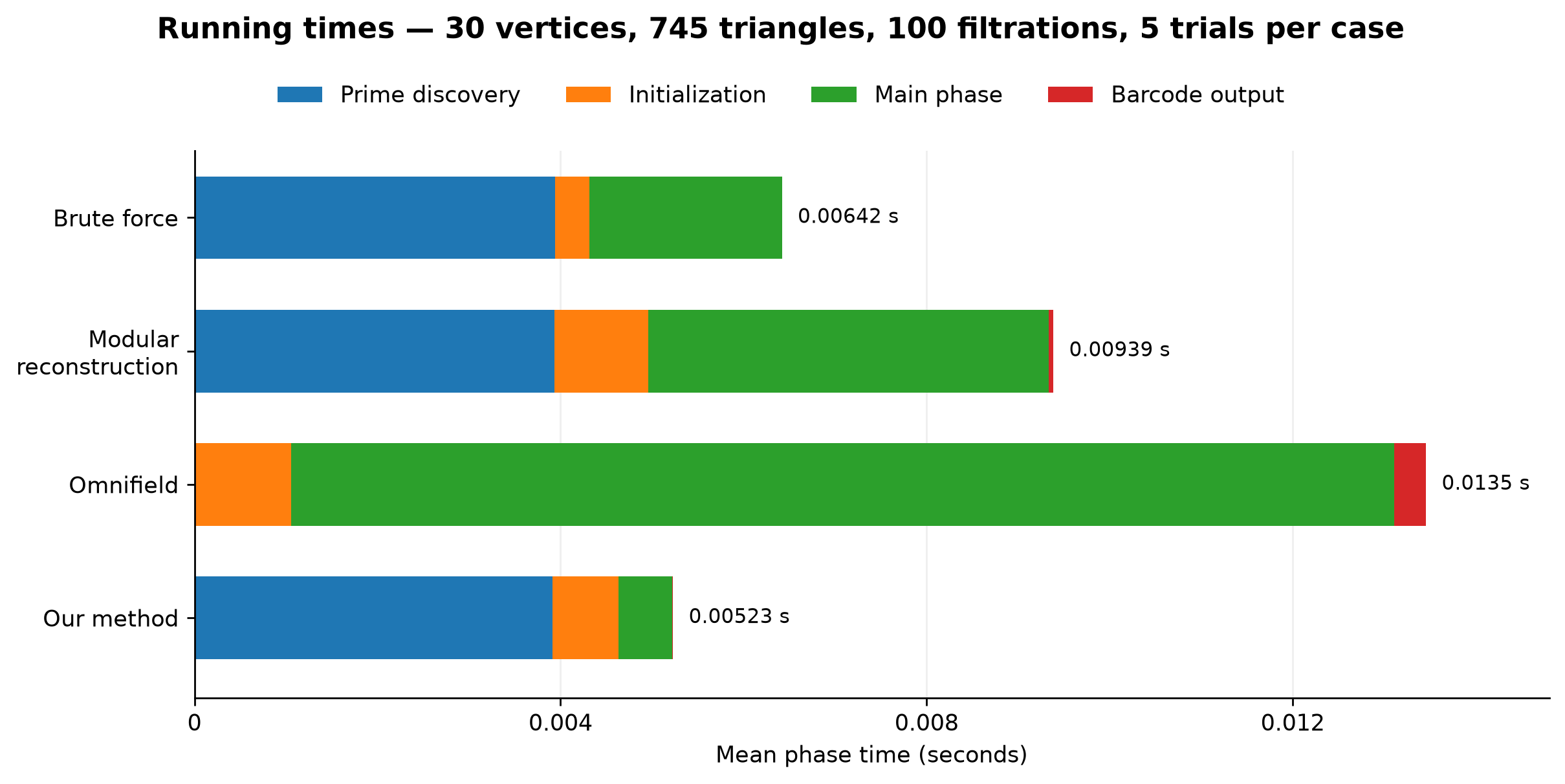}
    \\\vspace{.05cm}
    \includegraphics[width=1\linewidth]{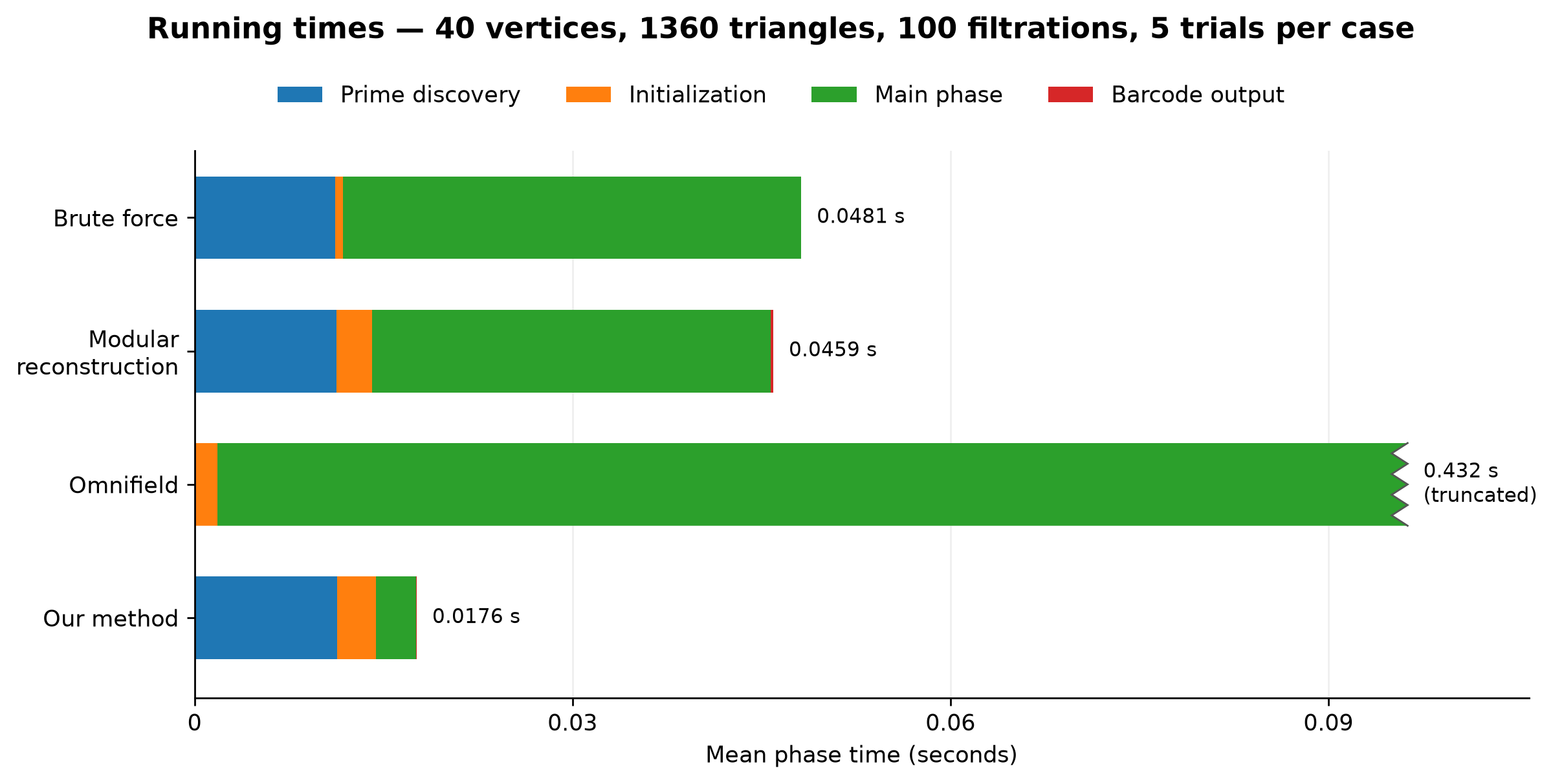}
    \\\vspace{.05cm}
    \includegraphics[width=1\linewidth]{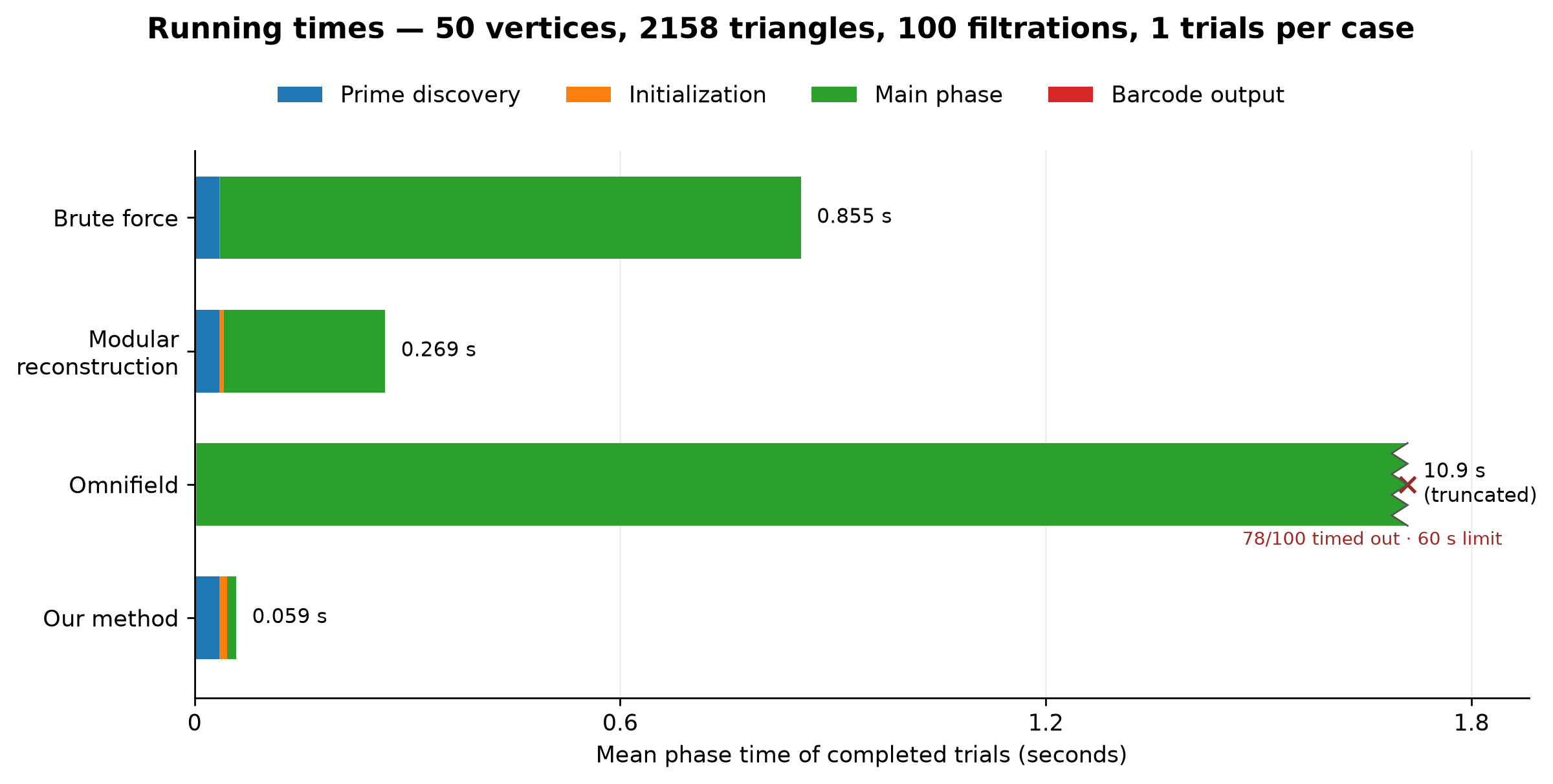}
    \caption{
        Average running time for all methods and $n\in\{30,40,50\}$, split by phase (as in \Cref{tab:benchmark_phases}).
    }
    \label{fig:numerical_experiments_phases}
\end{figure}

\begin{figure}[htbp]
    \centering
    \includegraphics[width=.98\linewidth]{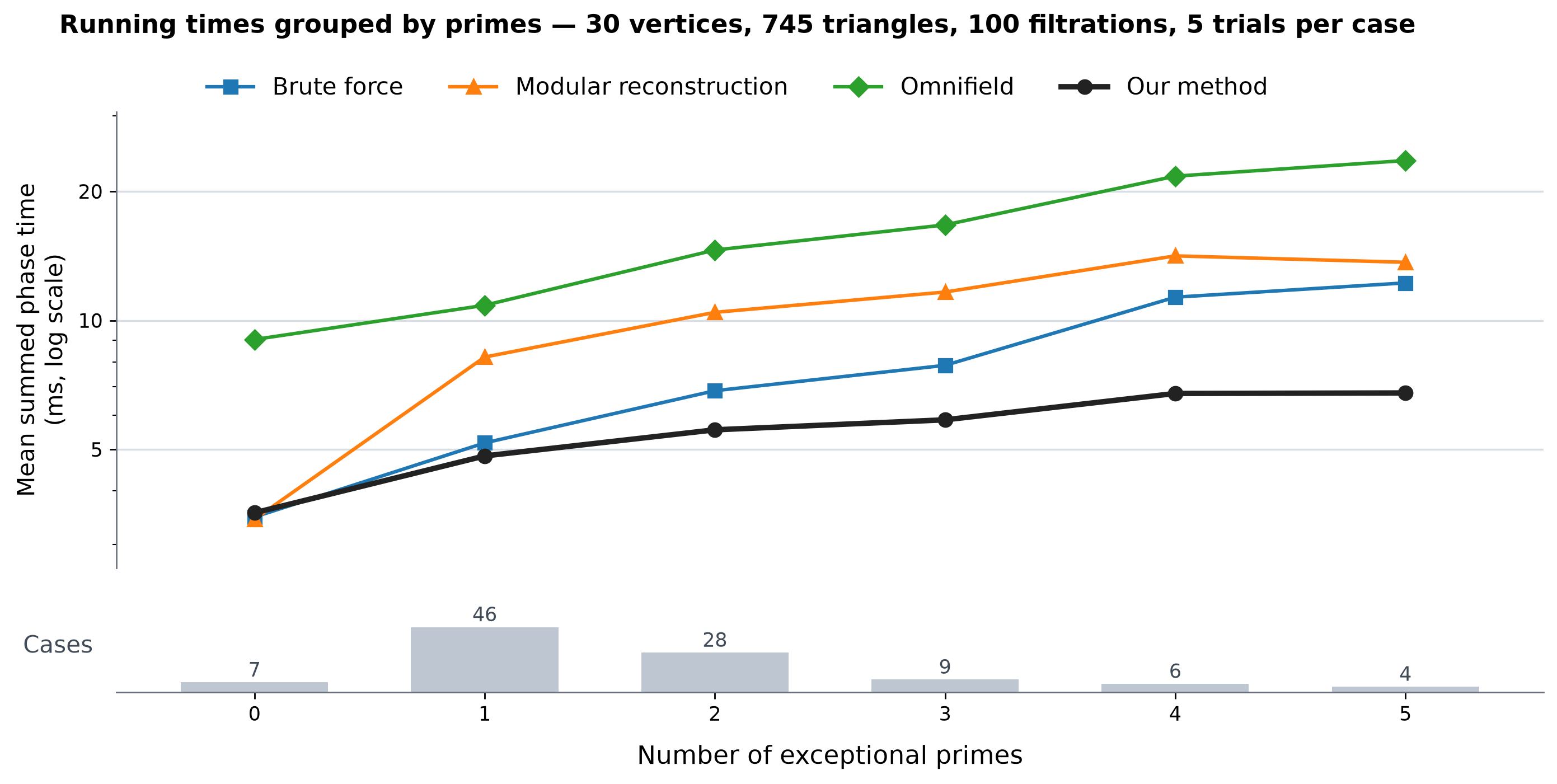}
    \includegraphics[width=.98\linewidth]{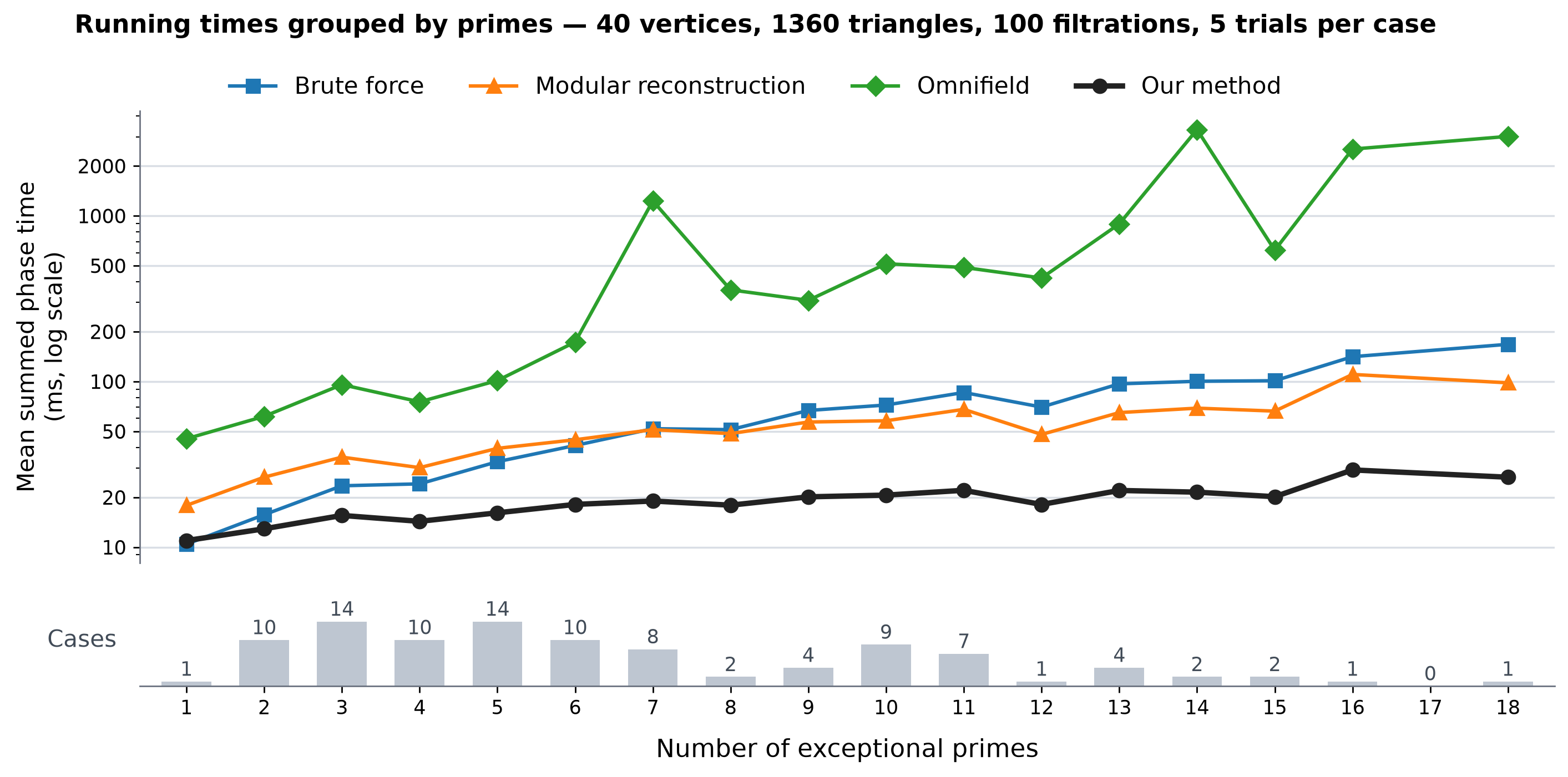}
    \includegraphics[width=.98\linewidth]{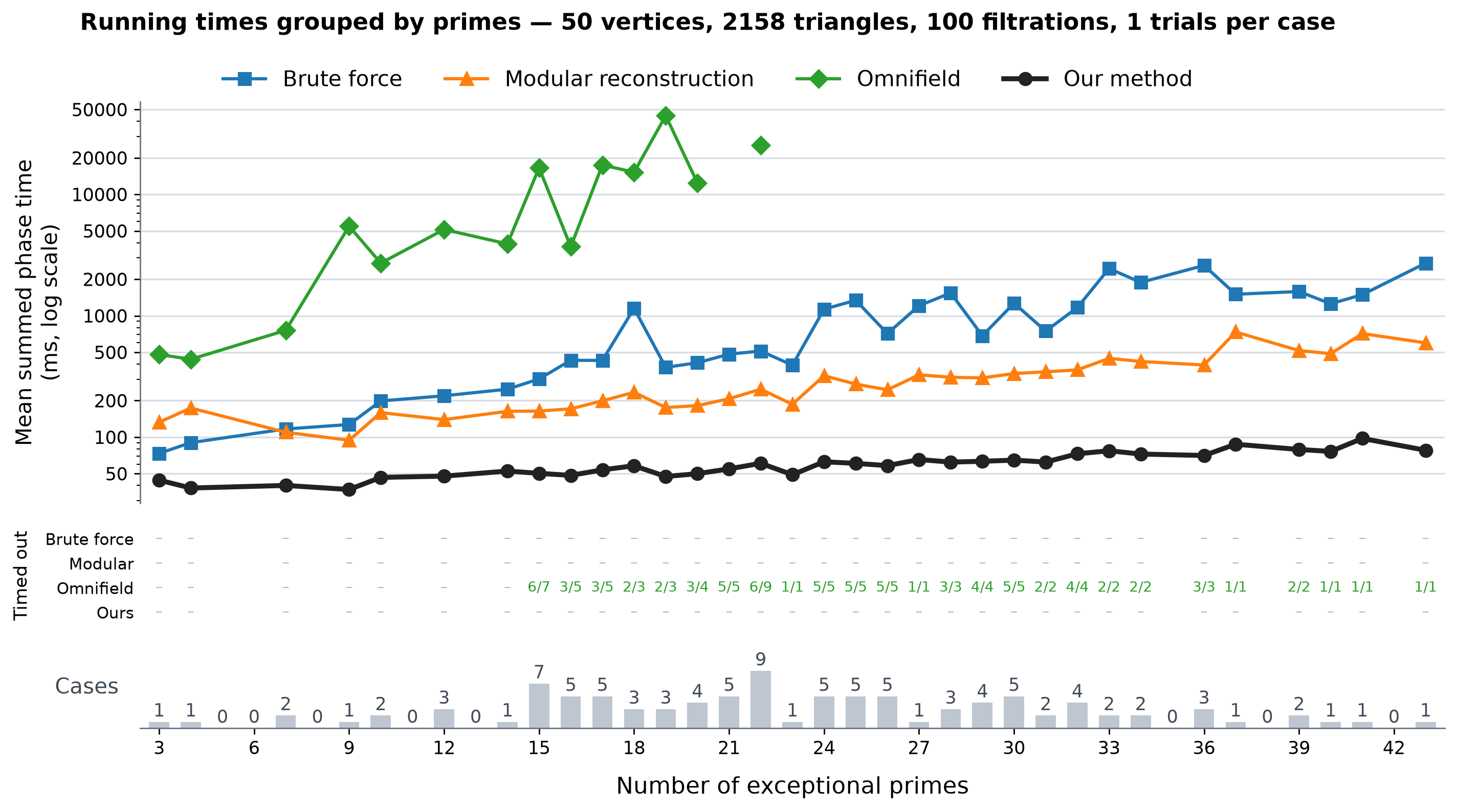}
    \caption{
        Average running time grouped by number of vertices $n$, with a time limit of $60$~s.
    }
    \label{fig:numerical_experiments_primes}
\end{figure}

For each $n\in\{30,40,50\}$, we sample 100 Linial--Meshulam filtrations and run all methods, averaging up to 5 trials per filtration to reduce timing variability.
We discard and replace filtrations with exceptional primes above $46337$, which \texttt{GUDHI} does not support.
The timings are broken down by phase in \Cref{fig:numerical_experiments_phases} and by the number of primes~in~\Cref{fig:numerical_experiments_primes}.

Our method outperforms the others in all cases with exceptional primes.
When none are present, it coincides with brute force and modular reconstruction.
The comparison with these two methods is particularly interesting for $n\in\{40,50\}$, where their main phase, rather than prime discovery, accounts for most of the running time.
Our main phase (Step~\ref{it:main_algorithm_step4}) is substantially faster because it reuses computations from the previous phases (Steps~\ref{it:main_algorithm_step1}--\ref{it:main_algorithm_step3}).

For example, the second sampled filtration with $n=40$ has 13 exceptional primes.
Our main phase corrects only 779 triangle columns, whereas brute force processes 17,680 triangle columns across its 13 field computations, and modular reconstruction processes 1,360 triangle columns using large integers.
Omnifield identifies a superset of the exceptional primes, containing 209 primes in this example, which increases the cost of subsequent computations.

\subparagraph{Second experiment.}

\begin{figure}[htbp]
    \centering
    \includegraphics[width=1\linewidth]{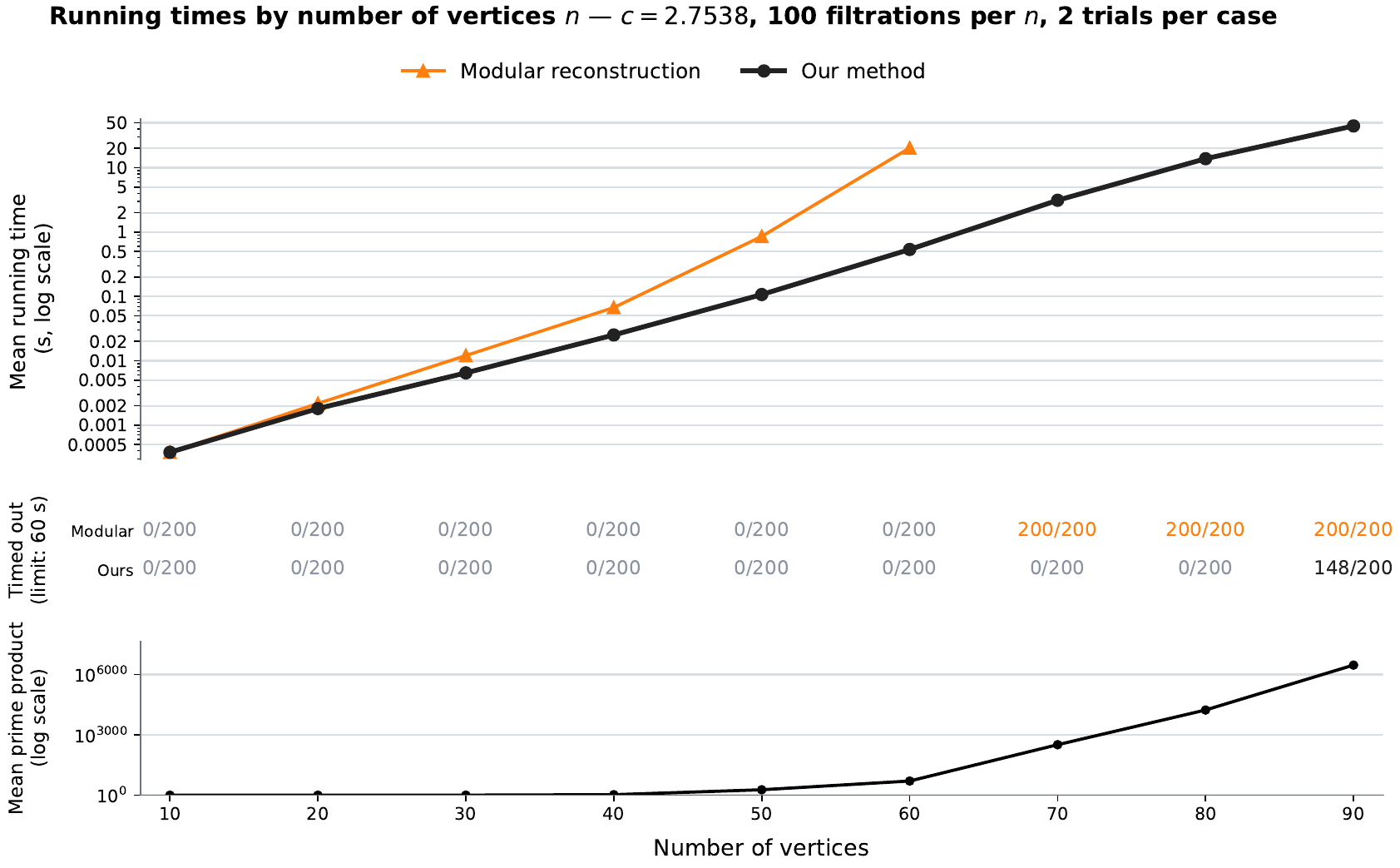}
    \caption{
        Average running time and product of exceptional primes found, with a time limit of~$60$~s.
    }
    \label{fig:numerical_experiments_vs_modular}
\end{figure}

The previous experiment cannot be extended beyond $n=50$: most filtrations have exceptional primes larger than \texttt{GUDHI}'s limit and omni-field becomes too slow.
We therefore compare our method with modular reconstruction alone for $n\in\{10,20,\dots,90\}$.
In particular, the last complexes have $11,\!284$ simplices.
The results are presented in \Cref{fig:numerical_experiments_vs_modular}.

On all $100$ filtrations tested at $n=70$, modular reconstruction exceeds the time limit of $60$ seconds.
This is because these instances have many exceptional primes, with products exceeding $2510$ digits.
One reached the prime $p = 88823020457549174488262651$.
Besides, our algorithm handled all filtrations up to $n=80$, taking on average $14$~seconds at that~size.%

\subsection{Torsion in random Rips complexes}
\label{subsec:torsion_random}

\subparagraph{Dataset.}

We extend Obayashi and Yoshiwaki's experiments on random Rips filtrations \cite{ObayashiYoshiwaki2023}. 
For each trial, we sample $1000$ random points $X\subset\R^n$ uniformly in $[0,1]^n$.
Its filtration is
\[
    K^\bullet = (K^t)_{t\geq 0},
    \qquad
    K^t =\operatorname{Rips}^t(X)
    = \big\{\sigma\subset X \mid \operatorname{diam}(\sigma)\leq t\big\}.
\]
We are interested in the degree-1 homology module $H_1(K^\bullet;\F_p)$ only, hence we restrict the complexes to their $2$-skeleton.
Moreover, instead of inserting all edges, we only insert the~first
\[
    \left\lfloor c \binom{1000}{2} \right\rfloor,
    \qquad
    c = 1/10.
\]
We test whether these filtrations exhibit exceptional primes, in the sense of~\Cref{eq:definition_exceptional_primes}.

The Rips filtration is not simplex-wise: an edge and several cofaces may enter at the same time.
We obtain a simplex-wise refinement by breaking ties by lexicographic order.
In our experiments, we compute the exceptional primes for both the Rips filtration and this refinement.
Note that refinement preserves existing exceptional primes but may introduce new ones.
This comparison addresses a question raised by Luo and Henselman-Petrusek in~\cite{LuoHenselman-Petrusek2025,LuoHenselman-Petrusek2026}: how often does simplex-wise refinement introduce spurious exceptional primes?


\subparagraph{First experiment.}

\begin{figure}[htbp]
    \centering
    \includegraphics[width=1\linewidth]{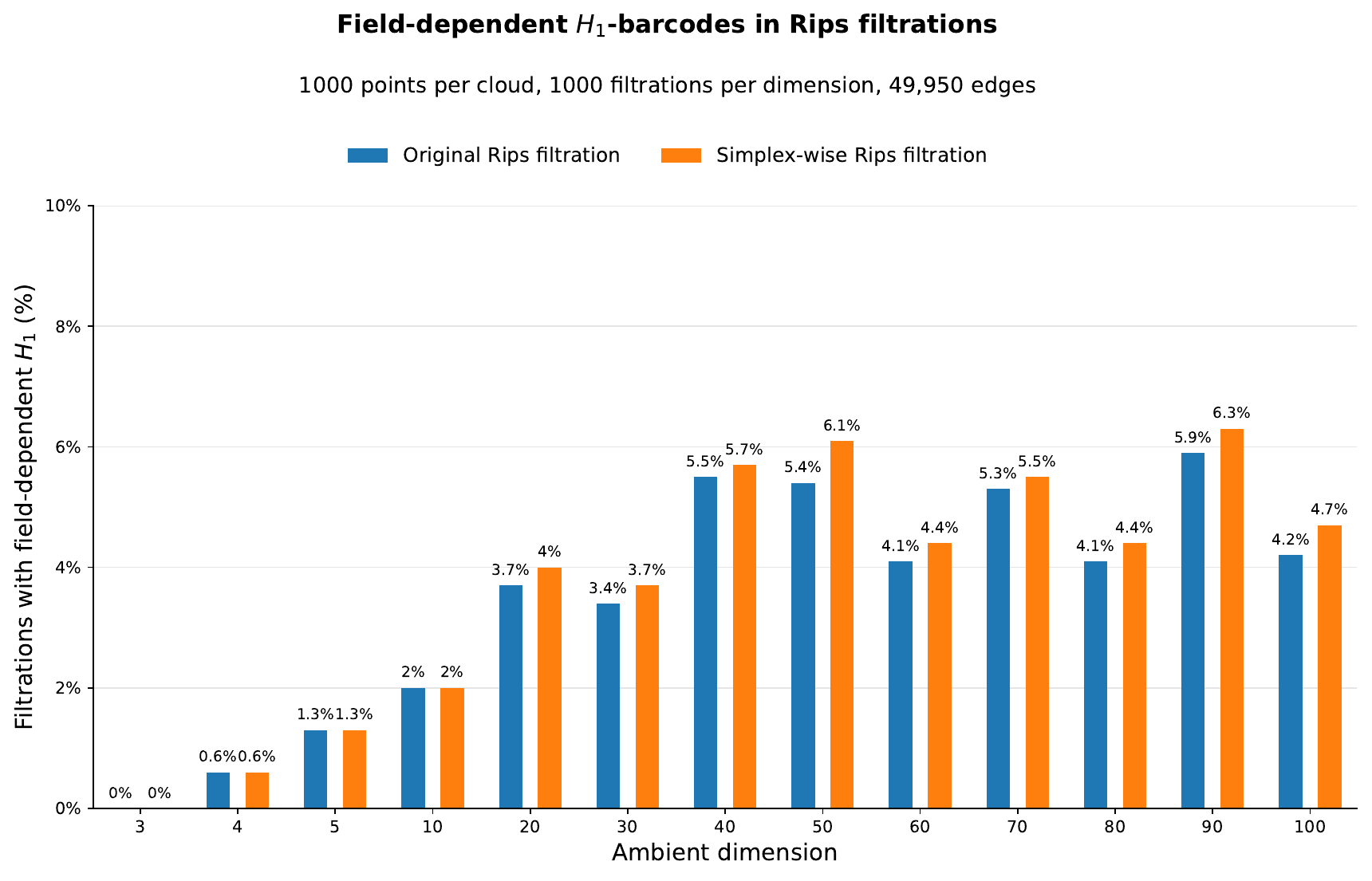}
    \caption{
        Frequency of random Rips filtrations in $\R^n$ with $\barcode(H_1(K^\bullet;\F_p))$ depending on $p$.
    }
    \label{fig:random_rips_simplexwise}
\end{figure}

We generate $1000$ clouds $X\subset\R^n$ for each $n\in\{3,4,5\}\cup\{10,20,\dots,100\}$ and compute the exceptional primes for both the original filtration and its simplex-wise refinement.
The results are summarized in \Cref{fig:random_rips_simplexwise}.

The proportion of filtrations with exceptional primes appears to increase up to ambient dimension $50$.
This is consistent with the trend observed in \cite{ObayashiYoshiwaki2023}.
However, beyond this dimension, it appears to stabilize.
Overall, field dependence affects $3.50\%$ of the original Rips filtrations and $3.75\%$ of their simplex-wise refinements.
Thus, our experiment shows that field dependence is not merely an artifact of simplex-wise refinement.%

\subparagraph{Map torsion.}

Exceptional primes witness the presence of \emph{homological torsion} in the persistence module.
For example, a Klein bottle $K$ has integral homology $H_1(K;\Z)\cong\Z\oplus\Z/2\Z$, yielding different Betti numbers over different fields: $H_1(K;\F_2)\cong\F_2^2$ but $H_1(K;\F_3)\cong\F_3$.

However, this is not the only reason why $\barcode(H_d(K^\bullet;\F_p))$ may depend on $p$.
In \Cref{sec:torsion_on_barcodes}, we distinguish three torsion phenomena, called map, tensor, and Tor torsion.
The Klein bottle is an example of tensor torsion, coming from the component $\Z/2\Z$ in $H_1(K;\Z)$.

Map torsion arises when the Betti numbers $\dim H_d(K^s;\F_p)$ do not depend on $p$ yet the maps $H_d(K^s;\F_p)\to H_d(K^t;\F_p)$ have rank that depends on $p$.
Equivalently, $H_d(K^s;\Z)\to H_d(K^t;\Z)$ has torsion in its cokernel.
To isolate map torsion, we mod out pointwise torsion:
\[
    H_d^{\mathrm{free}}(K^t;\Z)
    =
    H_d(K^t;\Z)/H_d^{\mathrm{tors}}(K^t;\Z).
\]
Tensoring these groups and their induced maps with $\F_p$ yields
the \emph{free module} of \Cref{eq:persistence_module_free}:
\begin{equation*}
    H_d^\mathrm{free}(K^\bullet;\Z)\otimes\F_p\colon\quad
    H_d^\mathrm{free}(K^0;\Z)\otimes\F_p
    \to H_d^\mathrm{free}(K^1;\Z)\otimes\F_p
    \to \cdots
    \to H_d^\mathrm{free}(K^T;\Z)\otimes\F_p.
\end{equation*}
Its barcode may depend on $p$, but only through map torsion.
Passing from the initial~module $H_d(K^\bullet;\F_p)$ to the free module may remove
exceptional primes but cannot introduce new~ones.%

\subparagraph{Second experiment.}

\begin{figure}[htbp]
    \centering
    \includegraphics[width=1\linewidth]{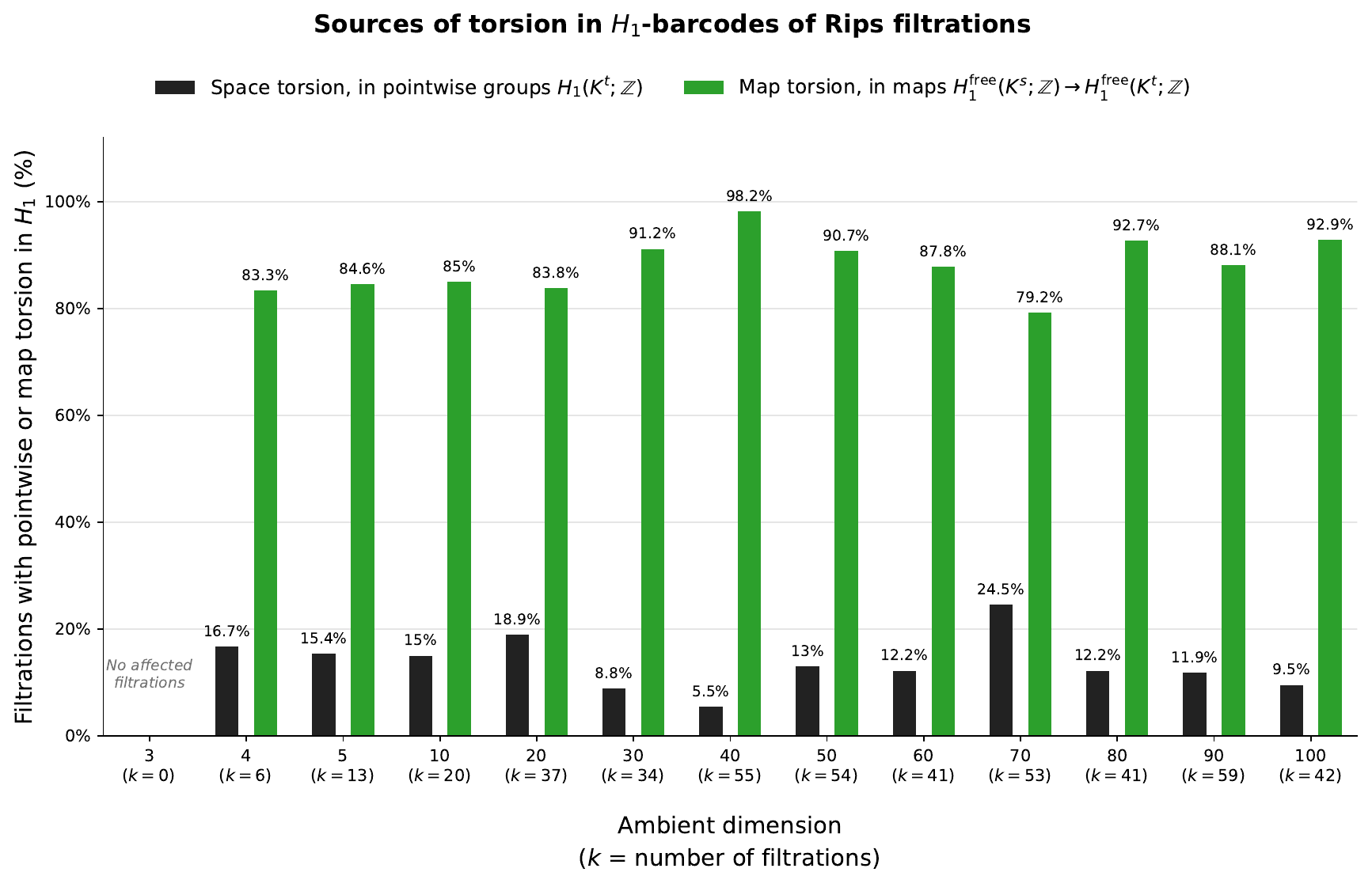}
    \caption{
        Type of torsion in the original Rips filtrations (not simplex-wise) with torsion of~\Cref{fig:random_rips_simplexwise}.%
    }
    \label{fig:random_rips_torsion}
\end{figure}

For each filtration $K^\bullet$ found to have exceptional primes in the first experiment, we test for \textbf{(1)} pointwise \emph{space torsion} in the groups $H_1(K^t;\Z)$ and \textbf{(2)} \emph{map torsion} in $H_1^\mathrm{free}(K^s;\Z)\to H_1^\mathrm{free}(K^t;\Z)$.
Note that these phenomena may coexist.
The results are given in \Cref{fig:random_rips_torsion}.
We refer to \Cref{sec:torsion_on_barcodes} for the theory and \Cref{sec:computation_presentation_matrices} for the~algorithms.

Interestingly, map torsion occurs in most of these filtrations ($88.13\%$) while space torsion is less common ($13.63\%$).
Of the $60$ filtrations with space torsion, $10$ also have map torsion.
Thus field dependence is mostly a \emph{persistence} effect here, invisible to pointwise homology~groups.

For map torsion, the only exceptional primes observed were $2$ alone in $394$ filtrations, $3$ alone in $4$, and both in $7$.
Pointwise torsion involved only the prime $2$.
This contrasts sharply with the large exceptional primes observed in the Linial--Meshulam process of \Cref{subsec:benchmark}.

\FloatBarrier








\section{Effect of torsion on persistence barcodes}
\label{sec:torsion_on_barcodes}

This theoretical section studies how torsion affects the persistence module $H_d(K^\bullet;\kk)$.
In \Cref{subsec:three_sources_of_torsion}, we recall the universal coefficient theorem and identify three types of torsion that we call \emph{Tor torsion}, \emph{tensor torsion}, and \emph{map torsion}.
In \Cref{subsec:three_persistence_modules}, we isolate their effects one by one, yielding a hierarchy of modules (see Equations \eqref{eq:persistence_module_field_coefficient}, \eqref{eq:persistence_module_tensor}, and \eqref{eq:persistence_module_free})
\[
    H_d(K^\bullet;\kk),
    \qquad H_d(K^\bullet;\Z)\otimes\kk,
    \qquad H_d^\mathrm{free}(K^\bullet;\Z)\otimes\kk.
\]
Our main result, \Cref{th:birth_death_profiles}, describes the link between the barcodes of these modules.

\subsection{Three sources of torsion}
\label{subsec:three_sources_of_torsion}

\subparagraph{Universal coefficient theorem.}

Let $K$ be a finite simplicial complex and let $\kk$ be a field.
The universal coefficient theorem (UCT) gives a short exact sequence of $\kk$-vector spaces
\begin{equation}
    \label{eq:uct}
    0 
    \xrightarrow[]{} H_d(K;\Z) \otimes \kk
    \xrightarrow[]{} H_d(K;\kk)
    \xrightarrow[]{} \Tor(H_{d-1}(K;\Z), \kk)
    \xrightarrow[]{} 0.
\end{equation}
Moreover, the sequence splits.
This implies that there exists an isomorphism
\begin{equation*}
    H_d(K;\kk) \simeq \big(H_d(K;\Z) \otimes \kk\big)\oplus\Tor(H_{d-1}(K;\Z), \kk).
\end{equation*}

The first term $H_d(K;\Z) \otimes \kk$ represents the \emph{tensor product} with $\kk$.
Each \emph{free summand} $\Z$ of $H_d(K;\Z)$ contributes a copy of $\kk$ in $H_d(K;\Z) \otimes \kk$.
Moreover, each \emph{cyclic summand} $\Z/p^k\Z$ contributes a copy of $\kk$ provided that the characteristic of $\kk$ is $p$.
Note that the characteristic of the field of rationals $\Q$ is zero, hence only free components can produce a $\Q$~component.

The second term, $\Tor(H_{d-1}(K;\Z), \kk)$, represents the \emph{Tor functor}.
By contrast, only cyclic summands $\Z/p^k\Z$ in $H_{d-1}(K;\Z)$ contribute a copy of $\kk$ in $\Tor(H_{d-1}(K;\Z), \kk)$, provided that the characteristic of $\kk$ is $p$.
See \Cref{rem:torsion_in_uct} for an explanation of this phenomenon.

\begin{example}    
    Consider the Klein bottle $K$.
    Its integral homology is $H_0(K;\Z)=\Z$, $H_1(K;\Z)=\Z\oplus\Z/2\Z$, and $H_2(K;\Z)=0$.
    By the UCT, its homology over $\F_2$ and $\F_3$ is
    \[
        H_1(K;\F_2) \simeq \F_2^2 \oplus 0,
        \quad 
        H_1(K;\F_3) \simeq \F_3 \oplus 0,
        \quad 
        H_2(K;\F_2) = 0 \oplus \F_2,
        \quad 
        H_2(K;\F_3) = 0 \oplus 0.
    \]
\end{example}

\subparagraph{Pointwise torsion in persistence modules.}

Consider the persistence module $H_d(K^\bullet;\kk)$, as defined in \Cref{eq:persistence_module_field_coefficient}.
As discussed above, the UCT identifies a first reason for the barcode to depend on the field $\kk$.
Let us focus on finite fields $\kk=\F_p$ with $p$ prime.
For a fixed $t\in\{0,\dots,T\}$, each cyclic component $\Z/p^k\Z$ in $H_d(K^t;\Z)$ or $H_{d-1}(K^t;\Z)$ yields an \emph{exceptional summand} in $H_d(K^t;\F_p)$.
When $p\neq q$, this torsion does not appear in $H_d(K^t;\F_q)$.

We refer to torsion coming from $H_d(K^t;\Z)$ as \emph{tensor torsion} and from $H_{d-1}(K^t;\Z)$ as \emph{Tor torsion}.
\Cref{th:birth_death_profiles} shows that they have distinct effects on the barcode: tensor torsion can only extend the bars to the right, and Tor torsion to the left; see \Cref{ex:torsion_relation,ex:torsion_cycle}.

\begin{remark}
    \label{rem:torsion_in_uct}
    We illustrate, at the level of homology classes, the phenomena of tensor and Tor torsion in the UCT.
    Consider a class $x\in H_d(K;\Z)$ of order $p^k$, i.e., such that $p^{k-1}x\neq0$ and $p^kx=0$.
    It is represented by a cycle $z\in Z_d(K;\Z)$ whose multiple $p^kz$ is a boundary:
    \[
        x = [z] \qquad\text{and}\qquad p^kz = b ~~\text{for some}~~ b\in B_d(K;\Z).
    \]
    Over a field $\kk$ of characteristic other than $p$, the scalar $p$ is invertible, making $z$ a~boundary:
    \[
        p\neq 0 \implies z = p^{-k} b \in B_d(K;\kk).
    \]
    In this case, the class $x$ vanishes in $H_d(K;\kk)$.
    For the second case, consider a class $x\in H_{d-1}(K;\Z)$ of order $p^k$ represented by a cycle $z\in Z_{d-1}(K;\Z)$.
    Again, $p^kz$ is a boundary:
    \[
        x = [z] \qquad\text{and}\qquad p^kz = \partial c ~~\text{for some}~~ c\in C_d(K;\Z).
    \]
    Over a field $\kk$ of characteristic $p$, the cycle $p^kz$ vanishes in $Z_{d-1}(K;\kk)$. 
    In this case, the chain $c$ is promoted to a cycle in $Z_d(K;\kk)$, thus defining a nonzero homology class $[c] \in H_d(K;\kk)$.%
\end{remark}

\subparagraph{Torsion in pointwise-free persistence modules.}

Next, suppose that all integral homology groups $H_d(K^t;\Z)$ and $H_{d-1}(K^t;\Z)$ are free.
In this case, for a fixed $t$, all field-coefficient homology groups $H_d(K^t;\kk)$ have the same dimension, equal to the rank $r$ of $H_d(K^t;\Z)\simeq\Z^r$.
Still, the barcodes may vary with $\kk$.
This occurs when a relative integral homology group 
\[
    H_d(K^t,K^s;\Z)
\]
has torsion, or equivalently, when a map $H_d(K^s;\Z)\to H_d(K^t;\Z)$ has torsion in its cokernel.

A typical example is multiplication by a nonzero $\ell\in\Z$, which we denote by $\phi\colon\Z\xrightarrow[]{\times\ell}\Z$.
Generically, the tensored map $\phi\otimes\F_p$ is an isomorphism.
However, it is zero when $p$ divides~$\ell$:%
\[
    (\phi\otimes \F_p)\colon \F_p\xrightarrow[]{~0~} \F_p\quad
    \text{ if $p\mid\ell$,}\qquad
    (\phi\otimes \F_p)\colon \F_p\xrightarrow[]{\;\cong\;} \F_p\quad
    \text{otherwise}.
\]
We refer to this phenomenon as \emph{map torsion}; it is illustrated in \Cref{ex:torsion_map} below.
On the barcodes, it has the effect of exchanging the death times of nested bars; see  \Cref{th:birth_death_profiles}.

\begin{example}[Map torsion]
    \label{ex:torsion_map}
    Consider the following subset of $\R^4$:
    \[
        X = \left\{ \begin{pmatrix} \cos\theta\\ \sin\theta\\ \cos3\theta\\ \sin3\theta \end{pmatrix} \mid \theta \in [0,2\pi)\right\}.
    \]
    It is a closed curve; more precisely, it is the \emph{torus knot} $(1,3)$ seen on the Clifford torus $S^1\times S^1\subset \R^4$.
    For $t\geq0$, let $X^t$ denote the $t$-thickening of $X$, that is, the set of points of $\R^4$ at distance at most $t$ from $X$.
    We consider the sublevel-set filtration $(X^t)_{t=0}^{+\infty}$.
    For simplicity, we work directly with this set filtration. 
    Although it is not simplicial, the same qualitative behavior can be recovered from a sufficiently dense finite sample using the associated \v{C}ech filtration.
    As $t$ varies, the thickenings are homotopy equivalent to:
    \[
        X^t \simeq
        \begin{cases} 
          S^1 & \text{if } 0 \leq t < t_1, \\
          S^1\times S^1 & \text{if } t_1 \leq t < t_2, \\
          S^1 & \text{if } t_2 \leq t < t_3, \\
          S^3 & \text{if } t_3 \leq t < t_4, \\
          \{\mathrm{pt}\} & \text{if } t_4 \leq t,
       \end{cases}
       \qquad\text{where}\qquad
       \left\{
       \begin{aligned}
       t_1 &= \sqrt{1-2/(3\sqrt{6})},\\ 
       t_2 &= 1,\\
       t_3 &= \sqrt{1+2/(3\sqrt{6})},\\
       t_4 &= \sqrt{2}.                  
       \end{aligned}
       \right.
    \]
    This computation is similar to \cite[Example 5]{Tinarrage2022}.
    In particular, the homology groups $H_1(X^t;\Z)$ are all free.
    Moreover, the induced maps between the five homotopy types above are
    \[    
        H_1(X^\bullet;\Z)\colon\;
        \Z
        \xrightarrow[]{\left(\begin{smallmatrix}1\\3\end{smallmatrix}\right)}
        \Z^2
        \xrightarrow[]{\left(\begin{smallmatrix}0&1\end{smallmatrix}\right)}
        \Z
        \xrightarrow[]{~~}
        0
        \xrightarrow[]{~~}
        0.
    \]
    Since the homology groups involved are free, the $\F_2$ and $\F_3$-modules are obtained by tensoring:
    \[    
        H_1(X^\bullet;\F_2)\colon\;
        \F_2
        \xrightarrow[]{\left(\begin{smallmatrix}1\\1\end{smallmatrix}\right)}
        \F_2^2
        \xrightarrow[]{\left(\begin{smallmatrix}0&1\end{smallmatrix}\right)}
        \F_2
        \xrightarrow[]{~~}
        0,
        \qquad
        H_1(X^\bullet;\F_3)\colon\;
        \F_3
        \xrightarrow[]{\left(\begin{smallmatrix}1\\0\end{smallmatrix}\right)}
        \F_3^2
        \xrightarrow[]{\left(\begin{smallmatrix}0&1\end{smallmatrix}\right)}
        \F_3
        \xrightarrow[]{~~}
        0.
    \]
    A direct computation shows that their persistence barcodes differ:
    \[
        \barcode\big(H_1(X^\bullet;\F_2)\big) 
        = \big\{[0,t_3),\, [t_1,t_2)\big\},
        \qquad
        \barcode\big(H_1(X^\bullet;\F_3)\big) 
        = \big\{[0,t_2),\, [t_1,t_3)\big\}.
    \]
    Moreover, the $\F_p$-barcodes coincide for all $p\neq3$.
    The prime $p=3$ is exceptional due to \emph{map torsion}: the composition becomes zero modulo $3$, causing the death times to be \emph{exchanged}.
\end{example}

\begin{example}[Tensor torsion]
    \label{ex:torsion_relation}
    Let $K^0 = S^1$ be the circle, on which we glue a disk along a degree-2 map $f$, resulting in the projective plane $K^1 = S^1 \cup_f D^2$.
    Their degree-1 homology is
    \[
        H_1(K^0;\Z)\cong\Z,
        \qquad 
        H_1(K^1;\Z)\cong\Z/2\Z.
    \]
    Consider the filtration $(K^t)_{t=0}^1$.
    The corresponding persistence modules over $\F_2$ and $\F_3$ are
    \[    
        H_1(K^\bullet;\F_2)\colon\;
        \F_2
        \xrightarrow[]{\;\text{id}\;}
        \F_2, 
        \qquad
        H_1(K^\bullet;\F_3)\colon\;
        \F_3
        \xrightarrow[]{\;\;\;}
        0.
    \]
    Their persistence barcodes are
    \[
        \barcode\big(H_1(K^\bullet;\F_2)\big) 
        = \big\{[0,\infty)\big\},
        \qquad
        \barcode\big(H_1(K^\bullet;\F_3)\big) 
        = \big\{[0,1)\big\}.
    \]
    In fact, the barcodes of $H_1(K^\bullet;\F_p)$ are identical for all primes $p\neq 2$.
    The only exceptional prime is $2$, which exhibits \emph{tensor torsion}: at $t=1$, the cyclic component $\Z/2\Z$ in $H_1(K^1;\Z)$ yields an exceptional class in $H_1(K^1;\F_2)$.
    As a result, the corresponding death is \emph{delayed}.
\end{example}

\begin{example}[Tor torsion]
    \label{ex:torsion_cycle}
    We take the Klein bottle $K^0$ and glue a disk along the torsion loop $a \in \pi_1(K^0) = \langle a,b\mid bab^{-1}=a^{-1}\rangle$.
    The resulting space $K^1 = K^0 \cup_{a} D^2$ is homotopy equivalent to the wedge $S^1 \vee S^2$.
    The integral homology groups of $K^0$ and $K^1$ are
    \[
        H_1(K^0;\Z)\cong\Z\oplus\Z/2\Z,
        \qquad 
        H_2(K^0;\Z)\cong0,
        \qquad 
        H_1(K^1;\Z)\cong\Z,
        \qquad 
        H_2(K^1;\Z)\cong\Z.
    \]    
    Although $H_2(K^0;\Z)$ is trivial, $H_2(K^0;\F_2)$ is not, due to the 2-torsion in $H_1(K^0;\Z)$.
    The persistence modules with coefficients in $\F_2$ and $\F_3$ associated with the filtration $(K^t)_{t=0}^1$ are
    \[    
        H_2(K^\bullet;\F_2)\colon\;
        \F_2
        \xrightarrow[]{\;\text{id}\;}
        \F_2, 
        \qquad
        H_2(K^\bullet;\F_3)\colon\;
        0
        \xrightarrow[]{\;\;\;}
        \F_3.
    \]
    One deduces the persistence barcodes
    \[
        \barcode\big(H_2(K^\bullet;\F_2)\big) 
        = \big\{[0,\infty)\big\},
        \qquad
        \barcode\big(H_2(K^\bullet;\F_3)\big) 
        = \big\{[1,\infty)\big\}.
    \]
    Among the barcodes of $H_2(K^\bullet;\F_p)$, only the prime $p=2$ is exceptional.
    Via \emph{Tor torsion}, $H_2(K^0;\F_2)$ gains an additional cycle, causing the corresponding class to be born \emph{earlier}.
\end{example}

\subsection{Three persistence modules and their barcodes}
\label{subsec:three_persistence_modules}

\subparagraph{Hierarchy of persistence modules.}

We still consider a filtration $(K^t)_{t=0}^N$ and the persistence modules $H_d(K^\bullet;\kk)$.
The discussion above prompts us to distinguish three phenomena: \emph{Tor torsion}, arising from the Tor functor $\Tor(H_{d-1}(K^t;\Z), \kk)$; \emph{tensor torsion}, from the tensor products $H_d(K^t;\Z) \otimes \kk$; and \emph{map torsion} in the homomorphisms $H_d(K^s;\Z)\to H_d(K^t;\Z)$.

To distinguish these effects, we successively peel off Tor torsion and tensor torsion.
Let us refer to the initial persistence module $H_d(K^\bullet;\kk)$ as the \emph{field-coefficient module}.
We produce another persistence module, called the \emph{tensor module}, by tensoring integral homology:%
\begin{equation}
    \label{eq:persistence_module_tensor}
    H_d(K^\bullet;\Z)\otimes\kk\colon\quad
    H_d(K^0;\Z)\otimes\kk \to H_d(K^1;\Z)\otimes\kk \to \cdots \to H_d(K^N;\Z)\otimes\kk.
\end{equation}
This persistence module can be seen as the field-coefficient module with the Tor effect~removed.%

We build a third persistence module by quotienting out the torsion components in $H_d(K^t;\Z)$.
More precisely, let $H_d^\mathrm{tors}(K^t;\Z)$ denote its \emph{torsion subgroup}---i.e., the classes $x\in H_d(K^t;\Z)$ such that $qx=0$ for some nonzero integer $q$---and let its \emph{free part} be the~quotient%
\begin{equation}
    \label{eq:free_part}
    H_d^\mathrm{free}(K^t;\Z) = H_d(K^t;\Z) / H_d^\mathrm{tors}(K^t;\Z).    
\end{equation}
Tensoring the free parts forms a persistence module, which we call the \emph{free module} and denote%
\begin{equation}
    \label{eq:persistence_module_free}
    H_d^\mathrm{free}(K^\bullet;\Z)\otimes\kk\colon\quad
    H_d^\mathrm{free}(K^0;\Z)\otimes\kk \to H_d^\mathrm{free}(K^1;\Z)\otimes\kk \to \cdots \to H_d^\mathrm{free}(K^N;\Z)\otimes\kk.
\end{equation}
These three persistence modules are linked through morphisms.
We refer to \cite{BauerLesnick2015} for definitions.

\begin{lemma}[Proof p.~\pageref{proof:three_persistence_modules}]
    \label{lem:three_persistence_modules}
    For every field $\kk$ there exist an epimorphism and a monomorphism%
    \[
        H_d(K^\bullet;\Z)\otimes\kk \twoheadrightarrow H_d^\mathrm{free}(K^\bullet;\Z)\otimes\kk,
        \qquad
        H_d(K^\bullet;\Z) \otimes \kk \hookrightarrow H_d(K^\bullet;\kk).
    \]
\end{lemma}

\subparagraph{Structure result.}

By \emph{generic module} we refer to the persistence module with coefficients in $\kk=\Q$.
In this case, all three persistence modules under consideration are isomorphic:
\[
    H_d(K^\bullet;\Q) \simeq H_d(K^\bullet;\Z)\otimes\Q \simeq H_d^\mathrm{free}(K^\bullet;\Z)\otimes\Q.
\]
Their common persistence barcode is called the \emph{generic barcode} and is denoted
\[
    \barcode\big(H_d(K^\bullet;\Q)\big) 
    = \big\{\!\big\{ \big[ b_i(\Q), d_{i}(\Q) \big) \mid 1\leq i\leq r \big\}\!\big\}.
\]
The next theorem shows how to build all other $\kk$-barcodes starting from this one.
For clarity, we assume that the filtration is simplex-wise.
Arbitrary filtrations are handled in \Cref{rem:birth_death_profiles}.

\begin{theorem}[Proof p.~\pageref{proof:birth_death_profiles}]
    \label{th:birth_death_profiles}
    Let $(K^t)_{t=0}^N$ be a simplex-wise filtration with $K^0=\emptyset$, let $d\geq0$, and suppose that the last integral homology group $H_{d-1}(K^N;\Z)$ is torsion free.
    Let $r$ be the rank of $Z_d(K^N;\Z)$.
    For every field $\kk$, there exists a permutation of $\{1,\dots,r\}$, denoted by
    \[
        \sigma_i(\kk),\quad 1\leq i\leq r,
    \]
    and a collection of $r$ pairs of elements
    \[
        \epsilon_i^-(\kk)\in\{0,\dots,N\},
        \qquad \epsilon_i^+(\kk)\in \{0,\dots,N\}\cup\{+\infty\},
        \qquad 1\leq i\leq r,
    \]
    such that the free, tensor, and field-coefficient modules of Equations \eqref{eq:persistence_module_free}, \eqref{eq:persistence_module_tensor}, and \eqref{eq:persistence_module_field_coefficient}~satisfy
    \begin{align*}
        \barcode\big(H_d^\mathrm{free}(K^\bullet;\Z)\otimes\kk\big)
        &= \big\{\!\big\{
                \phantom{{}-\epsilon_i^-(\kk)}
            \big[
                b_i(\Q),\,
                d_{\sigma_i(\kk)}(\Q)
            \big)
            \phantom{{}+\epsilon_i^+(\kk)}
            \mid 1\leq i\leq r
        \big\}\!\big\},
        \\
        \barcode\big(H_d(K^\bullet;\Z)\otimes\kk\big)
        &= \big\{\!\big\{
            \phantom{{}-\epsilon_i^-(\kk)}
            \big[
                b_i(\Q),\,
                d_{\sigma_i(\kk)}(\Q)+\epsilon_i^+(\kk)
            \big)
            \mid 1\leq i\leq r
        \big\}\!\big\},
        \\
        \barcode\big(H_d(K^\bullet;\kk)\big)
        &= \big\{\!\big\{
            \big[
                b_i(\Q)-\epsilon_i^-(\kk),\,
                d_{\sigma_i(\kk)}(\Q)+\epsilon_i^+(\kk)
            \big)
            \mid 1\leq i\leq r
        \big\}\!\big\}.
    \end{align*}
    Moreover, the values of $\sigma_i(\kk)$, $\epsilon_i^-(\kk)$, and $\epsilon_i^+(\kk)$ can be chosen to depend only on the characteristic of $\kk$, and there are only finitely many characteristics for which they differ from
    \[
        \sigma_\bullet(\Q) = \mathrm{id},\quad 
        \epsilon_i^-(\Q)=0,\quad
        \epsilon_i^+(\Q)=0.
    \]
    Last, $\sigma_\bullet(\kk)$ is obtained by successive swaps of the death times of nested intervals via
    \begin{equation}
        \label{eq:death_times_swap}
        [b,d),\; [b',d') \longrightarrow [b,d'),\; [b',d),
        \qquad
        b < b' < d' < d.        
    \end{equation}
\end{theorem}

Thus, one goes from the $\Q$-barcode to the barcode of the free module $H_d^\mathrm{free}(K^\bullet;\Z)\otimes\kk$ by swapping the death times of certain nested intervals; this is the effect of \emph{map torsion} shown in \Cref{ex:torsion_map}.
Then, the barcode of the tensor module $H_d(K^\bullet;\Z)\otimes\kk$ is obtained by extending certain intervals to the right.
This is due to \emph{tensor torsion}, illustrated in \Cref{ex:torsion_relation}.
Finally, the barcode of the field-coefficient module $H_d(K^\bullet;\kk)$ is obtained by extending certain intervals to the left, as a consequence of \emph{Tor torsion}; see \Cref{ex:torsion_cycle}.

\begin{remark}
    \label{rem:birth_death_profiles}
    In \Cref{th:birth_death_profiles}, the freeness hypothesis on $H_{d-1}(K^N;\Z)$ can be removed.
    In this case, the formula for $\barcode(H_d(K^\bullet;\kk))$ must be supplemented with
    \[
        \dim_{\kk}\Tor(H_{d-1}(K^N;\Z),\kk)
    \]
    additional intervals of the form $[b_j(\kk),+\infty)$.
    Moreover, the theorem extends to arbitrary filtrations by 
    \textbf{(1)} choosing an arbitrary simplex-wise refinement that starts from the empty complex $\emptyset$;
    \textbf{(2)} mapping back the interval endpoints to their original filtration indices, leaving $+\infty$ unchanged;
    and \textbf{(3)} discarding intervals whose endpoints coincide.
\end{remark}

\begin{remark}
    The theorem suggests a possible hierarchical algorithm: starting from the barcode of $H_d(K^\bullet;\Q)$, one could successively recover those of $H_d^\mathrm{free}(K^\bullet;\Z)\otimes\kk$, $H_d(K^\bullet;\Z)\otimes\kk$, and $H_d(K^\bullet;\kk)$, reusing previous computations.
    We reserve this perspective for future work.%
\end{remark}

\section{Computing presentation matrices}
\label{sec:computation_presentation_matrices}

The barcodes of the modules $H_d(K^\bullet;\kk)$, $H_d(K^\bullet;\Z)\otimes\kk$, and $H_d^\mathrm{free}(K^\bullet;\Z)\otimes\kk$, introduced in \Cref{sec:torsion_on_barcodes}, can be computed by our main algorithm (\Cref{alg:main_algorithm}).
It takes as an input a graded integer matrix $D$.
For the first module, we use the boundary matrix, as in classical persistence.
For the two others, we instead use \emph{presentation matrices}, whose pivot sequences determine the corresponding barcode.
In \Cref{subsec:presentation_matrix_tensor,subsec:presentation_matrix_free}, respectively, we build these matrices in $O(N^3)$ time, where $N$ is the number of simplices (\Cref{th:presentation_matrix_tensor_module,th:presentation_matrix_free_module}).

We mention that these constructions rely on standard tools.
For the tensor module, we use the exact sequence in \Cref{eq:exact_sequence_tensor} to construct a presentation matrix by incremental column reduction, as in Lambán et al.~\cite{LambanEtAl2016}.
For the free module, with exact sequence in \Cref{eq:exact_sequence_free}, we start with the previous matrix and replace the column lattice by its saturation.
Saturation via Hermite reduction is discussed by Pernet and Stein~\cite[Section~8]{PernetStein2010} and incremental reduction via the extended Euclidean algorithm by Storjohann~\cite[Chapter~3]{Storjohann2000}.%

\subsection{Presentation matrix for the tensor module}
\label{subsec:presentation_matrix_tensor}

We show in \Cref{subsubsec:presentation_matrix_tensor_overview} that the tensor module is presented by the $(d+1)$-boundary map%
\[
    \partial_{d+1}\colon C_{d+1}(K^t;\Z) \to Z_d(K^t;\Z).
\]
Note that the codomain is the group of integral cycles, rather than the chain group used in classical persistence.
To represent this map as a matrix, we need a cycle basis compatible
with the filtration.
We construct this basis in
\Cref{subsubsec:presentation_matrix_tensor_cycles}
using the degree-$d$ boundary map
\[
    \partial_d\colon C_d(K^t;\Z) \to C_{d-1}(K^t;\Z).
\]
The construction thus involves two matrices.
The full procedure is given in
\Cref{subsubsec:presentation_matrix_tensor_full}.

\subsubsection{Overview of the construction and the degree-\texorpdfstring{$(d+1)$}{(d+1)} matrix}
\label{subsubsec:presentation_matrix_tensor_overview}

\subparagraph{Presentation.}

Consider a filtration $K^\bullet=(K^t)_{t=0}^N$, which we can assume simplex-wise for now.
Let $t\in\{0,\dots,N\}$.
We start with the definition of integral homology as the quotient $H_d(K^t;\Z) = Z_d(K^t;\Z)/B_d(K^t;\Z)$.
Equivalently, it can be written via the exact sequence%
\begin{equation}
    \label{eq:exact_sequence_integral}
    C_{d+1}(K^t;\Z) 
    \xrightarrow{~\partial_{d+1}^t~} Z_d(K^t;\Z) 
    \xrightarrow{~\pi^t~}  H_d(K^t;\Z) 
    \xrightarrow{~~~~} 0.
\end{equation}
Next, let $\kk$ be a field.
Tensoring with $\kk$ is a right-exact functor, thus we get an exact~sequence%
\begin{equation}
    \label{eq:exact_sequence_tensor}
    C_{d+1}(K^t;\Z)\otimes\kk 
    \xrightarrow{~\partial_{d+1}^t\otimes\kk~} Z_d(K^t;\Z)\otimes\kk 
    \xrightarrow{~\pi^t\otimes\kk~} H_d(K^t;\Z)\otimes\kk 
    \xrightarrow{~~~~} 0.
\end{equation}
In particular, the tensored homology group $H_d(K^t;\Z)\otimes\kk$ can be written as a quotient
\begin{equation}
    \label{eq:tensor_module_as_quotient}
    H_d(K^t;\Z)\otimes\kk \cong Z_d(K^t;\Z)\otimes\kk\, / \im \bigl(\partial_{d+1}^t\otimes\kk\bigr).    
\end{equation}

\begin{remark}
    Under the natural identification $C_d(K^t;\Z)\otimes\kk\cong C_d(K^t;\kk)$, the numerator $Z_d(K^t;\Z)\otimes\kk$ in \Cref{eq:tensor_module_as_quotient} is a subspace
    of the $\kk$-cycles $Z_d(K^t;\kk)$.
    However, they need not coincide, because of Tor torsion (see \Cref{subsec:three_sources_of_torsion,ex:torsion_cycle}).
    The denominator $\im(\partial_{d+1}^t\otimes\kk)$, however, coincides with the $\kk$-boundaries $B_d(K^t;\kk)$, by definition.
\end{remark}

\subparagraph{Persistent presentation.}

To study the persistence module $H_d(K^\bullet;\Z)\otimes\kk$ as a whole, we need a persistent version of these facts.
\Cref{eq:exact_sequence_integral} is natural, thus yields a commutative~diagram%
\begin{equation}
    \label{eq:persistent_presentation_two_rows_integral}
    \begin{tikzcd}[
        column sep=huge,
        row sep=large,
        continue above/.style={
            dashed,
            start anchor={[yshift=1.8em]north},
            end anchor=north
        },
        continue below/.style={
            dashed, dash,
            start anchor=south,
            end anchor={[yshift=-1.8em]south}
        }
    ]
        C_{d+1}(K^t;\Z)
            \arrow[continue above]
            \arrow[r, "\partial_{d+1}^t"]
            \arrow[d]
        & Z_d(K^t;\Z)
            \arrow[continue above]
            \arrow[r]
            \arrow[d]
        & H_d(K^t;\Z)
            \arrow[continue above]
            \arrow[r]
            \arrow[d, "\phi^t"]
        & 0
        \\
        C_{d+1}(K^{t+1};\Z)
            \arrow[r, "\partial_{d+1}^{t+1}"]
            \arrow[continue below]
        & Z_d(K^{t+1};\Z)
            \arrow[r]
            \arrow[continue below]
        & H_d(K^{t+1};\Z)
            \arrow[r]
            \arrow[continue below]
        & 0
    \end{tikzcd}
\end{equation}
where the left and middle vertical maps are inclusions.
Again, tensoring with $\kk$ yields 
\begin{equation}
    \label{eq:persistent_presentation_two_rows_tensor}
    \begin{tikzcd}[
        column sep=large,
        row sep=large,
        continue above/.style={
            dashed,
            start anchor={[yshift=1.8em]north},
            end anchor=north
        },
        continue below/.style={
            dashed, dash,
            start anchor=south,
            end anchor={[yshift=-1.8em]south}
        }
    ]
        C_{d+1}(K^t;\Z)\otimes\kk
            \arrow[continue above]
            \arrow[r, "\partial_{d+1}^t\otimes\kk"]
            \arrow[d]
        & Z_d(K^t;\Z)\otimes\kk
            \arrow[continue above]
            \arrow[r]
            \arrow[d]
        & H_d(K^t;\Z)\otimes\kk
            \arrow[continue above]
            \arrow[r]
            \arrow[d, "\phi^t\otimes\kk"]
        & 0
        \\
        C_{d+1}(K^{t+1};\Z)\otimes\kk
            \arrow[r, "\partial_{d+1}^{t+1}\otimes\kk"]
            \arrow[continue below]
        & Z_d(K^{t+1};\Z)\otimes\kk
            \arrow[r]
            \arrow[continue below]
        & H_d(K^{t+1};\Z)\otimes\kk
            \arrow[r]
            \arrow[continue below]
        & 0
    \end{tikzcd}
\end{equation}

The third column of \Cref{eq:persistent_presentation_two_rows_tensor} is precisely the tensor module $H_d(K^\bullet;\Z)\otimes\kk$.
Thus, to encode it, it suffices to express the boundary maps $\partial_{d+1}^t\colon C_{d+1}(K^t;\Z)\to Z_d(K^t;\Z)$ in bases compatible with the filtration.
For the integral $(d+1)$-chains, we use the simplex basis
\[
    \B_{d+1}^C = (\sigma_1,\dots,\sigma_m)
\]
where $m$ is the number of $(d+1)$-simplices in $K^\bullet$.
For the integral $d$-cycles, we build a~basis%
\begin{equation}
    \label{eq:incremental_basis_cycle}
    \B_d^Z = (z_1,\dots,z_n)    
\end{equation}
where $n=\rank Z_d(K^N;\Z) $. 
In these bases, the boundary maps are encoded by the matrix
\begin{equation}
    \label{eq:graded_matrix_tensor}
    \setlength{\arraycolsep}{2pt}
    D_{d+1}^\mathrm{tens}
    \;=\; 
    \Matrix\big(\partial_{d+1}^N;\B_{d+1}^C,\B_d^Z\big)
    \;=\;
    \begin{pNiceArray}{ccccc}[first-row,first-col]
        &
        \sigma_1
        & \sigma_2
        & \sigma_3
        & \cdots
        & \sigma_m
        \\
        \begin{array}{c}
        z_1\\
        z_2\\
        \vdots\\
        z_{n}
        \end{array}
        &
        \begin{array}{c}
        \rule{0.4pt}{2em}\\[-0.3em]
        \partial\sigma_1\\[0.3em]
        \rule{0.4pt}{2em}
        \end{array}
        &
        \begin{array}{c}
        \rule{0.4pt}{2em}\\[-0.3em]
        \partial\sigma_2\\[0.3em]
        \rule{0.4pt}{2em}
        \end{array}
        &
        \begin{array}{c}
        \rule{0.4pt}{2em}\\[-0.3em]
        \partial\sigma_3\\[0.3em]
        \rule{0.4pt}{2em}
        \end{array}
        &
        \cdots
        &
        \begin{array}{c}
        \rule{0.4pt}{2em}\\[-0.3em]
        \partial\sigma_m\\[0.3em]
        \rule{0.4pt}{2em}
        \end{array}
    \end{pNiceArray}.
\end{equation}
We grade row $i$ by the birth time of $z_i$ and column $j$ by that of $\sigma_j$.
Thus, after tensoring with $\kk$, this graded matrix presents $H_d(K^\bullet;\Z)\otimes\kk$, and its barcode can therefore be computed by our main algorithm.
The rest of this section shows how to build $\B_d^Z$ and $D_{d+1}^\mathrm{tens}$ efficiently.%

\subsubsection{Identifying cycles and the degree-\texorpdfstring{$d$}{d} matrix}
\label{subsubsec:presentation_matrix_tensor_cycles}

\subparagraph{Degree-$d$ boundary matrix.}

Unlike the boundary operator $C_{d+1}(K^t;\Z) \to C_d(K^t;\Z)$ used in classical persistence, the matrix $D_{d+1}^\mathrm{tens}$ in \Cref{eq:graded_matrix_tensor} represents $C_{d+1}(K^t;\Z) \to Z_d(K^t;\Z)$.
The challenge is to maintain a cycle basis and compute the coordinates of each boundary~$\partial\sigma_i$.

To identify the $d$-cycles, we must consider the boundary matrix in degree $d$.
We denote it%
\begin{equation}
    \label{eq:graded_matrix_tensor_degree_d}
    \setlength{\arraycolsep}{2pt}
    D_d
    \;=\; 
    \Matrix\big(\partial_{d}^N;\B_d^C,\B_{d-1}^C\big)
    \;=\;
    \begin{pNiceArray}{ccccc}[first-row,first-col]
        &
        \tau_1
        & \tau_2
        & \tau_3
        & \cdots
        & \tau_\ell
        \\
        \begin{array}{c}
        \nu_1\\
        \nu_2\\
        \vdots\\
        \nu_k
        \end{array}
        &
        \begin{array}{c}
        \rule{0.4pt}{2em}\\[-0.3em]
        \partial\tau_1\\[0.3em]
        \rule{0.4pt}{2em}
        \end{array}
        &
        \begin{array}{c}
        \rule{0.4pt}{2em}\\[-0.3em]
        \partial\tau_2\\[0.3em]
        \rule{0.4pt}{2em}
        \end{array}
        &
        \begin{array}{c}
        \rule{0.4pt}{2em}\\[-0.3em]
        \partial\tau_3\\[0.3em]
        \rule{0.4pt}{2em}
        \end{array}
        &
        \cdots
        &
        \begin{array}{c}
        \rule{0.4pt}{2em}\\[-0.3em]
        \partial\tau_\ell\\[0.3em]
        \rule{0.4pt}{2em}
        \end{array}
    \end{pNiceArray},
\end{equation}
where $\B_d^C=\{\tau_1,\dots\}$, $\B_{d-1}^C=\{\nu_1,\dots\}$ are the simplex bases of $C_d(K^N;\Z)$, $C_{d-1}(K^N;\Z)$.

\subparagraph{Incremental bases, field case.}

Suppose for now that $D_d$ has coefficients in a field $\kk$.
A cycle basis can be computed using the classical persistence algorithm, described in \Cref{subsubsec:classical_reduction}.
Through column operations, it transforms $D_d$ into a reduced matrix whose nonzero columns have distinct pivot indices.
Moreover, the reduction is \emph{incremental}: it processes columns from left to right, never modifying a previous column.
For instance, the final state could be
\[
    \setlength{\arraycolsep}{2pt}
    D_d
    =
    \begin{pNiceArray}{ccccc}[first-row,first-col]
        &
        \tau_1
        & \tau_2
        & \tau_3
        & \cdots
        & \tau_\ell
        \\
        \begin{array}{c}
        \nu_1\\
        \nu_2\\
        \vdots\\
        \nu_k
        \end{array}
        &
        \begin{array}{c}
        \rule{0.4pt}{2em}\\[-0.3em]
        \partial\tau_1\\[0.3em]
        \rule{0.4pt}{2em}
        \end{array}
        &
        \begin{array}{c}
        \rule{0.4pt}{2em}\\[-0.3em]
        \partial\tau_2\\[0.3em]
        \rule{0.4pt}{2em}
        \end{array}
        &
        \begin{array}{c}
        \rule{0.4pt}{2em}\\[-0.3em]
        \partial\tau_3\\[0.3em]
        \rule{0.4pt}{2em}
        \end{array}
        &
        \cdots
        &
        \begin{array}{c}
        \rule{0.4pt}{2em}\\[-0.3em]
        \partial\tau_\ell\\[0.3em]
        \rule{0.4pt}{2em}
        \end{array}
    \end{pNiceArray}
    \longrightarrow
    \setlength{\arraycolsep}{6pt}
    \begin{pNiceArray}{ccccc}[first-row,first-col]
        & c_1 & c_2 & c_3 & \cdots & c_\ell \\
        \firstcol{\nu_1}    & p_1 & 0 & *   &  & * \\
        \firstcol{\nu_2}    & 0   & 0 & * &  & p_\ell \\
        \firstcol{\vdots}    & \vdots & \vdots & \vdots & \cdots & \vdots \\
        \firstcol{\nu_k} & 0 & 0 &  p_3 &  & 0
    \end{pNiceArray}.
\]

After reduction, the elements $c_1,\dots,c_\ell$ form a basis of $C_d(K^N;\kk)$.
Elements corresponding to zero columns are called \emph{positive}; they form a basis of $Z_d(K^N;\kk)$.
Conversely, nonzero columns correspond to \emph{negative} elements and form a basis of a complement $Y_d$ to $Z_d(K^N;\kk)$.
In this sense, the persistence algorithm maintains bases $\B_d^Z(t)$ and $\B_d^Y(t)$ for a~decomposition%
\[
    C_d(K^t;\kk) = Z_d(K^t;\kk) \oplus Y_d^t.
\]
Moreover, the bases are \emph{incremental}: new elements are appended without altering the others.%

\subparagraph{Incremental bases, integral case.}

The situation is subtler over $\Z$.
We can still decompose
\begin{equation}
    \label{eq:integral_decomposition_cycles}
    C_d(K^t;\Z) = Z_d(K^t;\Z) \oplus Y_d^t,    
\end{equation}
but only the cycle basis $\B_d^Z(t)$ can be built incrementally.
This is because the quotient
\[
    Z_d(K^{t+1};\Z) / Z_d(K^t;\Z)
\]
is free.
The complementary basis $\B_d^Y(t)$, however, may need to be updated when a $d$-simplex $\tau_{t+1}\in K^{t+1}$ is inserted.
This occurs when its boundary generates a nonzero \emph{torsion} class
\[
    [\partial\tau_{t+1}]\in H_{d-1}(K^t;\Z),
\]
i.e., such that $[\partial\tau_{t+1}]\neq0$ and $q[\partial\tau_{t+1}]=0$ for some $q>1$.
We refer to \Cref{eq:homology_class_boundary_new_simplex} in the proof of \Cref{th:birth_death_profiles} for details.
In particular, if there exists an element $c\in \B_d^Y(t)$ such~that 
\[
    q \, \partial\tau_{t+1} = \partial c,
    \qquad
    q>1,
\]
then $c$ must be replaced by $\tau_{t+1}$ in $\B_d^Y(t+1)$ and $q\tau_{t+1}-c$ must be added to $\B_d^Z(t+1)$.

\begin{example}
    The following integral matrix is reduced up to its penultimate column. 
    Yet, completing the reduction (using invertible column operations) requires modifying the first column.
    Indeed, the operation $\tau_{t+1} \gets \tau_{t+1}-\frac{1}{2}c_1$ is not allowed as 2 is not invertible over $\Z$.
    \begin{equation}
        \label{eq:integral_column_echelon_form}
        \setlength{\arraycolsep}{6pt}
        \begin{pNiceArray}{cccc}[first-row,first-col]
            & c_1 & \cdots & c_t & \tau_{t+1}  \\
            \firstcol{\nu_1}    & 2 & & * & 1 \\
            \firstcol{\vdots}    & \vdots & \cdots & \vdots & \vdots \\
            \firstcol{\nu_k} & 0 &  & * & 0 
        \end{pNiceArray}
        \longrightarrow
        \begin{pNiceArray}{cccc}[first-row,first-col]
            & c_1' & \cdots & c_t & c_{t+1}  \\
            \firstcol{\nu_1}    & 1 & & * & 0 \\
            \firstcol{\vdots}    & \vdots & \cdots & \vdots & \vdots \\
            \firstcol{\nu_k} & 0 &  & * & 0 
        \end{pNiceArray},
        \qquad
        \left\{
            \begin{aligned}
                c_1'      &\gets \tau_{t+1},\\
                c_{t+1}  &\gets -c_1 + 2\tau_{t+1}.
            \end{aligned}
        \right.
    \end{equation}
\end{example}

\subparagraph{Full chain basis.}

Let $\B_d^Z(t),\B_d^Y(t)$ be bases adapted to the decomposition of $C_d(K^t;\Z)$ in \Cref{eq:integral_decomposition_cycles}.
Their coordinates in the simplex basis form the columns of an invertible~matrix%
\begin{equation}
    \label{eq:full_chain_basis_matrix}
    U^t =
    \Bigl(
        \begin{matrix}
            \B_d^Z(t) & \B_d^Y(t)
        \end{matrix}
    \Bigr)
    \in \GL_{r(t)}(\Z), 
    \qquad 
    r(t) =  \rank C_d(K^t;\Z).
\end{equation}
Note that the boundary map sends $\B_d^Z(t)$ to zero and $\B_d^Y(t)$ to an integral basis of $B_{d-1}(K^t;\Z)$, which we denote $H^t$.
Thus, writing $D_d^t$ for the boundary matrix at time $t$, we obtain
\begin{equation}
    \label{eq:graded_matrix_tensor_degree_d_boundary}
    D_d^t\, U^t = \begin{pmatrix}0 & H^t\end{pmatrix}.
\end{equation}
During the algorithm, we keep $H^t$ in column echelon form and maintain the inverse $(U^t)^{-1}$.
We use $H^t$ to determine whether a new $d$-simplex creates a cycle (see \Cref{lem:presentation_order}), and $(U^t)^{-1}$ to convert simplex coordinates into coordinates in the basis adapted to $Z_d(K^t;\Z)\oplus Y_d^t$.

\subparagraph{Euclidean transformations.}

When a new column $c_j$ is appended to $H^t$, we restore column echelon form in a quadratic number of arithmetic operations.
We perform column~operations%
\begin{equation}
    \label{eq:euclidean_reduction_iterated}
    \begin{pmatrix} \vphantom{\dfrac{1}{1}} H^t & c_j \end{pmatrix}Q 
\end{equation}
encoded by an invertible matrix 
\begin{equation}
    \label{eq:euclidean_reduction_iterated_Q}
    Q\in\GL_{s(t)+1}(\Z), \qquad 
    s(t) = \rank Y_d^t.
\end{equation}
If the lowest nonzero entry of $c_j$ lies in a row already owned by a column $c_i$, we resolve the conflict by applying a \emph{Euclidean transformation} to columns $i$ and~$j$:
\begin{equation}
    \label{eq:euclidean_reduction}
    \setlength{\arraycolsep}{6pt}
    \begin{pNiceArray}{cc}[first-row]
        c_i & c_j \\
        * & * \\
        a & b \\
        0 & 0
    \end{pNiceArray}
    \xrightarrow[]{~~\begin{pmatrix}x&-b/g\\y&a/g\end{pmatrix}~~}
    \begin{pNiceArray}{cc}[first-row]
        c_i' & c_j' \\
        * & * \\
        g & 0 \\
        0 & 0
    \end{pNiceArray}
    \quad\text{where}\quad
    xa + yb = g = \gcd(a, b).
\end{equation}
This operation replaces the pivot entry of $c_i$ by $g$ and zeroes out the entry in $c_j$, hence it eliminates the conflict.
We repeat until $c_j$ is zero or its pivot is distinct from all existing~pivots:%
\begin{equation*}
    c_j = 
    \begin{pmatrix}
    \vdots\\
    *\\
    *\\
    * 
    \end{pmatrix}
    \longrightarrow
    \begin{pmatrix}
    \vdots\\
    *\\
    *\\
    0 
    \end{pmatrix}
    \longrightarrow
    \begin{pmatrix}
    \vdots\\
    *\\
    0\\
    0 
    \end{pmatrix}
    \longrightarrow 
    \cdots
\end{equation*}
The composition of all Euclidean transformations gives the matrix $Q$ of \Cref{eq:euclidean_reduction_iterated}.

The Bézout coefficients $x,y$ in \Cref{eq:euclidean_reduction} are found by the \emph{extended Euclidean~algorithm} to which we assign constant time.
There is at most one Euclidean transformation per pivot, each requiring linearly many arithmetic operations.
Updating $H^t$ thus takes quadratic~time.%

\subparagraph{Basis updates.}

Suppose that we introduce a new $d$-simplex $\tau_{t+1}\in K^{t+1}$.
To decide whether it produces a new $d$-cycle, we extend the basis matrix $U^t$ of     \Cref{eq:full_chain_basis_matrix} via
\[
    \widehat U =\;
    \begin{pNiceArray}{cc}[first-row,first-col]
        & (\B_d^Z(t), \B_d^Y(t)) & \tau_{t+1} \\
        & U^t    & 0 \\
        \tau_{t+1} & 0    & 1
    \end{pNiceArray},
    \qquad
    D_d^{t+1}\, \widehat U =
    \begin{pNiceArray}{ccc}[first-row]
        \B_d^Z(t) & \B_d^Y(t) & \tau_{t+1} \\
        \vphantom{\dfrac{1}{1}}0 & H^t & \partial\tau_{t+1}
    \end{pNiceArray}.
\]
As in \Cref{eq:euclidean_reduction_iterated}, we restore column echelon form via column operations:
\begin{equation}
    \label{eq:euclidean_reduction_full_matrices}
    \begin{pmatrix} \vphantom{\dfrac{1}{1}} H^t & \partial\tau_{t+1}\end{pmatrix}Q 
    = \begin{pmatrix} \vphantom{\dfrac{1}{1}} \widetilde H & \beta\end{pmatrix}.
\end{equation}
See \Cref{ex:cycle_euclidean_reduction} for an illustration.
Applying the change of basis induced by $Q$ to $\widehat U$, we get
\begin{equation}
    \label{eq:update_full_matrices}
    \begin{pmatrix} U^t & 0 \\ 0 & 1 \end{pmatrix}
    \begin{pmatrix} I_{z(t)} & 0 \\ 0 & Q \end{pmatrix}
    =
    \begin{pmatrix} \vphantom{\dfrac{1}{1}} \B_d^Z(t) & \widetilde \B_d^Y(t) & \gamma\end{pmatrix},
    \qquad
    z(t) = \rank Z_d(K^t;\Z).
\end{equation}
Here, $\gamma$ is a $d$-chain whose boundary is $\beta$.
Moreover, the modified complementary chains $\widetilde \B_d^Y(t)$ will be used for updates below.
We can now identify whether $\tau_{t+1}$ creates a new~cycle.

\begin{lemma}[Proof p.~\pageref{proof:presentation_order}]
    \label{lem:presentation_order}
    If $\beta=0$, then $\gamma\in Z_d(K^{t+1};\Z)$ is a $d$-cycle independent from~$\B_d^Z(t)$. 
    In this case, let $q$ be the absolute value of the bottom-right entry of $Q$.
    Then $q$ is the order of $[\partial\tau_{t+1}]$ in $H_{d-1}(K^t;\Z)$, where the zero class has order one by convention.
\end{lemma}

Thus, we distinguish two cases.
If $\beta=0$, a cycle is created and we perform the updates 
\begin{equation}
    \label{eq:update_bases_cycle}
    \begin{alignedat}{2}
        \B_d^Z(t+1) &= \B_d^Z(t) \cup \{\gamma\},
        \qquad
        & U^{t+1} &= \begin{pmatrix} \vphantom{\dfrac{1}{1}} \B_d^Z(t) & \gamma & \widetilde{\B}_d^Y(t) \end{pmatrix},
        \\
        \B_d^Y(t+1) &= \widetilde{\B}_d^Y(t),
        \qquad
        & H^{t+1} &= \widetilde{H},
    \end{alignedat}
\end{equation}
where $\widetilde{H}$ and $\widetilde \B_d^Y(t)$ are defined in \Cref{eq:euclidean_reduction_full_matrices,eq:update_full_matrices}.
If $\beta\neq0$, the updates are
\begin{equation}
    \label{eq:update_bases_no_cycle}
    \begin{alignedat}{2}
        \B_d^Z(t+1) &= \B_d^Z(t),
        \qquad
        & U^{t+1} &= \begin{pmatrix} \vphantom{\dfrac{1}{1}} \B_d^Z(t) & \widetilde \B_d^Y(t) & \gamma\end{pmatrix},
        \\
        \B_d^Y(t+1) &= \widetilde \B_d^Y(t) \cup \{\gamma\},
        \qquad
        & H^{t+1} &= \begin{pmatrix} \widetilde{H} & \beta\end{pmatrix}.
    \end{alignedat}
\end{equation}

\begin{example}
    \label{ex:cycle_euclidean_reduction}
    The reduction of \Cref{eq:euclidean_reduction_full_matrices} applied to the matrix in \Cref{eq:integral_column_echelon_form} yields%
    \[
        \setlength{\arraycolsep}{6pt}
        \underbracket{\begin{pNiceArray}{cccc}[first-row]
            c_1 & \cdots & c_t & \tau_{t+1}  \\
            2 & & * & 1 \\
            \vdots & \cdots & \vdots & \vdots \\
            0 &  & * & 0 
        \end{pNiceArray}}_{\begin{pmatrix}H^t&\partial\tau_{t+1}\end{pmatrix}}
        \underbracket{\begin{pNiceArray}{cccc}
            0 & & * & -1 \\
            \vdots & \cdots & \vdots & \vdots \\
            1 &  & * & 2 
        \end{pNiceArray}}_{Q}
        =
        \underbracket{\begin{pNiceArray}{cccc}[first-row]
            c_1' & \cdots & c_t & \gamma  \\
            1 & & * & 0 \\
            \vdots & \cdots & \vdots & \vdots \\
            0 &  & * & 0 
        \end{pNiceArray}}_{\begin{pmatrix}\widetilde H & \beta\end{pmatrix}},
        \qquad
        \left\{
            \begin{aligned}
                c_1' &= \tau_{t+1},\\
                \gamma &= -c_1 + 2\tau_{t+1},\\
                \beta &=  0.
            \end{aligned}
        \right.
    \]
\end{example}

\subsubsection{Full procedure with the degree-\texorpdfstring{$(d+1)$}{(d+1)} and degree-\texorpdfstring{$d$}{d} matrices}
\label{subsubsec:presentation_matrix_tensor_full}

\subparagraph{Maintained data.}

We denote by $D_{d+1}^\mathrm{tens}(t)$ the presentation matrix built after processing $K^t$.
Its rows correspond to the cycle basis $\B_d^Z(t)$ and its columns to the $(d+1)$-simplices already inserted.
The procedure uses the degree-$d$ boundary matrix $D_d^t$ of \Cref{eq:graded_matrix_tensor_degree_d}~and%
\[
    D_d^t\; U^t = \begin{pmatrix}0 & H^t\end{pmatrix},
    \qquad
    W^t = (U^t)^{-1}.
\]
The columns of $U^t$ encode the bases $\B_d^Z(t)$ and $\B_d^Y(t)$.
Thus, it is enough to store%
\[
    D_{d+1}^\mathrm{tens}(t),
    \quad H^t,
    \quad\text{and}\quad W^t.
\]

\subparagraph{Full procedure.}

We describe the procedure for a simplex-wise filtration starting from the empty complex.
An arbitrary filtration is handled by ordering its simplices by filtration time, breaking ties by increasing dimension, and retaining their original times as grades.

For each insertion $K^{t+1}=K^t\cup\{\sigma\}$, we proceed as follows.

\begin{enumerate}
    \item
    \label{it:algorithm_tensor_case_d+1}
    If $\dim\sigma=d+1$, we compute the coordinates $r_\sigma$
    of $\partial\sigma$ in $\B_d^Z(t)$:
    \[
        \begin{pmatrix}r_\sigma\\0\end{pmatrix}
        = W^t\,\partial\sigma.
    \]
    The lower block vanishes because $\partial\sigma$ is a cycle.
    We update the matrix as
    \[
        D_{d+1}^\mathrm{tens}(t+1) = \begin{pmatrix} D_{d+1}^\mathrm{tens}(t) & r_\sigma \end{pmatrix}.
    \]
    The other data are left unchanged.
 
    \item
    If $\dim\sigma=d$, the degree-$d$ boundary matrix becomes
    \[
        D_d^{t+1}
        = \begin{pmatrix}D_d^t & \partial\sigma\end{pmatrix}.
    \]
    We perform the reduction $\left(\begin{matrix} \vphantom{\dfrac{1}{1}} H^t & \partial\sigma\end{matrix}\right)Q 
    = \left(\begin{matrix} \vphantom{\dfrac{1}{1}} \widetilde H & \beta\end{matrix}\right)$ of \Cref{eq:euclidean_reduction_full_matrices} and compute
    \begin{equation}
        \label{eq:update_chain_basis_inverse}
        \widehat W = 
        \begin{pmatrix}I_{z(t)}&0\\0&Q^{-1}\end{pmatrix}
        \begin{pmatrix}W^t&0\\0&1\end{pmatrix}.
    \end{equation}
    For efficiency, this update is done through the corresponding elementary
    row operations.

    \begin{enumerate}
        \item
        \label{it:algorithm_tensor_case_d_a}
        If $\beta=0$, a new cycle is created.
        We update $\B_d^Z(t),\B_d^Y(t),H^t$ as in \Cref{eq:update_bases_cycle}, and obtain $W^{t+1}$ by moving the last row of $\widehat W$ to position $z(t)+1$.
        Moreover, we set
        \[
            D_{d+1}^\mathrm{tens}(t+1)
            = \begin{pmatrix}D_{d+1}^\mathrm{tens}(t)\\0\end{pmatrix}.
        \]

        \item
        \label{it:algorithm_tensor_case_d_b}
        If $\beta\neq0$, no cycle is created.
        We update $\B_d^Z(t),\B_d^Y(t),H^t$ as in \Cref{eq:update_bases_no_cycle}, and set
        \[
            W^{t+1}=\widehat W,
            \qquad
            D_{d+1}^\mathrm{tens}(t+1)=D_{d+1}^\mathrm{tens}(t).
        \]
    \end{enumerate}

    \item
    \label{it:algorithm_tensor_case_d-1}
    If $\dim\sigma=d-1$, we append a zero row to the
    degree-$d$ boundary matrix and to $H^t$:
    \[
        D_d^{t+1}=\begin{pmatrix}D_d^t\\0\end{pmatrix},
        \qquad
        H^{t+1}=\begin{pmatrix}H^t\\0\end{pmatrix}.
    \]
\end{enumerate}

\noindent
Simplices of other dimensions trigger no update.
The full procedure is given in~\Cref{alg:presentation_matrix_tensor_module}.

\begin{theorem}[Proof p.~\pageref{proof:presentation_matrix_tensor_module}]
    \label{th:presentation_matrix_tensor_module} 
    \Cref{alg:presentation_matrix_tensor_module} takes as an input a simplicial filtration $(K^t)_{t=0}^T$ and returns the graded integer matrix $D_{d+1}^\mathrm{tens}$ of \Cref{eq:graded_matrix_tensor}.
    In particular, for every field $\kk$, the tensored matrix $D_{d+1}^\mathrm{tens}\otimes\kk$ is a presentation matrix for the tensor module $H_d(K^\bullet;\Z)\otimes\kk$.
    The algorithm performs $O(N^3)$ arithmetic operations, where $N$ is the number of simplices.
\end{theorem}

\subsection{Presentation matrix for the free module}
\label{subsec:presentation_matrix_free}

We now consider the free module $H_d^\mathrm{free}(K^\bullet;\Z)\otimes\kk$.
We follow the same strategy as in \Cref{subsec:presentation_matrix_tensor}: we derive a presentation (\Cref{eq:persistent_presentation_two_rows_free}) in \Cref{subsubsec:presentation_matrix_free_overview} and encode it as a matrix (\Cref{eq:graded_matrix_free}).
Building this matrix requires a \emph{saturation} routine that we present in \Cref{subsubsec:presentation_matrix_free_saturation}.
We state our complexity result in \Cref{subsubsec:presentation_matrix_free_full}.
In this section, we suppose that \Cref{alg:presentation_matrix_tensor_module} has already been run on $(K^t)_{t=0}^N$.
In particular, we assume that we have access to the presentation matrix $D_{d+1}^\mathrm{tens}$.
Its rows are expressed in the incremental cycle basis%
\[
    \B_d^Z = (z_1,\dots,z_n),
    \qquad 
    n=\rank Z_d(K^N;\Z).
\]

\subsubsection{Presentation via saturated boundaries}
\label{subsubsec:presentation_matrix_free_overview}

\subparagraph{Saturated boundaries.}

Let $t\in\{0,\dots,N\}$ be a filtration index.
In \Cref{eq:free_part}, we have defined the free homology group $H_d^\mathrm{free}(K^t;\Z)$ as the quotient
\[
    H_d^\mathrm{free}(K^t;\Z) = H_d(K^t;\Z) / H_d^\mathrm{tors}(K^t;\Z),
\]
where $H_d^\mathrm{tors}(K^t;\Z)$ denotes the torsion subgroup.
Equivalently, it can be defined as
\begin{equation}
    \label{eq:free_homology_via_saturation}
    H_d^{\mathrm{free}}(K^t;\Z) \cong Z_d(K^t;\Z)/B_d^{\mathrm{sat}}(K^t;\Z),    
\end{equation}
where the \emph{saturation} $B_d^{\mathrm{sat}}(K^t;\Z)$ of $B_d(K^t;\Z)$, seen as a subset of $Z_d(K^t;\Z)$, is
\[
    B_d^{\mathrm{sat}}(K^t;\Z)
    = \big\{z\in Z_d(K^t;\Z) \mid
        qz\in B_d(K^t;\Z)\text{ for some integer }q\geq1\big\}.
\]
Note that the saturation is the smallest subgroup of $Z_d(K^t;\Z)$ containing $B_d(K^t;\Z)$ such~that%
\[
    Z_d(K^t;\Z)/B_d^{\mathrm{sat}}(K^t;\Z)
    \text{ is torsion-free}.
\]

\subparagraph{Persistent presentation.}

By \Cref{eq:free_homology_via_saturation}, the free module fits in the short exact sequence
\[
    0 
    \xrightarrow{~~~~} 
    B_d^{\mathrm{sat}}(K^t;\Z)
    \xrightarrow{~\iota^t~} 
    Z_d(K^t;\Z)
    \xrightarrow{~~~~} 
    H_d^{\mathrm{free}}(K^t;\Z)
    \xrightarrow{~~~~}
    0,
\]
where $\iota^t$ denotes the inclusion.
Tensoring with $\kk$ again yields a short exact sequence
\begin{equation}
    \label{eq:exact_sequence_free}
    0 
    \xrightarrow{~~~~}
    B_d^{\mathrm{sat}}(K^t;\Z)\otimes\kk
    \xrightarrow{~\iota^t\otimes\kk~} 
    Z_d(K^t;\Z)\otimes\kk
    \xrightarrow{~~~~}
    H_d^{\mathrm{free}}(K^t;\Z)\otimes\kk 
    \xrightarrow{~~~~}
    0.
\end{equation}
The fact that the sequence remains exact follows from $H_d^{\mathrm{free}}(K^t;\Z)$ being a free abelian~group.

Both sequences above are natural.
In particular, the first one yields a commutative~diagram%
\begin{equation*}
    \begin{tikzcd}[
        column sep=normal,
        row sep=large,
        continue above/.style={
            dashed,
            start anchor={[yshift=1.8em]north},
            end anchor=north
        },
        continue below/.style={
            dashed, dash,
            start anchor=south,
            end anchor={[yshift=-1.8em]south}
        }
    ]
        0 \arrow[r]
        & B_d^{\mathrm{sat}}(K^t;\Z)
            \arrow[continue above]
            \arrow[r, "\iota^t"]
            \arrow[d]
        & Z_d(K^t;\Z)
            \arrow[continue above]
            \arrow[r]
            \arrow[d]
        & H_d^{\mathrm{free}}(K^t;\Z)
            \arrow[continue above]
            \arrow[r]
            \arrow[d, "\psi^t"]
        & 0
        \\
        0 \arrow[r]
        & B_d^{\mathrm{sat}}(K^{t+1};\Z)
            \arrow[r, "\iota^{t+1}"]
            \arrow[continue below]
        & Z_d(K^{t+1};\Z)
            \arrow[r]
            \arrow[continue below]
        & H_d^{\mathrm{free}}(K^{t+1};\Z)
            \arrow[r]
            \arrow[continue below]
        & 0
    \end{tikzcd}
\end{equation*}
After tensorization, we obtain a commutative diagram of $\kk$-vector spaces:
\begin{equation}
    \label{eq:persistent_presentation_two_rows_free}
    \begin{tikzcd}[
        column sep=large,
        row sep=large,
        continue above/.style={
            dashed,
            start anchor={[yshift=1.8em]north},
            end anchor=north
        },
        continue below/.style={
            dashed, dash,
            start anchor=south,
            end anchor={[yshift=-1.8em]south}
        }
    ]
        0 \arrow[r]
        &[-2em] B_d^{\mathrm{sat}}(K^t;\Z)\otimes\kk
            \arrow[continue above]
            \arrow[r, "\iota^t\otimes\kk"]
            \arrow[d]
        & Z_d(K^t;\Z)\otimes\kk
            \arrow[continue above]
            \arrow[r]
            \arrow[d]
        &[-1.75em] H_d^{\mathrm{free}}(K^t;\Z)\otimes\kk
            \arrow[continue above]
            \arrow[r]
            \arrow[d, "\psi^t\otimes\kk"]
        &[-2em] 0
        \\
        0 \arrow[r]
        & B_d^{\mathrm{sat}}(K^{t+1};\Z)\otimes\kk
            \arrow[r, "\iota^{t+1}\otimes\kk"]
            \arrow[continue below]
        & Z_d(K^{t+1};\Z)\otimes\kk
            \arrow[r]
            \arrow[continue below]
        & H_d^{\mathrm{free}}(K^{t+1};\Z)\otimes\kk
            \arrow[r]
            \arrow[continue below]
        & 0
    \end{tikzcd}
\end{equation}
We recognize the free module $H_d^{\mathrm{free}}(K^\bullet;\Z)\otimes\kk$ in the last column.

Note that these two diagrams are analogous to \Cref{eq:persistent_presentation_two_rows_integral,eq:persistent_presentation_two_rows_tensor} obtained for the tensor module in \Cref{subsubsec:presentation_matrix_tensor_overview}.
Following our strategy for the tensor module, it suffices to express the inclusion maps $\iota^t\colon B_d^{\mathrm{sat}}(K^{t};\Z)\to Z_d(K^{t};\Z)$ in bases compatible with the filtration.
For the integral $d$-cycles, we consider the basis $\B_d^Z = (z_1,\dots,z_n)$ already built by \Cref{alg:presentation_matrix_tensor_module}.
For the saturated boundaries, we build in \Cref{eq:incremental_basis_saturated_boundaries} an incremental basis
\[
    \B_d^{B} = (b_1,\dots,b_{m'}),
    \qquad
    m' = \rank B_d(K^N;\Z).
\]
The desired presentation matrix is obtained by expressing this inclusion in the chosen bases:
\begin{equation}
    \label{eq:graded_matrix_free}
    \setlength{\arraycolsep}{3pt}
    D_d^\mathrm{free} 
    \;=\; 
    \Matrix\big(\iota^N;\B_d^B,\B_d^Z\big)
    \;=\;
    \begin{pNiceArray}{ccccc}[first-row,first-col]
        &
        b_1
        & b_2
        & b_3
        & \cdots
        & b_{m'}
        \\
        \begin{array}{c}
        z_1\\
        z_2\\
        \vdots\\
        z_{n}
        \end{array}
        &
        \begin{array}{c}
        \rule{0.4pt}{2em}\\[-0.3em]
        b_1\\[0.3em]
        \rule{0.4pt}{2em}
        \end{array}
        &
        \begin{array}{c}
        \rule{0.4pt}{2em}\\[-0.3em]
        b_2\\[0.3em]
        \rule{0.4pt}{2em}
        \end{array}
        &
        \begin{array}{c}
        \rule{0.4pt}{2em}\\[-0.3em]
        b_3\\[0.3em]
        \rule{0.4pt}{2em}
        \end{array}
        &
        \cdots
        &
        \begin{array}{c}
        \rule{0.4pt}{2em}\\[-0.3em]
        b_{m'}\\[0.3em]
        \rule{0.4pt}{2em}
        \end{array}
    \end{pNiceArray}.
\end{equation}

\subsubsection{Iterative saturation of the boundary lattice}
\label{subsubsec:presentation_matrix_free_saturation}

\subparagraph{Saturated lattice.}

A convenient feature of the saturated boundaries is that the quotients
\[
    B_d^{\mathrm{sat}}(K^{t+1};\Z) / B_d^{\mathrm{sat}}(K^t;\Z)
\]
are all free.
In particular, the basis $\B_d^{B}$ can be built incrementally.
We recall that such an incremental basis need not exist for ordinary boundaries, as discussed in \Cref{subsubsec:presentation_matrix_tensor_cycles}.

To compute a basis for $B_d^{\mathrm{sat}}(K^t;\Z)$, we use the presentation matrix $D_{d+1}^{\mathrm{tens}}$ (\Cref{eq:graded_matrix_tensor}) returned by \Cref{alg:presentation_matrix_tensor_module}.
It is an integral matrix of shape $n\times m$, where 
\[
    n = \rank Z_d(K^N;\Z),
    \qquad 
    m = \rank C_{d+1}(K^N;\Z).
\]
We denote its $j$th column by
\[
    a_j = D_{d+1}^{\mathrm{tens}}[\colon,j].
\]
It is seen as an element of $\Z^n$.
Over $\Z$, the first $j$ columns span the integer lattice
\[
    \Lambda_j = \Span_{\Z}(a_1,\dots,a_j)\subset \Z^n.
\]
Its saturation is obtained by taking all rational linear combinations and intersecting with $\Z^n$:%
\begin{equation}
    \label{eq:saturated_relation_prefix}
    \Lambda^\mathrm{sat}_j = \Span_{\Q}(a_1,\dots,a_j)\cap\Z^n.
\end{equation}

At filtration time $t$, let $j$ be the number of relation columns whose grades are at most $t$.
Then, under the identification through $\B_d^Z$, the lattice
$\Lambda_j^{\mathrm{sat}}$ represents $B_d^{\mathrm{sat}}(K^t;\Z)$.

\begin{example}
    \label{ex:saturation_lattices}
    Consider the following $2\times3$ integer matrix and define two vectors of $\Z^2$:
    \[
        D_{d+1}^{\mathrm{tens}} =
        \begin{pNiceArray}{ccc}[first-row]
            a_1&a_2&a_3\\
            1&1&1\\
            -1&1&0
        \end{pNiceArray},
        \qquad
        x=\begin{pmatrix}1\\-1\end{pmatrix},
        \qquad
        y=\begin{pmatrix}0\\1\end{pmatrix}.
    \]
    The successive integer spans and their saturations are
    \[
        \begin{alignedat}{3}
            \Lambda_1 &= \Z x,
            &\qquad \Lambda_2 &= \Z x\oplus2\Z y,
            &\qquad \Lambda_3 &= \Z x\oplus\Z y,\\
            \Lambda^{\mathrm{sat}}_1 &= \Z x,
            &\qquad \Lambda^{\mathrm{sat}}_2 &= \Z x\oplus\Z y,
            &\qquad \Lambda^{\mathrm{sat}}_3 &= \Z x\oplus\Z y.
        \end{alignedat}
    \]
\end{example}

\subparagraph{Maintained data.}

During the algorithm, we process the columns of $D_{d+1}^{\mathrm{tens}}$ successively and update a basis $\B_d^B(t)$ of $B_d^{\mathrm{sat}}(K^t;\Z)$.
More precisely, we maintain an invertible matrix
\begin{equation}
    \label{eq:change_of_basis_free_module}
    U_j\in\GL_n(\Z),
    \qquad
    n = \rank Z_d(K^N;\Z),
\end{equation}
whose first $s(j)$ columns form a basis of $\Lambda^{\mathrm{sat}}_{j}$, where 
\[
    s(j) = \rank \Lambda^{\mathrm{sat}}_j.
\]
We identify $Z_d(K^N;\Z)$ with $\Z^n$ via the basis $\B_d^Z$.
At filtration time $t$, let $j$ be the number of relation columns
whose grades are at most $t$.
The incremental basis of $B_d^{\mathrm{sat}}(K^t;\Z)$ is then
\begin{equation}
    \label{eq:incremental_basis_saturated_boundaries}
    \B_d^B(t) = \big(U_j[:,1],\dots,U_j[:,s(j)]\big),
\end{equation}
and the working basis of $Z_d(K^N;\Z)$ is  $(U_j[:,1],\dots,U_j[:,n])$.
During the algorithm, we also maintain the inverse $W_j=U_j^{-1}$.
By convention, we start from $U_0 = I_n$ and $s(0)=0$.

\subparagraph{Saturation routine.}

Suppose we introduce a new column $a_{j+1}$ of $D_{d+1}^{\mathrm{tens}}$.
We must check whether it increases the rank of $\Lambda^{\mathrm{sat}}_j$.
To do so, we compute its decomposition in the working basis and extract its last coordinates:
\begin{equation}
    \label{eq:saturation_routine_coordinates}
    w=W_j\,a_{j+1},
    \qquad v=w\big[(s(j)+1) : n\big].    
\end{equation}
Note that $v$ can be understood as the coordinates of the class of $a_{j+1}$ in the quotient lattice
\[
    \Z^n / \Lambda^{\mathrm{sat}}_j.
\]
If $v=0$, then $a_{j+1}$ belongs to $\Lambda^{\mathrm{sat}}_j$.
Thus, $\Lambda^{\mathrm{sat}}_{j+1}=\Lambda^{\mathrm{sat}}_j$ and no update is needed.

Otherwise, the saturated basis must be extended.
First, we take the gcd of $v$'s entries:%
\[
    g = \gcd\big(v_1,\dots,v_{n-s(j)}\big) > 0.
\]
Note that $g=1$ if and only if $v$ is \emph{primitive} in $\Z^{n-s(j)}$, i.e., if it has no nontrivial divisors.
Via the extended Euclidean algorithm (see \Cref{eq:euclidean_reduction}), we reduce $v$ to $\left(\begin{matrix} g & 0 & \cdots & 0\end{matrix}\right)^\top$:
\begin{equation*}
    v = 
    \begin{pmatrix}
    v_{1}\\
    \vdots\\
    v_{n-s(j)-2}\\
    v_{n-s(j)-1}\\
    v_{n-s(j)\phantom{-1}} 
    \end{pmatrix}
    \longrightarrow
    \begin{pmatrix}
    v_{1}\\
    \vdots\\
    v_{n-s(j)-2}\\
    *\\
    0 
    \end{pmatrix}
    \longrightarrow
    \begin{pmatrix}
    v_{1}\\
    \vdots\\
    *\\
    0\\
    0 
    \end{pmatrix}
    \longrightarrow 
    ~\cdots~
    \longrightarrow 
    \begin{pmatrix}
    g\\
    \vdots\\
    0\\
    0\\
    0 
    \end{pmatrix}
    = g e_1,
    \qquad
    e_1 = 
    \begin{pmatrix}
    1\\
    \vdots\\
    0\\
    0\\
    0 
    \end{pmatrix}.
\end{equation*}
Let $Q\in\GL_{n-s(j)}(\Z)$ be the composition of all these Euclidean transformations.
In particular,%
\begin{equation}
    \label{eq:free_saturation_euclidean_reduction}
    Qv=g e_1.
\end{equation}
Then, we update the change-of-basis matrices as
\begin{equation}
    \label{eq:free_saturation_basis_update}
    U_{j+1}=U_j\begin{pmatrix}I_{s(j)}&0\\0&Q^{-1}\end{pmatrix}
    \qquad
    W_{j+1}=\begin{pmatrix}I_{s(j)}&0\\0&Q\end{pmatrix} W_j.
\end{equation}
The first $s(j)$ columns of $U_j$ are preserved, hence the basis is incremental.
The next column~is%

\begin{lemma}[Proof p.~\pageref{proof:free_saturation_primitive_relation}]
    \label{lem:free_saturation_primitive_relation}
    $u = U_{j+1}\big[:\,,s(j)+1\big] = \big( a_{j+1}-U_j\big[:\,,1:s(j)\big]w\big[1:s(j)\big] \big)/g$.
\end{lemma}
Since $v/g$ is primitive in $\Z^{n-s(j)}$, the class of $u$ is primitive in $\Z^n/\Lambda^{\mathrm{sat}}_j$.
Therefore,
\[
    \Lambda^{\mathrm{sat}}_{j+1}=\Lambda^{\mathrm{sat}}_j \oplus \Z u,
    \qquad
    s(j+1)=s(j)+1.
\]

\begin{example}
    We apply the saturation routine to the matrix $D^{\mathrm{tens}}_{d+1} = \left(\begin{smallmatrix}1&1&1\\-1&1&0\end{smallmatrix}\right)$ of \Cref{ex:saturation_lattices}.
    We maintain the invertible matrix $U_j$, its inverse $W_j$, and the rank $s(j)=\rank \Lambda^{\mathrm{sat}}_j$.
    At the start, $U_0=W_0=I_2$ and $s(0)=0$. 
    For the first column $a_1=(1,-1)^\top$, the computations are
    \[
        w=W_0\,a_1=\begin{pmatrix}1\\-1\end{pmatrix},
        \qquad
        v=\begin{pmatrix}1\\-1\end{pmatrix},
        \qquad g=1,
        \qquad
        Q_1v=\begin{pmatrix}1\\0\end{pmatrix},
        \qquad
        Q_1=\begin{pmatrix}1&0\\1&1\end{pmatrix}.
    \]
    The updated matrices are
    \[
        U_1=Q_1^{-1}=
        \left(\begin{array}{c|c}
            1&0\\
            -1&1
        \end{array}\right),
        \qquad
        W_1=Q_1=
        \begin{pmatrix}
            1&0\\
            1&1
        \end{pmatrix},
        \qquad s(1)=1.
    \]
    For the second column $a_2=(1,1)^\top$, the routine computes
    \[
        w=W_1\,a_2=\begin{pmatrix}1\\2\end{pmatrix},
        \qquad
        v=\begin{pmatrix}2\end{pmatrix},
        \qquad g=2,
        \qquad Q_2v=\begin{pmatrix}2\end{pmatrix},
        \qquad Q_2=\begin{pmatrix}1\end{pmatrix}.
    \]
    We update the change-of-basis matrices via
    \[
        U_2=U_1
        \begin{pmatrix}
            1&0\\
            0&Q_2^{-1}
        \end{pmatrix}
        =
        \begin{pmatrix}
            1&0\\
            -1&1
        \end{pmatrix},
        \qquad
        W_2=
        \begin{pmatrix}
            1&0\\
            0&Q_2
        \end{pmatrix}
        W_1=
        \begin{pmatrix}
            1&0\\
            1&1
        \end{pmatrix}.
    \]
    The matrices remain unchanged, but the rank increases to $s(2)=2$.
    The second column of $U_2$ is now kept as a saturated relation.
    For the last column $a_3=(1,0)^\top$, we obtain
    \[
        w=W_2\,a_3=\begin{pmatrix}1\\1\end{pmatrix}.
    \]
    Since $s(2)=2$, there are no remaining coordinates, and $v$ is the zero vector in $\Z^0$.
    Thus, no update is needed.
    The final matrix is $D^{\mathrm{free}}_d
     = \left(\begin{smallmatrix}1&0\\-1&1\end{smallmatrix}\right)$.
     Its columns inherit the times of~$a_1,a_2$.%
\end{example}

\subsubsection{Correctness and complexity of the full procedure}
\label{subsubsec:presentation_matrix_free_full}

The full procedure is given in \Cref{alg:presentation_matrix_free_module}.
It processes the columns of $D_{d+1}^{\mathrm{tens}}$ in filtration order, maintaining the invertible matrix $U_j$ (\Cref{eq:change_of_basis_free_module}), its inverse $W_j$, and the rank $s(j)$ of the saturated relation lattice.
The first $s(j)$ columns of $U_j$ form an incremental basis of this lattice.
Whenever the rank increases, we produce a new primitive relation via Euclidean transformations.
These relations form $D_d^{\mathrm{free}}$, with appearance times inherited from $D_{d+1}^{\mathrm{tens}}$.

\begin{theorem}[Proof p.~\pageref{proof:presentation_matrix_free_module}]
    \label{th:presentation_matrix_free_module}
    Given the graded matrix $D_{d+1}^\mathrm{tens}$ returned by \Cref{alg:presentation_matrix_tensor_module}, \Cref{alg:presentation_matrix_free_module} returns the graded matrix $D_d^\mathrm{free}$ of \Cref{eq:graded_matrix_free}.
    In particular, for every field $\kk$, the tensored matrix $D_d^\mathrm{free}\otimes\kk$ is a presentation matrix for the free module $H_d^\mathrm{free}(K^\bullet;\Z)\otimes\kk$.
    The algorithm performs $O(N^3)$ arithmetic operations, where $N$ is the number of simplices.
\end{theorem}

\section{Filtrations with large torsion}
\label{sec:filtrations_with_large_torsion}

In this section, we build 2-dimensional simplex-wise filtrations $(K^t)_{t=1}^N$ with many exceptional primes $p\in\PP$, as defined in \Cref{eq:definition_exceptional_primes}.
In particular, these primes change the $H_1$-barcode:
\[
    p\in\PP 
    \iff
    \barcode\big(H_1(K^\bullet;\F_p)\big) \neq \barcode\big(H_1(K^\bullet;\Q)\big).
\]
We first recall Newman's construction of simplicial complexes with large torsion (\Cref{sec:filtrations_with_large_torsion_newman}).
We then adapt it to filtrations and prove lower bounds on the complexity of the algorithms discussed in \Cref{subsec:intro_algorithms}: brute force (\Cref{sec:filtrations_with_large_torsion_brute_force}), modular reconstruction (\Cref{sec:filtrations_with_large_torsion_modular_reconstruction}), and omni-field persistence (\Cref{sec:filtrations_with_large_torsion_omnifield}).
Another key ingredient is Morozov's construction of a filtration which attains the worst-case running time of the persistence algorithm~\cite{Morozov2005}.

\subsection{Sphere-and-telescope construction}
\label{sec:filtrations_with_large_torsion_newman}

Let $M\geq 2$ be an integer.
In \cite[Lemma 3.1]{Newman2019}, Newman builds a small 2-dimensional simplicial complex $X$ whose first homology group is
\begin{equation*}
    H_{1}(X;\Z) \cong \Z^{k-1} \oplus \Z/M\Z    
\end{equation*}
for some $k\geq1$.
The construction goes as follows.
First, write the binary expansion of $M$ as
\[
    M = 2^{n_1} + \cdots + 2^{n_k}, 
    \qquad
    0 \leq n_1 < n_2 < \cdots < n_k.
\]
Second, build two complexes $X_\mathrm{tors}$ and $X_\mathrm{base}$ whose first homology groups can be presented~as
\begin{align}
    H_1\left(X_\mathrm{tors};\Z\right) &\cong \langle \gamma_0,\dots, \gamma_{n_k} \mid 2\gamma_0=\gamma_1,\;\dots,\; 2\gamma_{n_k-1}=\gamma_{n_k} \rangle, \label{eq:newman_complex_L1}\\
    H_1\left(X_\mathrm{base};\Z\right) &\cong \langle \tau_1,\dots,\tau_k \mid \tau_1 + \dots +\tau_k = 0 \rangle. \notag
\end{align}
In addition, these complexes are such that the generators $\gamma_0,\dots, \gamma_{n_k}$ and $\tau_1,\dots,\tau_k$ of these presentations are represented by triangular loops (three edges).
Last, take the simplicial~gluing%
\begin{equation}
    \label{eq:Newman_complex}
    X = X_\mathrm{base} \sqcup_f X_\mathrm{tors}
\end{equation}
where the simplicial map $f\colon A\subset X_\mathrm{tors}\to X_\mathrm{base}$ sends the triangular loops 
\[
    \gamma_{n_i} \to \tau_i, \qquad i\in\{1,\dots,k\}.
\]
In particular, only $k$ of the $n_k+1$ domain triangular loops $\gamma_i$ are affected.
Crucially, Newman builds the complexes so that $X$ has only $O(\log M)$ simplices.
We summarize this construction.

\begin{lemma}[{\cite[Lemma 3.1]{Newman2019}}]
    \label{lem:Newman_complex}
    For $M\geq2$, Newman's complex $X$ built in \Cref{eq:Newman_complex} has $O(\log M)$ simplices, and its first homology group has torsion $H_1^\mathrm{tors}(X;\Z) \cong \Z/M\Z$.
\end{lemma}

\begin{figure}[!htbp]
	\centering
	\subcaptionbox{
        \label{subfig:newman_construction_sphere}
		$X_\mathrm{base}$ is a sphere with $k$ holes.
        The corresponding cycles satisfy $\tau_1+\cdots+\tau_k=0$.
	}[0.29\textwidth]{%
		\includegraphics[width=1\linewidth]{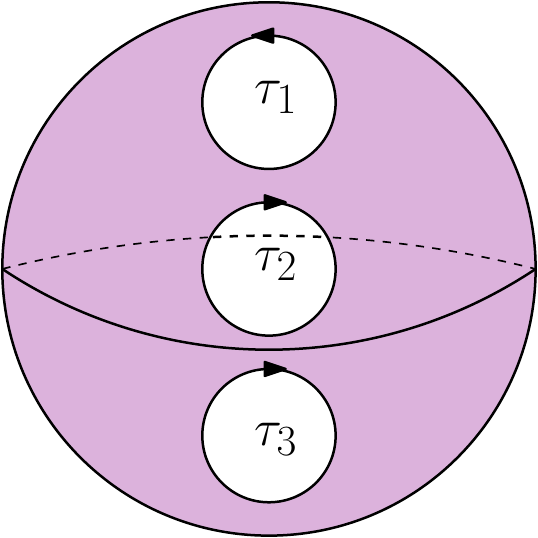}
		}
	\hfill
	\subcaptionbox{
        \label{subfig:newman_construction_telescope}
		$X_\mathrm{tors}$ is the successive gluing of $n_k+1$ annuli along degree-2 maps.
        The corresponding homology cycles satisfy $2\gamma_0=\gamma_1,\;\dots,\; 2\gamma_{n_k-1}=\gamma_{n_k}$.
	}[0.70\textwidth]{%
		\includegraphics[width=1\linewidth]{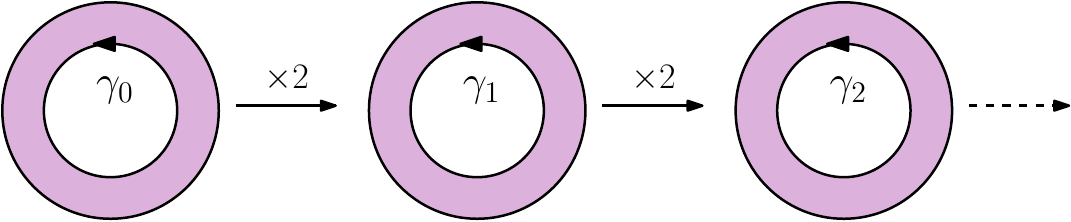}}
	\caption{
		Newman's complex is obtained by gluing $X_\mathrm{tors}$ on $X_\mathrm{base}$ via the relations $\gamma_{n_i} = \tau_i$ for all $i\in\{1,\dots,k\}$.
        It has few simplices but a large torsion subgroup; see \Cref{lem:Newman_complex}.
		}
\end{figure}

\begin{remark}
    Concretely, the complex $X_\mathrm{base}$ is chosen as a sphere with $k$ holes, representing the generators $\tau_1,\dots,\tau_k$ with the correct orientation; see \Cref{subfig:newman_construction_sphere}.
    Moreover, $X_\mathrm{tors}$ is obtained as the mapping telescope of $n_k$ degree-2 maps
    \[
        S^1 \xrightarrow[]{~\times2~} S^1 \xrightarrow[]{~\times2~} \cdots \xrightarrow[]{~\times2~} S^1.
    \]
    In other words, $X_\mathrm{tors}$ is the successive gluing of $n_k+1$ annuli via degree-2 maps, inducing the relations $2\gamma_0=\gamma_1,\;\dots,\;2\gamma_{n_k-1}=\gamma_{n_k}$; see \Cref{subfig:newman_construction_telescope}.
\end{remark}

\subsection{Many exceptional primes and the brute-force algorithm}
\label{sec:filtrations_with_large_torsion_brute_force}

Consider the brute force algorithm described in \Cref{subsec:intro_algorithms}.
Given a simplicial filtration $K^\bullet$ with exceptional primes $\PP$, the algorithm runs an independent reduction over all exceptional fields $\F_p$, $p\in\PP$.
To build a costly instance, we take Newman's complex on many primes.

\subparagraph{Base construction.}

Let $N > 0$ be sufficiently large and define
\[
    L=\left\lfloor\frac{N}{\log N}\right\rfloor
    \qquad\text{and}\qquad
    M=p_1\cdots p_L,
\]
where $p_1,\dots,p_L$ are the first $L$ primes.
By the prime number theorem, $p_L=O(L\log L)$,~hence
\[
    \log M
    \leq L\log p_L
    = O(L\log L)
    = O(N).
\]
Let $X_N=X$ be the corresponding Newman complex, and let $X_N^\bullet$ be any simplex-wise filtration on it.
By \Cref{lem:Newman_complex}, the $H_1$-torsion subgroup of $X_N$ is $\Z/M\Z$ and it has 
\[
    O(\log M) = O(N)
\]
simplices.
Since a filtration can only introduce new exceptional primes, $X_N^\bullet$ satisfies 
\[
    |\PP| \geq L = \Omega\left( \frac{N}{\log N} \right).
\]
Thus, $X_N^\bullet$ has many exceptional primes.
However, this construction alone does not guarantee a costly instance of the brute-force algorithm.
We also seek a filtration on which the persistence algorithm attains its worst-case~complexity.

\subparagraph{Adding a costly reduction.}

To this end, we use Morozov's worst-case filtration $\widetilde{Y}_N^\bullet$~\cite{Morozov2005}, described in \Cref{sec:filtrations_with_large_torsion_omnifield}.
Here, we only need to know that it has $\Theta(N)$ simplices, and that the classical persistence algorithm requires
\[
    \Omega(N^3)
\]
field operations, regardless of the field.
Taking disjoint vertex sets, we append $\widetilde{Y}_N^\bullet$ to $X_N^\bullet$:
\begin{equation}
    \label{eq:instance_brute_force}
    K_N^\bullet = X_N^\bullet \sqcup \widetilde{Y}_N^\bullet.
\end{equation}
In particular, the boundary matrix $D(K_N^\bullet)$ of $K_N^\bullet$ is block diagonal:
\[
    D(K_N^\bullet) = \begin{pmatrix} D(X_N^\bullet) & 0 \\ 0 & D(\widetilde{Y}_N^\bullet)\end{pmatrix}.
\]
The filtration $K_N^\bullet$ still has $\Theta(N)$ simplices, $\Omega(N/\log N)$ exceptional primes, and the persistence algorithm still spends $\Omega(N^3)$ time on it.
We directly obtain the following bound.

\begin{proposition}
    \label{prop:number_exceptional_primes_newman}
    The filtration $K_N^\bullet$ of \Cref{eq:instance_brute_force} has $\Theta(N)$ simplices and 
    \[
        \Omega\left( \frac{N}{\log N} \right)
    \]   
    exceptional primes.
    Moreover, on this instance, the brute-force algorithm has complexity
    \[
        \Omega\biggl(\frac{N^4}{\log N}\biggr).
    \]
\end{proposition}

\subsection{Many exceptional bars and the modular reconstruction algorithm}
\label{sec:filtrations_with_large_torsion_modular_reconstruction}

\subparagraph{Complexity.}

Consider a simplex-wise filtration $K^\bullet$ and $L$ fields $\F_{p_1},\dots,\F_{p_L}$.
As explained in \Cref{subsec:intro_algorithms}, Boissonnat and Maria's algorithm is sensitive to the number of exceptional primes in $K^\bullet$ but also to the total number of bars (without multiplicity) found in the~barcodes:%
\[
    \QQ_L = \bigl|\bigl\{ [i,j) \text{ such that } [i,j) \in \barcode(H_d(K^\bullet;\kk)) \text{ for some } \kk \in \big\{\F_{p_1},\dots,\F_{p_L}\big\} \text{ and } d\geq0 \bigr\}\bigr|.
\]
More precisely, they give in \cite[§5.2]{Boissonnat_2019} the complexity
\begin{equation}
    \label{eq:complexity_modular_reconstruction}
    O\bigl(L\times(\QQ_L-\QQ_*)+\QQ_L^3\bigr)    
\end{equation}
where, for simplicity, we count an arithmetic operation as $O(1)$ instead of $A_L$.

\begin{figure}[!htbp]
	\centering
    \includegraphics[width=.825\linewidth]{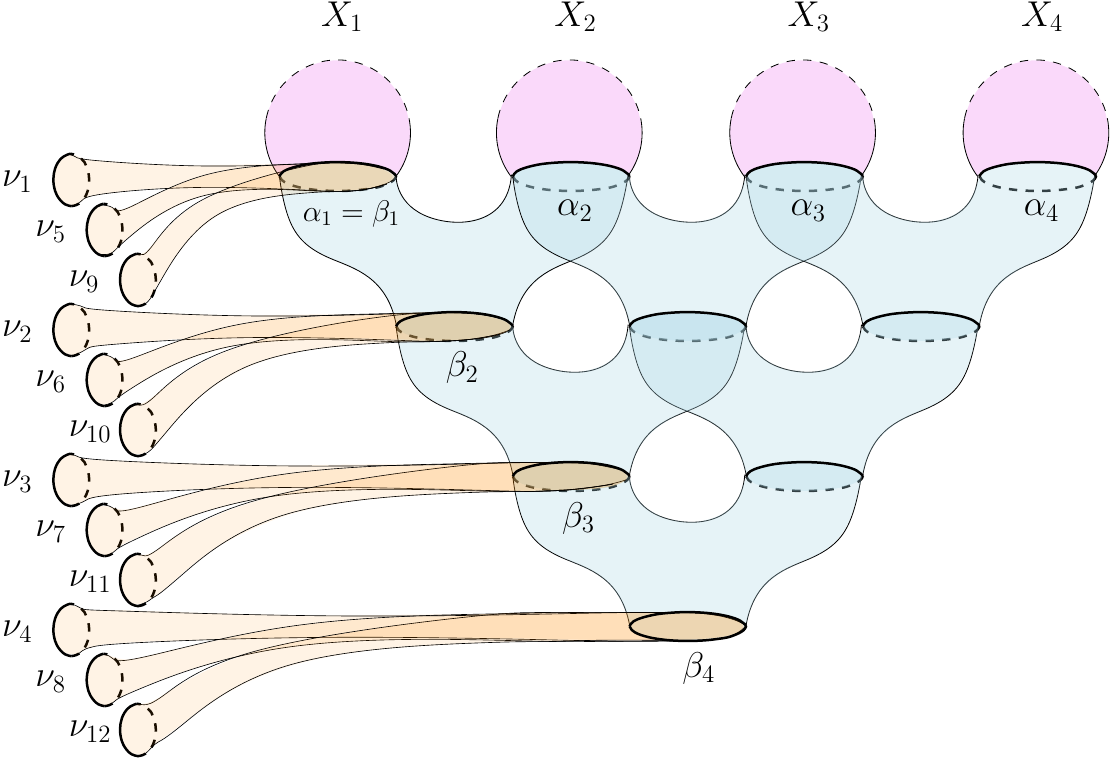}
	\caption{
        In \Cref{prop:number_exceptional_bars}, we build a filtration $K^\bullet$ with many exceptional bars. 
        It depends on two parameters $L,R>0$. 
        First, we choose $L$ primes $p_1,\dots,p_L$ and Newman complexes $X_1,\dots,X_L$ with distinguished classes $[\alpha_i]\in H_1(X_i;\Z)$ with the property that, over $\F_{p_j}$, $\alpha_i$ vanishes when $i>j$ and is nonzero when $i\leq j$.
        Next, using the coefficients of Pascal's triangle and triangulated pairs of pants, we combine these classes into triangular loops $\beta_1,\dots,\beta_L$ with the property that, over $\F_{p_i}$, any $i$ of them are linearly independent.
        Last, we connect $R$ new loops $\nu_1,\dots,\nu_R$ to the $\beta_i$ periodically using triangulated cylinders.
        A certain filtration of this complex has the required property.
		}
	\label{fig:complex_many_bars}
\end{figure}

Suppose now that the $p_1,\dots,p_L$ are chosen as all the exceptional primes $\PP$ of $(K^t)_{t=0}^N$, hence that $L=|\PP|$.
By \Cref{prop:number_exceptional_primes_newman}, there are filtrations for which $L=\Omega(N/\log N)$.
However, this alone gives only a weak lower bound on the expression in \Cref{eq:complexity_modular_reconstruction}.
To obtain a stronger bound, we focus on the cubic term $\QQ_L^3$.
We will build filtrations for which%
\[
    \QQ_L = \Omega\left(\frac{N^{3/2}}{(\log N)^{1/2}}\right),
\]
see \Cref{prop:number_exceptional_bars}.
An overview of the construction is given in \Cref{fig:complex_many_bars}.

We emphasize that our result concerns only the complexity bound stated in \Cref{eq:complexity_modular_reconstruction}.
It does not establish a lower bound on the algorithm's worst-case running time.

\subparagraph{Base complex.}

Our construction starts from a combination of Newman complexes.
Let $L>0$ be a large integer that will be chosen later.
Choose $L$ primes such that
\[
    L \leq p_1 < \cdots < p_L \leq L^2.
\]
This is possible if $L$ is sufficiently large by the prime number theorem.
For $i\in\{1,\dots,L\}$,~put%
\[
    M_i=\prod_{\ell=i}^L p_\ell,
\]
and let $X_i$ be the corresponding Newman complex, as in \Cref{sec:filtrations_with_large_torsion_newman}.
Each $X_i$ has $O(\log M_i)$ simplices, and has $H_1$-torsion $\Z/M_i\Z$, generated by a triangular cycle called $\gamma_0$ in \Cref{eq:newman_complex_L1}, which we will denote by $\alpha_i$ here.
Take the disjoint union of these complexes:%
\begin{equation}
    \label{eq:exceptional_bars_base_complex}
    X = \bigsqcup_{i=1}^L X_i.
\end{equation}
Since $X$ is a disjoint union, its homology is the direct sum of each contribution:
\[
    H_1(X;\Z) \cong \Z^k \oplus \bigoplus_{i=1}^L \Z/M_i\Z
\]
for a certain $k\geq0$.
Each $\Z/M_i\Z$ is generated by the class $[\alpha_i]\in H_1(X_i;\Z)$ seen in $H_1(X;\Z)$ via the inclusion $X_i\hookrightarrow X$.
Moreover, given $i,j \in\{1,\dots,L\}$, the definition of $M_i$~yields%
\[
    p_j \text{ divides } M_i \iff i \leq j.
\]
Thus, over $\F_{p_i}$, $[\alpha_1],\dots,[\alpha_i]$ are independent and $[\alpha_{i+1}],\dots,[\alpha_L]$ vanish. 
In particular,
\begin{equation}
    \label{eq:exceptional_bars_base_complex_dimension}
    \dim H_1(X;\F_{p_i}) = k+i.
\end{equation}
Here and below, we use the same notation for an integral class and its image under the map
\[
    H_1(X;\Z) \to H_1(X;\F_{p_i})
\]
induced by changing coefficients; see the universal coefficient theorem in \Cref{subsec:three_sources_of_torsion}.
In other words, the classes $[\alpha_1],\dots,[\alpha_i]$ form a basis of the subspace of $H_1(X;\F_{p_i})$ arising from integral torsion.
The contribution from the free part of $H_1(X;\Z)$ is not used in our~construction.%

\begin{lemma}[Proof p.~\pageref{proof:exceptional_bars_base}]
    \label{lem:exceptional_bars_base}
    The complex $X$ of \Cref{eq:exceptional_bars_base_complex} has $O(L^2\log L)$ simplices.
\end{lemma}

\subparagraph{Pascal basis.}

We introduce a more convenient family than $\alpha_1,\dots,\alpha_L$.
For $i\in\{1,\dots,L\}$,~let%
\begin{equation}
    \label{eq:exceptional_bars_pascal}
    [\beta_i]
    =
    \sum_{\ell=1}^i
    \binom{i-1}{\ell-1} [\alpha_\ell]
    \; \in H_1(X;\Z).
\end{equation}
These coefficients can be read on Pascal's triangle.
As an illustration, the first classes are:
\begin{align*}
    [\beta_1] = [\alpha_1],
    \qquad
    [\beta_2] = [\alpha_1]+[\alpha_2],
    \qquad
    [\beta_3] = [\alpha_1]+2[\alpha_2]+[\alpha_3].
\end{align*}
These new classes have two notable features.
First, they ``symmetrize'' the previous family:

\begin{lemma}[Proof p.~\pageref{proof:exceptional_bars_pascal}]
    \label{lem:exceptional_bars_pascal}
    Let $i\in\{1,\dots,L\}$.
    Under the map $H_1(X;\Z) \to H_1(X;\F_{p_i})$, the images of any $i$ distinct classes among $[\beta_1],\dots,[\beta_L]$ are linearly independent.
    Moreover, any collection of more than $i$ of these images is linearly dependent.
\end{lemma}

Secondly, they can be realized as triangular loops efficiently.

\begin{lemma}[Proof p.~\pageref{proof:exceptional_bars_pascal_construction}]
    \label{lem:exceptional_bars_pascal_construction}
    There exists a $2$-dimensional simplicial complex $\widetilde{X}\supset X$, obtained by adding $O(L^2)$ simplices, such that each class $[\beta_i]$ is represented in $\widetilde X$ by a triangular loop.
    Moreover, the inclusion $X\hookrightarrow \widetilde{X}$ induces an injection on $H_1$ over $\Z$ and every field.
\end{lemma}

\begin{remark}    
    Our construction exploits Pascal's triangle to reduce the number of simplices.
    We point out that realizing each sum in \Cref{eq:exceptional_bars_pascal} separately is possible but would give a larger construction.
    For instance, one could represent each term $\binom{i-1}{\ell-1}[\alpha_\ell]$ by a triangular loop using a telescope as in Newman's construction.
    Such a telescope requires $O\big(1+\log\binom{i-1}{\ell-1}\big)$ simplices.
    Then, these loops could be combined via successive additions, using for instance triangulated pairs of pants.
    For this construction, the number of added simplices is
    \[
        O\biggl(
            i+\sum_{\ell=1}^{i}
            \bigg(1+\log\binom{i-1}{\ell-1}\bigg)
        \biggr)
        = O\big(i^2\big)
    \]
    for each $\beta_i$.
    Therefore, constructing all $L$ classes independently would use $O(L^3)$ simplices.
\end{remark}

\subparagraph{Adding relations.}

We now exploit the structure of the $\beta_1,\dots,\beta_L$ to create many exceptional bars.
Let $R>L^2$ be an integer to be chosen later, consider $R$ new triangular loops $\nu_1,\dots,\nu_R$, and form their disjoint union:
\[
    Y = \bigsqcup_{i=1}^R \nu_i.
\]
For every $i\in\{1,\dots,R\}$, connect the loop $\beta_i\subset \widetilde{X}$ of \Cref{lem:exceptional_bars_pascal_construction} to $\nu_i$, where we read $\beta_i$ periodically: $\beta_{L+i} = \beta_i$.
More precisely, we insert a triangulated cylinder $C_i$ whose ends are 
\[
    \partial C_i = \beta_i \cup \nu_i,
    \qquad i\in\{1,\dots,R\}.
\]
Homologically, this corresponds to imposing the relation $[\beta_i]=[\nu_i]$.
Our final complex is
\begin{equation}
    \label{eq:exceptional_bars_final_complex}
    K = \big(Y \sqcup \widetilde{X}\big) \cup \bigcup_{i=1}^R C_i.
\end{equation}

Finally, we define a filtration $K^\bullet$ of $K$.
We introduce
\textbf{(1)} the new loops $\nu_R, \nu_{R-1}, \dots, \nu_1$ in reverse order,
\textbf{(2)} the complex $\widetilde X$, 
\textbf{(3)} the cylinders $C_1,\dots,C_R$ in order.
Choose any simplex-wise refinement at each stage.
Let $b_i$ denote the birth time of $\nu_i$ and $d_i$ that of $C_i$:
\[
    b_i = \max_{\sigma\in\nu_i} t(\sigma), 
    \qquad d_i = \max_{\sigma\in C_i} t(\sigma), 
    \qquad i\in\{1,\dots,R\}.
\]
Thus, $b_R < \cdots < b_1 < d_1 < \cdots < d_R$.
This is illustrated in \Cref{fig:complex_many_bars} for $L=4$ and $R=12$.

Let us inspect this filtration.
After all new loops $\nu_i$ and Pascal loops $\beta_i$ have been introduced, the relations $C_i$ successively impose
\[
    [\beta_1]=[\nu_1], \qquad [\beta_2]=[\nu_2], \qquad [\beta_3]=[\nu_3], \qquad \dots
\]
Let $i\in\{1,\dots,L\}$.
Over $\F_{p_i}$, we have seen in \Cref{lem:exceptional_bars_pascal} that $[\beta_1],\dots,[\beta_{i}]$ are linearly independent, and that $[\beta_{i+1}]$ is a linear combination of those.
Thus, the $(i+1)$th relation $C_{i+1}$ introduces a linear dependency among $[\nu_1],\dots,[\nu_{i+1}]$ which ends the bar created by the youngest class $[\nu_1]$ (the $\nu_i$ have been introduced in reverse order).
Consequently, the barcode over $\F_{p_i}$ contains the bar $[b_1, d_{i+1})$.
This pattern repeats at the introduction of each~relation.%

\begin{lemma}[Proof p.~\pageref{proof:exceptional_bars_filtration}]
    \label{lem:exceptional_bars_filtration}
    For every $i\in\{1,\dots,L\}$ and $t\in\{1,\dots,R-i\}$,
    \[
        \left[b_t,d_{t+i}\right)
        \in
        \barcode\bigl(H_1(K^\bullet;\F_{p_i})\bigr).
    \]
    In particular, the union of the barcodes over $\F_{p_1},\dots,\F_{p_L}$ contains at least
    \begin{equation}
        \label{eq:exceptional_bars_count}
        L(R-L)
    \end{equation}
    distinct bars.
    Moreover, each of these barcodes differs from the rational barcode.    
\end{lemma}

By choosing suitable values for the parameters $L$ and $R$, we obtain our main result.

\begin{proposition}[Proof p.~\pageref{proof:number_exceptional_bars}]
    \label{prop:number_exceptional_bars}
    For large $L,R>0$, the complex $K$ of \Cref{eq:exceptional_bars_final_complex} has $O(L^2\log L + R)$ simplices.
    In particular, with $R = \lfloor L^2\log L\rfloor$ and $L = \lfloor \sqrt{N/\log N}\rfloor$, it has
    \[
        O(N)
    \]
    simplices.
    In this case, the corresponding filtration $K^\bullet$ has $\Omega(\sqrt{N/\log N})$ exceptional primes, and the number of distinct bars appearing in $\barcode(H_1(K^\bullet;\F_p))$ for some prime $p$ satisfies
    \[
        \QQ =
        \Omega\left(\frac{N^{3/2}}{\sqrt{\log N}}\right).
    \]
\end{proposition}

\subsection{Many exceptional reductions and omni-field persistence}
\label{sec:filtrations_with_large_torsion_omnifield}

Finally, we consider Morozov's omni-field persistence algorithm, mentioned in \Cref{subsec:intro_algorithms}.
Given an integer boundary matrix, the algorithm runs a reduction over $\Q$ and additional reductions over $\F_p$ when prime-specific behavior is detected.
During the reduction, a column is marked as $p$-exceptional when $p$ divides the numerator of its current pivot.
Whenever another column uses a previous $p$-exceptional column, it receives a separate reduction~over~$\F_p$.

We build a costly instance by combining Newman's complex from \Cref{sec:filtrations_with_large_torsion_newman} with (a slight modification of) Morozov's worst-case filtration \cite{Morozov2005}.
We first describe the latter~construction.

\subparagraph{Base construction.}

\begin{figure}[!htbp]
    \centering
	\subcaptionbox{
        \label{subfig:worst_case_Morozov_base_triangles}
		Base triangles $T_1,T_2,\dots$ are built successively.
        The last edge to be introduced in the filtration is~$E_2$.%
	}[0.49\textwidth]{%
		\includegraphics[width=.625\linewidth]{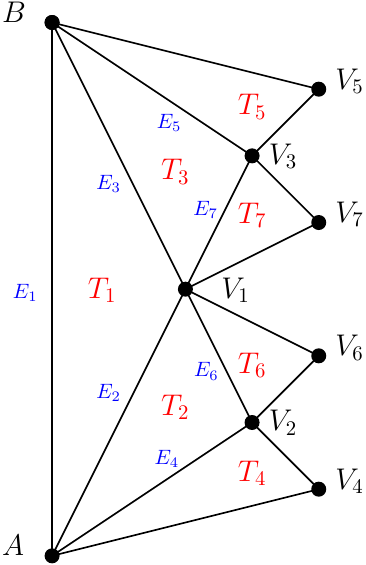}
		}
	\hfill
    \subcaptionbox{
        Fin triangles $T'_1,T'_2,\dots$ are all attached along the edge $E_2=AV_1$
        with a distinct additional vertex.%
        \label{subfig:worst_case_Morozov_fin_triangles}
    }[0.49\textwidth]{%
		\includegraphics[width=1\linewidth]{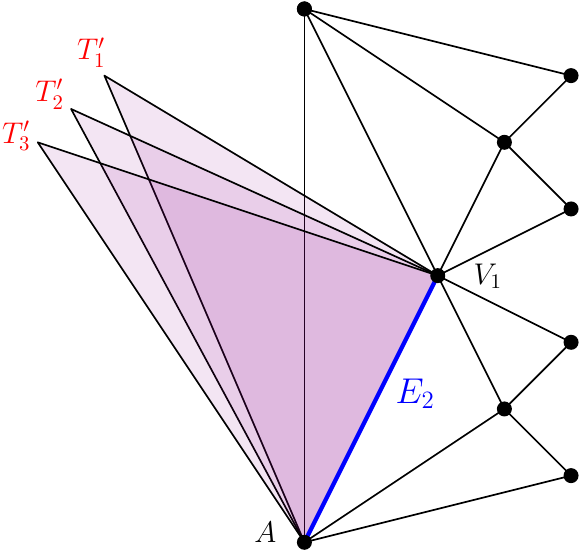}
    }
    \caption{
        Morozov's filtration, denoted $\widetilde{Y}_N^\bullet$, attains the $\Theta(N^3)$ worst-case running time of the persistence algorithm.
        It is built in two stages, starting with base triangles in the plane, followed by fin triangles.
        We label vertices in black, edges in blue, and triangles in red.
    }
\end{figure}

Fix a sufficiently large integer $N$ and choose $\ell$ so that
\[
    2^{\ell+1}-1=\Theta(N).
\]
Start with two points $A$ and $B$ vertically aligned, as in \Cref{subfig:worst_case_Morozov_base_triangles}.
Place a new vertex $V_1$ slightly to the right of the midpoint of $AB$. 
Next, place $V_2$ and $V_3$ slightly to the right of the midpoints of $AV_1$ and $BV_1$.
At the following stage, place $V_4,V_5,V_6,V_7$ near the midpoints of
\[
    AV_2,\quad BV_3,\quad V_1V_2,\quad V_1V_3.
\]
At the $i$th stage, $2^{i-1}$ vertices are introduced.
We continue until the number of vertices $V_i$ is
\[
    1+2+\cdots+2^\ell=2^{\ell+1}-1.
\]
For $i\in\{1,\dots,2^{\ell+1}-1\}$, let $E_i$ denote the edge used to define $V_i$.
The first edges are
\[
    E_1 = AB, \quad E_2 = AV_1, \quad E_3 = BV_1, \quad E_4 = AV_2, \quad E_5 = BV_3, \quad E_6 = V_1V_2, \quad E_7 = V_1V_3.
\]
In addition, we define $T_i$ as the triangle formed by $V_i$ and $E_i$.
Thus, the first $i$ triangles form a triangulated disk with $i+2$ boundary edges, as shown in \Cref{subfig:worst_case_Morozov_base_triangles}.
We point out that, for simplicity, we omit the initial vertex $C$ from Morozov's original construction~\cite{Morozov2005}.

Now we define a simplex-wise filtration $Y_N^\bullet$.
First, all $2+2^{\ell+1}-1$ vertices are introduced.
Second, the initial edge $E_1=AB$ is introduced, followed by all other edges in reverse order
\[
    \dots,\; E_5,\; E_4,\; E_3,\; E_2\;.
\]
After inserting the $1$-skeleton, insert the base triangles in order $T_1, T_2,\dots$

The key feature of this construction is that the persistence algorithm produces a cascade of reductions.
For example, the columns of the first three triangles reduce as follows:
\[
    \setlength{\arraycolsep}{6pt}
    \begin{pNiceArray}{ccc}[first-row,first-col]
                          & T_1 & T_2 & T_3 \\
        \firstcol{E_1}    & *   & 0   & 0   \\
        \firstcol{E_7}    & 0   & 0   & *   \\
        \firstcol{E_6}    & 0   & *   & 0   \\
        \firstcol{E_5}    & 0   & 0   & *   \\
        \firstcol{E_4}    & 0   & *   & 0   \\
        \firstcol{E_3}    & *   & 0   & *   \\
        \firstcol{E_2}    & *   & *   & 0
    \end{pNiceArray}
    \quad\xrightarrow{~~\text{persistence}~~}\quad
    \begin{pNiceArray}{ccc}[first-row,first-col]
                          & T_1       & T_2       & T_3       \\
        \firstcol{E_1}    & *         & *         & *         \\
        \firstcol{E_7}    & 0         & 0         & *         \\
        \firstcol{E_6}    & 0         & *         & *         \\
        \firstcol{E_5}    & 0         & 0         & *         \\
        \firstcol{E_4}    & 0         & *         & \boxed{*} \\
        \firstcol{E_3}    & *         & \boxed{*} & 0         \\
        \firstcol{E_2}    & \boxed{*} & 0         & 0
    \end{pNiceArray}.
\]
More generally, let $R_i$ denote the reduced column of $T_i$.
With suitable orientations, they~satisfy%
\[
    R_i = \partial T_i + R_{i-1} =\partial T_1 + \cdots + \partial T_i.
\]
Thus, the column $R_i$ represents the boundary of the triangulated disk formed by the first $i$ base triangles.
In particular, it has $i+2$ nonzero entries.

\subparagraph{Fin triangles.}

We add new triangles $T'_1,\dots,T'_N$, called \emph{fin triangles}, whose reduction is costly.
More precisely, each $T'_i$ is formed by joining a new vertex to the base edge
\[
    E_2 = AV_1.
\]
This vertex and the two new edges are shared with no other triangle; see \Cref{subfig:worst_case_Morozov_fin_triangles}.

We interleave these simplices with the base filtration $Y_N^\bullet$ described above as follows.
Let $\widetilde{Y}_N^\bullet$ be the filtration obtained by inserting \textbf{(1)} all vertices, \textbf{(2)} the non-shared fin edges, \textbf{(3)} the base edges followed by the base triangles as in $Y_N^\bullet$, and \textbf{(4)} the fin triangles.

We now apply the persistence algorithm to $\widetilde{Y}_N^\bullet$.
Morozov shows that reducing the column of a fin triangle $T'_i$ successively uses the reduced columns $R_1,R_2,\dots$ of \emph{all} base triangles.
For example, writing $E_2,u,v$ for the edges of $T'_1$, the first reduction steps are:
\[
    \setlength{\arraycolsep}{3pt}
    \begin{pNiceArray}{ccccc}[first-row,first-col]
                          & T_1       & T_2       & T_3       & \cdots & T_1' \\
        \firstcol{u}      & 0         & 0         & 0         & \cdots & * \\
        \firstcol{v}      & 0         & 0         & 0         & & * \\ \hline
        \firstcol{E_1}    & *         & *         & *         & & 0 \\
        \firstcol{E_7}    & 0         & 0         & *         & & 0 \\
        \firstcol{E_6}    & 0         & *         & *         & & 0 \\
        \firstcol{E_5}    & 0         & 0         & *         & & 0 \\
        \firstcol{E_4}    & 0         & *         & \boxed{*} & & 0 \\
        \firstcol{E_3}    & *         & \boxed{*} & 0         & & 0 \\
        \firstcol{E_2}    & \boxed{*} & 0         & 0         & & *
    \end{pNiceArray}
    \;\xrightarrow{~~~}\;
    \begin{pNiceArray}{ccccc}
        0         & 0         & 0         & \cdots & * \\
        0         & 0         & 0         & & * \\ \hline
        *         & *         & *         & & * \\
        0         & 0         & *         & & 0 \\
        0         & *         & *         & & 0 \\
        0         & 0         & *         & & 0 \\
        0         & *         & \boxed{*} & & 0 \\
        *         & \boxed{*} & 0         & & * \\
        \boxed{*} & 0         & 0         & & 0
    \end{pNiceArray}
    \;\xrightarrow{~~~}\;
    \begin{pNiceArray}{ccccc}
        0         & 0         & 0         & \cdots & * \\
        0         & 0         & 0         & & * \\ \hline
        *         & *         & *         & & * \\
        0         & 0         & *         & & 0 \\
        0         & *         & *         & & * \\
        0         & 0         & *         & & 0 \\
        0         & *         & \boxed{*} & & * \\
        *         & \boxed{*} & 0         & & 0 \\
        \boxed{*} & 0         & 0         & & 0
    \end{pNiceArray}
    \;\xrightarrow{~~~}\;
    \cdots
\]
Thus, reducing this column requires
\begin{equation}
    \label{eq:fin_triangle_cost}
    \Omega\left(\sum_{i=1}^{2^{\ell+1}-1}(i+2)\right)
    = \Omega\left( N^2 \right)    
\end{equation}
arithmetic operations.
Since we introduced $N$ fins, the full reduction of $\widetilde{Y}_N^\bullet$ costs $\Omega(N^3)$.

\subparagraph{Adding torsion.}

During reduction of Morozov's filtration $\widetilde{Y}_N^\bullet$, the column of each fin triangle $T'_i$ first encounters a pivot conflict at $E_2=AV_1$.
We will make the column owning this pivot $p$-exceptional for many primes, therefore yielding a costly instance for omni-field~persistence.%

As in \Cref{sec:filtrations_with_large_torsion_brute_force}, take the product $M$ of the first primes:
\[
    L=\left\lfloor\frac{N}{\log N}\right\rfloor,
    \qquad
    M=p_1\cdots p_L.
\]
Let $X_N$ be the corresponding Newman complex.
As already observed, it has $O(N)$ simplices.
Moreover, its torsion subgroup $\Z/M\Z$ is generated by the triangular cycle $\gamma_0$ of \Cref{eq:newman_complex_L1}.

Identify $\gamma_0$ with the boundary of the base triangle $T_1=ABV_1$.
The resulting complex is
\[
    K_N=X_N \cup_{\gamma_0=\partial T_1} \widetilde{Y}_N,
\]
where $\widetilde{Y}_N$ is the full Morozov complex.
We equip it with a simplex-wise filtration by inserting 
\textbf{(1)} all vertices,
\textbf{(2)} the non-shared fin Morozov edges,
\textbf{(3)} the base Morozov edges,
\textbf{(4)} the remaining Newman edges,
\textbf{(5)} the Newman triangles,
\textbf{(6)} the base Morozov triangles,
\textbf{(7)} the fin Morozov triangles.
The simplices corresponding to Morozov's filtration are introduced as in $\widetilde{Y}_N^\bullet$, and those of Newman's complex are introduced in any order.
We denote it
\begin{equation}
    \label{eq:instance_omnifield}
     K_N^\bullet.
\end{equation}

\begin{proposition}[Proof p.~\pageref{proof:number_exceptional_columns_omnifield}]
    \label{prop:number_exceptional_columns_omnifield}
    The filtration $K_N^\bullet$ of \Cref{eq:instance_omnifield} has $\Theta(N)$ simplices.
    Moreover, on this instance, the omni-field persistence algorithm has complexity
    \[
        \Omega\left(\frac{N^4}{\log N}\right).
    \]
\end{proposition}


\bibliography{bibliography}


\subparagraph{AI Statement.}

ChatGPT~6 (Chat mode and Codex) was used to assist with writing (grammar checks and rephrasing), mathematics (proof checking, literature exploration, and suggestions for arguments), and implementation.
In particular, it helped strengthen the examples of filtrations with large torsion presented in \Cref{sec:filtrations_with_large_torsion}.


\appendix

\section{Proofs}
\label{sec:proofs}

\subsection{Proofs for \texorpdfstring{\Cref*{sec:algorithmic_result}}{Section \ref*{sec:algorithmic_result}}}

We begin with a standard characterization of the pivots of a reduction $R=DV$ in terms of submatrix determinants.
It is found in the Pairing Uniqueness Lemma of \cite{CohenSteiner2006} or in the characterization of row rank profiles in \cite{Dumas2015}.
We formulate it for augmented matrices $\widetilde R=\widetilde D V$ and row priorities $\tau$ (see \Cref{def:reduction_row_priority}).
The rest of this section proves the results of \Cref{sec:algorithmic_result}.

\begin{lemma}[Proof p.~\pageref{proof:pivot_rank_profile}]
    \label{lem:pivot_rank_profile}
    Let $\kk$ be a field.
    Let $D\in\kk^{n\times m}$ be any matrix, let $\widetilde D=\left(\begin{smallmatrix}I_m\\D\end{smallmatrix}\right)$ be its augmentation,
    and let $\widetilde R=\widetilde D V$ be a $\tau$-reduction for a certain permutation
    \[
        \tau = (\tau_1,\dots,\tau_{n+m}),
        \quad\text{ meaning }\quad
        \tau_1 \succ \dots \succ \tau_{n+m}.
    \]
    For $j\in\{1,\dots,m\}$, denote the pivot index and entry by%
    \[
        p_j=\low(\widetilde R,j,\tau) \in \{\bot_1,\dots,\bot_m,1,\dots,n\},
        \qquad
        a_j=\piv(\widetilde R,j,\tau) \in\kk.
    \]
    Then the following statements hold.
    
    \begin{enumerate}
        \item
        \label{it:lem_pivot_rank_profile_1}
        For every $j\in\{1,\dots,m\}$ and $i\in\{1,\dots,n+m\}$,
        \[
            \rank\widetilde D[(\tau_1,\dots,\tau_i),1:j]
            =
            \bigl|\{\tau_1,\dots,\tau_i\}\cap\{p_1,\dots,p_j\}\bigr|.
        \]

        \item
        \label{it:lem_pivot_rank_profile_2}
        For every $j\in\{1,\dots,m\}$, the first $j$ pivot indices $\{p_1,\dots,p_j\}$, listed from highest to lowest $\tau$-priority, form the smallest tuple $(r_1,\dots,r_j)$ of distinct row indices such that
        \[
            \det\widetilde D[(r_1,\dots,r_j),1:j]\neq0.
        \]
        Here, tuples $(r_1,\dots,r_j)$ are compared lexicographically using the order $\tau_1<\cdots<\tau_{n+m}$ on indices $r_i$.
        Thus, higher-priority rows come first.

        \item
        \label{it:lem_pivot_rank_profile_3}
        For every $j\in\{1,\dots,m\}$,
        \[
            \det\widetilde D[(p_1,\dots,p_j),1:j]
            =
            \prod_{k=1}^{j}a_k.
        \]
    \end{enumerate}
\end{lemma}

In particular, the lemma proves that the pivot indices $p_j$ and entries $a_j$ depend only on $\widetilde D$ and $\tau$, and not on the chosen $\tau$-reduction.

\begin{proof}[Proof of \Cref{lem:pivot_rank_profile}]
    \phantomsection
    \label{proof:pivot_rank_profile}
    During the proof, we fix a column index $j\in\{1,\dots,m\}$ and define
    \[
        V_j=V[1:j,1:j] \in \kk^{j\times j}.
    \]
    Since $V$ is upper triangular with diagonal entries $1_\kk$, the matrix $V_j$ is invertible and satisfies $\det V_j = 1_\kk$. 
    Moreover, for every tuple $I$ of row indices, the relation $\widetilde{R}=\widetilde{D}V$ yields
    \[
        \widetilde{R}[I,1:j]
        =
        \widetilde{D}[I,1:j]\,V_j.
    \]
    In particular, their ranks coincide:
    \begin{equation}
        \label{eq:proof_pivot_rank_profile_rank}
        \rank \widetilde{R}[I,1:j] =\rank \widetilde{D}[I,1:j].
    \end{equation}
    In addition, when $|I| = j$, they satisfy
    \begin{equation}
        \label{eq:proof_pivot_rank_profile_det}
        \det \widetilde{R}[I,1:j] =\det \widetilde{D}[I,1:j].
    \end{equation}

    \proofsubparagraph{Statement~\ref{it:lem_pivot_rank_profile_1}.}

    Let $i\in\{1,\dots,n+m\}$.
    In $\widetilde R[(\tau_1,\dots,\tau_i),1:j]$, a column is nonzero precisely when its pivot belongs to $\{\tau_1,\dots,\tau_i\}$.
    Consequently,
    \[
        \rank\widetilde{R}[(\tau_1,\dots,\tau_i),1:j]
        =
        \bigl|\{\tau_1,\dots,\tau_i\}
        \cap\{p_1,\dots,p_j\}\bigr|.
    \]
    \Cref{eq:proof_pivot_rank_profile_rank} then proves the statement.

    \proofsubparagraph{Statement~\ref{it:lem_pivot_rank_profile_2}.}

    Consider the following greedy procedure: we inspect the rows of $\widetilde D[:,1:j]$ from highest to lowest $\tau$-priority, retaining each row if it is independent
    of the previously retained rows.
    By Statement~\ref{it:lem_pivot_rank_profile_1}, the rank of the inspected row span increases precisely at the pivot rows $\{p_1,\dots,p_j\}$.
    Thus this procedure selects the pivot rows in priority order, yielding a tuple $(r_1,\dots,r_j)$ with
    \begin{equation}
        \label{eq:proof_pivot_rank_profile_det_statement2}
        \det\widetilde{D}[(r_1,\dots,r_j),1:j]\neq0.
    \end{equation}
    Now, let $(s_1,\dots,s_j)$ be another tuple that is lexicographically smaller than $(r_1,\dots,r_j)$.
    Let $i$ be the first index at which they differ.
    In particular, $s_i$ has higher priority than $r_i$: $s_i\succ r_i$.
    Thus, the row $\widetilde D[s_i,1:j]$ belongs to the span of the previous rows 
    \[
        \widetilde D[s_1,1:j],\quad\dots,\quad\widetilde D[s_{i-1},1:j],
    \]
    otherwise it would have been chosen at the $i$th stage of the greedy procedure.
    Therefore,
    \[
        \det\widetilde{D}[(s_1,\dots,s_j),1:j] = 0.
    \]
    We deduce that $(r_1,\dots,r_j)$ is the smallest tuple to satisfy \Cref{eq:proof_pivot_rank_profile_det_statement2}, as required.
    
    \proofsubparagraph{Statement~\ref{it:lem_pivot_rank_profile_3}.}

    Consider the matrix
    \[
        S=\widetilde R[(p_1,\dots,p_j),1:j].
    \]
    There exists a permutation $P$ of $\{1,\dots,j\}$ that makes $PSP^{-1}$ lower triangular, with diagonal entries $a_{P(1)},\dots,a_{P(j)}$.
    \Cref{eq:proof_pivot_rank_profile_det}, proved at the beginning, then shows that
    \[
        \det\widetilde D[(p_1,\dots,p_j),1:j]
        =
        \det S
        =
        \prod_{k=1}^{j}a_k.
        \qedhere
    \]
\end{proof}

\begin{proof}[Proof of \Cref{lem:exhaustive_reduction}]
    \phantomsection
    \label{proof:exhaustive_reduction}
    The first two equalities are direct consequences of Statements~\ref{it:lem_pivot_rank_profile_2} and \ref{it:lem_pivot_rank_profile_3} of
    \Cref{lem:pivot_rank_profile}.
    The last equality follows from the definition of a $\tau_{\mathrm{piv}}$-reduction (\Cref{def:reduction_row_priority}).
\end{proof}

\begin{remark}[See \Cref{subsubsec:efficient_updates_priority}]
    \label{rem:vineyard_algorithm}
    Let $D$ be a boundary matrix over $\kk$.
    The vineyard algorithm of Cohen-Steiner, Edelsbrunner, and Morozov \cite{CohenSteiner2006} updates a reduction $R=D V$ after swapping two adjacent simplices $\sigma_i$ and $\sigma_{i+1}$.
    This amounts to swapping adjacent rows and columns of $D$.
    Thus, the algorithm produces a reduction of $ D' = P D P$, where $P$ is the permutation matrix that swaps $i$ and $i+1$.
    Our \Cref{alg:transposition_priority} is similar but only swaps two \emph{rows}.
    Indeed, a $\tau$-reduction of $\widetilde{D}$ is equivalent to an ordinary reduction of $T \widetilde{D}$, where $T$ is the permutation matrix of $\tau$.
    Thus, \Cref{alg:transposition_priority} can be seen as a row-only version of the vineyard algorithm: morally, it produces a reduction of $D'' = P D$.%
\end{remark}

\begin{proof}[Proof of \Cref{lem:full_correction_priority}]
    \phantomsection
    \label{proof:full_correction_priority}
    Let $\rho$ denote the current row priority processed by \Cref{alg:full_correction_priority}.
    Initially, it is equal to $\tau$.
    At the beginning of stage $i$, the first $i-1$ entries of $\rho$ agree with~$\tau'$:%
    \[
        (\rho_1,\dots,\rho_{i-1}) = (\tau'_1,\dots,\tau'_{i-1}).
    \]
    Then, the algorithm moves $\tau_i'$ to position $i$ by adjacent transpositions, without changing the prefix $(\rho_1,\dots,\rho_{i-1})$.
    Thus, after this stage, the first $i$ entries agree.
    At termination, we~have%
    \[
        \rho=\tau'.
    \]
    Consequently, the output is a $\tau'$-reduction of $\widetilde{D}$.

    Let us count the number of operations.
    During stage $i$, $\tau_i'$ only crosses indices $\rho_j$ such~that%
    \[
        \rho_j \succ_{\rho} \tau_i'
        \qquad\text{and}\qquad
        \rho_j \prec_{\tau'} \tau_i'.
    \]
    Thus, each adjacent transposition corrects one discordant pair in the sense of \Cref{eq:kendall_distance_discordant_pairs}.
    It decreases the Kendall distance $K(\rho,\tau')$ by exactly one.
    Since this distance starts at $K(\tau,\tau')$ and ends at zero, the algorithm performs exactly $K(\tau,\tau')$ adjacent transpositions.

    According to the cases of \Cref{alg:transposition_priority} described in
    \Cref{subsubsec:efficient_updates_priority}, each adjacent transposition requires at most one column operation.
    We deduce the result.
\end{proof}

\begin{remark}[See \Cref{subsubsec:efficient_updates_priority}]
    \label{rem:reduction_update_vs_from_scratch}
    A simpler way to meet the specifications of \Cref{alg:full_correction_priority} in \Cref{lem:full_correction_priority} would be to perform a full $\tau'$-reduction of $\widetilde{D}$ via \Cref{alg:persistence_algorithm_priority}.
    However, this approach can require significantly more operations.
    For instance, consider the $(n+m)\times m$ matrix $\widetilde{D}$ displayed below. 
    It is reduced for the canonical priority 
    \[
        \tau=(n,\dots,1,\bot_m,\dots,\bot_1).
    \]
    Consider the priority obtained by moving $1$ to the beginning: 
    \[
        \tau'=(1,n,n-1,\dots,2,\bot_m,\dots,\bot_1).
    \]
    The Kendall tau distance between these permutations is $K(\tau,\tau')=n-1$, so \Cref{alg:full_correction_priority} requires only $O(n)$ column operations. 
    By contrast, applying \Cref{alg:persistence_algorithm_priority} from scratch requires $n(n-1)/2 = O(n^2)$ column operations.
    This can be seen on the first steps:
    \begin{equation*}
        \widetilde{D} =
        \begin{pNiceArray}{cccc}
            1&0&0&\cdots\\
            0&1&0&\cdots\\
            0&0&1&\cdots\\
            \vdots&\vdots&\vdots&\\
            \hline
            \boxed{1}&1&1&\cdots\\
            \vdots&\vdots&\vdots&\\
            0&0&1&\cdots\\
            0&1&0&\cdots\\
            1&0&0&\cdots
        \end{pNiceArray}
        \to
        \begin{pNiceArray}{cccc}
            1&\mathllap{-}1&0&\cdots\\
            0&1&0&\cdots\\
            0&0&1&\cdots\\
            \vdots&\vdots&\vdots&\\
            \hline
            \boxed{1}&0&1&\cdots\\
            \vdots&\vdots&\vdots&\\
            0&0&1&\cdots\\
            0&1&0&\cdots\\
            1&\mathclap{\boxed{-1}}&0&\cdots
        \end{pNiceArray}
        \to
        \begin{pNiceArray}{cccc}
            1&\mathllap{-}1&\mathllap{-}1&\cdots\\
            0&1&0&\cdots\\
            0&0&1&\cdots\\
            \vdots&\vdots&\vdots&\\
            \hline
            \boxed{1}&0&0&\cdots\\
            \vdots&\vdots&\vdots&\\
            0&0&1&\cdots\\
            0&1&0&\cdots\\
            1&\mathclap{\boxed{-1}}&-1&\cdots
        \end{pNiceArray}
        \to
        \begin{pNiceArray}{cccc}
            1&\mathllap{-}1&\mathllap{-}1&\cdots\\
            0&1&\mathllap{-}1&\cdots\\
            0&0&1&\cdots\\
            \vdots&\vdots&\vdots&\\
            \hline
            \boxed{1}&0&0&\cdots\\
            \vdots&\vdots&\vdots&\\
            0&0&1&\cdots\\
            0&1&\mathclap{~~\boxed{-1}~}&\cdots\\
            1&\mathclap{\boxed{-1}~~}&0&\cdots
        \end{pNiceArray}
    \end{equation*}
\end{remark}

\begin{proof}[Proof of \Cref{lem:generic_reduction_identify_primes}]
    \phantomsection
    \label{proof:generic_reduction_identify_primes}
    Let $\widetilde{D}^{\F_p}$ denote the reduction modulo $p$ of $\widetilde{D}$.
    Moreover, put 
    \[
        p_j^\Q=\low(\Q,j),
        \qquad
        p_j^{\F_p}=\low(\F_p,j),
        \qquad 
        j\in\{1,\dots,m\}.
    \]
    Let us sort the pivot sequences from highest to lowest canonical priority.
    We obtain tuples
    \[
        \big(p_{\sigma(1)}^\Q,\dots,p_{\sigma(j)}^\Q\big),
        \qquad
        \big(p_{\sigma'(1)}^{\F_p},\dots,p_{\sigma'(j)}^{\F_p}\big)
    \]
    for some permutations $\sigma$ and $\sigma'$ of $\{1,\dots,j\}$.
    By Statement~\ref{it:lem_pivot_rank_profile_2}, these are the smallest tuples, with respect to the lexicographic order, such that 
    \[
        \det\widetilde D\big[\big(p_{\sigma(1)}^\Q,\dots,p_{\sigma(j)}^\Q\big),1:j\big]\neq0,
        \qquad
        \det\widetilde D^{\F_p}\big[\big(p_{\sigma'(1)}^{\F_p},\dots,p_{\sigma'(j)}^{\F_p}\big),1:j\big]\neq0.
    \]
    In particular, if
    \[
        \det\widetilde D^{\F_p}\big[\big(p_{\sigma(1)}^\Q,\dots,p_{\sigma(j)}^\Q\big),1:j\big]\neq0,
    \]
    then the sorted tuples must coincide.
    On the other hand, note that
    \[
        \det\widetilde D^{\F_p}\big[\big(p_{\sigma(1)}^\Q,\dots,p_{\sigma(j)}^\Q\big),1:j\big]
        = 
        \underbrace{\det\widetilde D\big[\big(p_{\sigma(1)}^\Q,\dots,p_{\sigma(j)}^\Q\big),1:j\big]}_{\pm\Delta_j} \pmod{p}.
    \]
    Thus, the tuples coincide precisely when $p$ does not divide $\Delta_j$.

    For the second statement, we use the definition in \Cref{eq_exceptional_primes}: a prime $p$ is exceptional~if%
    \[
        (p_1^\Q,\dots,p_m^\Q)
        \neq
        (p_1^{\F_p},\dots,p_m^{\F_p}).
    \]
    In particular, some prefixes, seen as sets, must differ: there exists $j\in\{1,\dots,m\}$ such that
    \[
        \{ p_1^\Q,\dots,p_j^\Q \}
        \neq
        \{ p_1^{\F_p},\dots,p_j^{\F_p} \}.
    \]
    By the previous statement, this occurs precisely when $p$ divides $\Delta_j$.
\end{proof}

\begin{proof}[Proof of \Cref{lem:number_exceptional_primes}]
    \phantomsection
    \label{proof:number_exceptional_primes}

    We prove the two statements in order.
    To simplify notation, we put $\mu=\|D\|_\infty$.
    The key step of the proof is to apply Hadamard's inequality to the columns of $D$, an approach also used in \cite[§3.5]{10.1145/800204.806288} and \cite[§3.5]{DUMAS200171}.

    \proofsubparagraph{First statement.}

    Consider the principal minors $\Delta_1,\Delta_2,\dots$ of $\widetilde{D}$, defined in \Cref{eq:principal_minor}.
    We recall that, if the $j$th column of $R$ (the non-augmented reduction) is zero, then
    \[
        \low(\Q,j) \in \{\bot_1,\dots,\bot_m\},
    \]
    and it is associated to a unit pivot $\piv(\Q,j)=1$.
    In this case, the consecutive minors~coincide: 
    \[
        \Delta_{j}=\Delta_{j-1}.
    \]
    For this proof, we ignore such columns.
    We are left with $r$ remaining columns, where $r=\rank_\Q(D)$.
    The corresponding minors are denoted
    \[
        \Delta_1,\dots,\Delta_r.
    \]

    Next, we bound these principal minors.
    Let $\ell_1<\cdots<\ell_r$ be the indices of the retained columns.
    Thus, for every $j\in\{1,\dots,r\}$,
    \[
        \Delta_j
        = \det D\big[(\low(\Q,\ell_1),\dots,\low(\Q,\ell_j)), (\ell_1,\dots,\ell_j) \big].
    \]
    By Hadamard's inequality, the determinant of a matrix is upper bounded by the product of the Euclidean norms of its columns.    
    Each entry of this $j\times j$ submatrix is at most $\mu$.
    Thus its columns have norm at most $j^{1/2}\mu$, and we deduce
    \begin{equation}
        \label{proof:number_exceptional_primes_statement1_minors}
        |\Delta_j|
        \leq (j^{1/2}\mu)^j
        \leq (r^{1/2}\mu)^j.
    \end{equation}

    To conclude, we deduce a bound on $\ell=|\PP|$.
    Denote the exceptional primes by
    \[
        p_1<\dots<p_\ell.
    \]
    Recall that \Cref{lem:generic_reduction_identify_primes} gives
    \[
        \PP = \{ p \text{ prime} \mid p \text{ divides } \Delta_i \text{ for some } i \in \{1,\dots,r\} \}.
    \]
    Together with \Cref{proof:number_exceptional_primes_statement1_minors}, we deduce that
    \[
        \prod_{i=1}^\ell p_i 
        \leq \prod_{j=1}^r |\Delta_j| 
        \leq \prod_{j=1}^r (r^{1/2}\mu)^j 
        \leq (r^{1/2}\mu)^{r(r+1)/2}.
    \]
    Since the exceptional primes are ordered we have $p_i \geq i+1$ for all $i\in\{1,\dots,\ell\}$.
    Therefore, 
    \[
        \prod_{i=1}^\ell p_i 
        \geq \prod_{i=1}^\ell (i+1) 
        \geq \prod_{i=r+1}^\ell\!(r+1) 
        = (r+1)^{\ell-r}.
    \]
    Combined with the previous inequality, we deduce that
    \[
        (r+1)^{\ell-r} \leq (r^{1/2}\mu)^{r(r+1)/2}.
    \]
    We take the logarithm on both sides:
    \[
        (\ell-r)\log(r+1) \leq \frac12r(r+1)\log(r^{1/2}\mu).
    \]
    This inequality is equivalent to
    \[
        \ell \leq 
        \underbrace{r + \frac{r(r+1)}{4}\frac{\log(r)}{\log(r+1)}}_{\leq r^2}
        + \underbrace{\frac{r(r+1)}{2}}_{\leq r^2} \frac{\log(\mu)}{\log(r+1)}.
    \]
    We deduce the first inequality of the lemma:
    \[
        \ell\leq r^2 \bigg( 1 + \frac{\log(\mu)}{\log(r+1)} \bigg).
    \]

    \proofsubparagraph{Second statement.}

    We now suppose that $D$ is a boundary matrix on $N$ simplices.
    In particular, its maximal entry is $\mu=1$, and the previous inequality yields
    \begin{equation}
        \label{proof:number_exceptional_primes_statement2}
        |\PP| \leq r^2.
    \end{equation}
    It remains to show that the rational rank $r = \rank_\Q D$ is upper bounded by $(N-1)/2$.
    Let us see $D\colon \Q^m\to\Q^n$ as a linear map.
    Its image $\im (D) \subset \Q^n$ satisfies
    \[
        r = \dim_\Q \im (D).
    \]    
    Besides, the kernel of $D$ corresponds to homology cycles and its image to boundaries.
    Thus,
    \[
        \sum_{k=0}^{+\infty} \dim H_k(K^N;\Q) = \sum_{k=0}^{+\infty} \bigg( \dim Z_k(K^N;\Q) - \dim B_k(K^N;\Q) \bigg) = \dim \ker (D) - \dim \im (D),
    \]
    where we compute dimensions over $\Q$.
    Since the left-hand side is at least 1, we deduce
    \[
        1 \leq \dim \ker (D) - \dim \im (D) = (N-r) - r = N - 2r.
    \]
    Thus, $r\leq (N-1)/2$, and \Cref{proof:number_exceptional_primes_statement2} yields
    \[
        |\PP|\leq \left( \frac{N-1}{2} \right)^2.
        \qedhere
    \]
\end{proof}

\begin{proof}[Proof of \Cref{lem:pivot_sequence_total}]
    \phantomsection
    \label{proof:pivot_sequence_total}
    Put $\tau=\tau_{\mathrm{piv}}^{\F_p}$ and pick $j\in\{1,\dots,m\}$.
    Statement~\ref{it:lem_pivot_rank_profile_3} of
    \Cref{lem:pivot_rank_profile}, applied to a canonical reduction over $\F_p$, gives
    \[
        \det\widetilde{D}^{\F_p}[(\tau_1,\dots,\tau_j),1:j]\neq0.
    \]
    Since $\widetilde{D}$ has integer entries, the corresponding integer determinant is nonzero, and hence
    \[
        \det\widetilde{D}^{\Q}[(\tau_1,\dots,\tau_j),1:j]\neq0.
    \]
    Now, let $\widetilde R^\Q$ be any rational $\tau$-reduction.
    The tuple $(\tau_1,\dots,\tau_j)$ is the lexicographically smallest tuple of $j$ distinct rows for this priority, and its corresponding minor is nonzero.
    Thus Statement~\ref{it:lem_pivot_rank_profile_2} of
    \Cref{lem:pivot_rank_profile} gives
    \[
        \big\{\low(\widetilde{R}^\Q,k,\tau)\mid k \in\{1,\dots,j\}\big\}
        =
        \{\tau_1,\dots,\tau_j\}.
    \]
    This shows that
    \[
        \low(\widetilde{R}^\Q,j,\tau)
        = \tau_j
        = \low(\F_p,j)
    \]
    for every $j\in\{1,\dots,m\}$, as required.
\end{proof}

\begin{proof}[Proof of \Cref{lem:pivot_sequence_prefix}]
    \phantomsection
    \label{proof:pivot_sequence_prefix}
    If $j=0$, the first column of $\widetilde R_0^\Q$ equals that of $\widetilde D$, and the statement follows.
    We now assume that $j\geq1$.
    Denote $\rho =\low(\F_p,\bullet)$ and consider the integer matrix
    \[
        B = \widetilde{D}[(\rho_1,\dots,\rho_j),1:j]\in\Z^{j\times j}.
    \]
    By Statement~\ref{it:lem_pivot_rank_profile_3}, its determinant is nonzero modulo $p$.

    By \ref{hyp:1}, we can write $\widetilde R_j^\Q=\widetilde D V$ where $V$ is upper
    triangular with diagonal entries one.~Put%
    \[
        x=V[1:j,j+1] \in \Q^j.
    \]
    The entry $V[j+1,j+1]$ is one.
    Thus,
    \begin{equation}
         \label{eq:proof_pivot_sequence_prefix_x}
        \widetilde{R}_j^\Q[:,j+1]
        =
        \widetilde{D}[:,1:j]\,x+\widetilde D[:,j+1].
    \end{equation}
    We now show that the denominators of the entries of $x$ are not divisible by $p$.

    The rows $\rho_1,\dots,\rho_j$ come first in $\tau_j$ and, by \ref{hyp:2}, are already the pivot rows of the
    first $j$ columns.
    Since $\widetilde R_j^\Q$ is $\tau_j$-reduced, its next column must vanish on these rows:
    \begin{equation}
        \label{eq:proof_pivot_sequence_prefix_vanish}
        \widetilde{R}_j^\Q[(\rho_1,\dots,\rho_j),j+1] = 0.
    \end{equation}
    Restricting \Cref{eq:proof_pivot_sequence_prefix_x} to the rows $\rho_1,\dots,\rho_j$ therefore gives
    \[
        0=\underbrace{\widetilde{D}[(\rho_1,\dots,\rho_j),1:j]}_{B}x+\widetilde D[(\rho_1,\dots,\rho_j),j+1],
    \]
    which is equivalent to
    \[
        x=-B^{-1}\widetilde D[(\rho_1,\dots,\rho_j),j+1].
    \]
    The adjugate matrix formula
    \[
        B^{-1}=\frac{\operatorname{adj}(B)}{\det(B)}
    \]
    and the fact that $p$ does not divide $\det(B)$ show that the denominators of the entries of $x$ are not divisible by $p$.
    The same holds for $\widetilde R_j^\Q[:,j+1]$ by \Cref{eq:proof_pivot_sequence_prefix_x} and the fact that $\widetilde D$ has integer entries.
    Together with \ref{hyp:3}, this proves that the denominators of entries of 
    \[
        \widetilde{R}^\Q_j[\colon\!,1\colon\!(j+1)]
    \]
    are not divisible by $p$.

    It remains to identify the next pivot.
    Let $\bar y$ be the reduction modulo $p$ of $\widetilde R_j^\Q[:,j+1]$, and let $\bar x$ denote that of $x$.
    Consider the $\F_p$-matrix $\widetilde{D}^{\F_p}[:,1:j]$ and a canonical reduction
    \[
        S=\widetilde{D}^{\F_p}[:,1:j]\,U
    \]
    where $U$ is upper triangular with diagonal entries one.
    Its pivot indices are $\rho_1,\dots,\rho_j$, and
    \[        
        \begin{pmatrix} S & \bar y \end{pmatrix}
        =
        \widetilde D^{\F_p}[:,1:(j+1)]
        \begin{pmatrix}
            U&\bar x\\
            0&1
        \end{pmatrix}.
    \]
    By \Cref{eq:proof_pivot_sequence_prefix_vanish}, the column $\bar y$ vanishes on the indices $\rho_1,\dots,\rho_j$, so its canonical pivot differs from every pivot of $S$.
    Thus $\begin{pmatrix} S & \bar y \end{pmatrix}$ is a canonical reduction of 
    \[
        \widetilde{D}^{\F_p}[:,1:(j+1)].
    \]
    Thus, the lowest nonzero entry of $\bar y$ has index $\low(\F_p,j+1)$, as required.
\end{proof}

\begin{proof}[Proof of \Cref{lem:full_correction_modp}]
    \phantomsection
    \label{proof:full_correction_modp}
    \ref{hyp:1}--\ref{hyp:3} hold after each stage of \Cref{alg:full_correction_modp}.
    This is shown by applying \Cref{lem:pivot_sequence_prefix} inductively.
    Therefore, at termination, $\widetilde{R}^\Q_m$ is a $\tau_{\mathrm{piv}}^{\F_p}$-reduction by \ref{hyp:1}, and its $\tau_{\mathrm{piv}}^{\F_p}$-pivot sequence is $\low(\F_p,\bullet)$ by \ref{hyp:2}.

    To conclude, we count the number of operations performed.
    At stage $i$, \Cref{alg:full_correction_modp} calls \Cref{alg:full_correction_priority} with priorities $\tau_i$ and $\tau_{i+1}$.
    Together, these calls perform exactly the adjacent transpositions occurring in the first $m$ stages of a single run of \Cref{alg:full_correction_priority} with priorities $\tau_{\mathrm{piv}}^\Q$ and $\tau_{\mathrm{piv}}^{\F_p}$.
    Therefore, by \Cref{lem:full_correction_priority}, the total number of
    column operations is at most
    \[
        K\bigl(\tau_{\mathrm{piv}}^\Q,\tau_{\mathrm{piv}}^{\F_p}\bigr).
        \qedhere
    \]
\end{proof}

\begin{remark}[see \Cref{subsubsec:full_algorithm}]
    \label{rem:implementation_details}
    In Step~\ref{it:main_algorithm_step4} of \Cref{alg:main_algorithm}, primes are processed independently, leading to redundant computations when they share the same rational columns.
    A more efficient approach is to scan the columns one by one and to partition $\PP$ into subsets of primes sharing the same pivot indices so far.
    We use this optimization in the pseudocode of \Cref{alg:main_algorithm} and in our implementation, but omit it from the theoretical analysis for~simplicity.

    For the complexity bound in \Cref{th:main_algorithm}, we also use an optimization that avoids reducing an entire column modulo $p$ when identifying its pivot.
    For a rational column $c\in\Q^{n+m}$, we denote by $a_1,\dots,a_{n+m}$ the numerators of its entries, from bottom to top, and we store
    \[
        g_k=\gcd(a_1,\dots,a_k),
        \qquad k\in\{1,\dots,n+m\}.
    \]
    By \Cref{lem:pivot_sequence_prefix}, when $c$ is queried modulo $p$, its pivot corresponds to the first index $k$ such that $p\nmid g_k$.
    We find this index by binary search; this uses $O(\log(n+m))$ divisibility tests.
    Computing the $g_k$ costs $O(n+m)$ gcd calls.
    They are recomputed after each column~operation.%
\end{remark}

\begin{proof}[Proof of \Cref{lem:full_correction_amortized}]
    \phantomsection
    \label{proof:full_correction_amortized}
    As in the proof of \Cref{lem:number_exceptional_primes}, we will use Hadamard's inequality to bound the number of exceptional events.
    For an exceptional prime $p\in\PP$, let
    \[
        T_p = K\bigl(\tau_\mathrm{piv}^{\Q\vphantom{\F_p}},\tau_\mathrm{piv}^{\F_p}\bigr).
    \]
    We are interested in
    \begin{equation}
        \label{eq:proof_full_correction_T}
        T = \sum_{p\in\PP} T_p.
    \end{equation}
    First, we bound the individual Kendall distances. 
    Fix $p\in\PP$ and, for $i\in\{1,\dots,n+m\}$, let
    \[
        A_i^\Q=\{\tau_\mathrm{piv}^{\Q}[1],\dots,
               \tau_\mathrm{piv}^{\Q}[i]\},
        \qquad
        A_i^{\F_p}=\{\tau_\mathrm{piv}^{\F_p}[1],\dots,
               \tau_\mathrm{piv}^{\F_p}[i]\}.
    \]
    By \cite[Theorem~2]{DiaconisGraham1977}, the Kendall distance is upper bounded by Spearman's footrule distance:
    \[
        T_p\leq 2\sum_{i=1}^{n+m}\bigl|A_i^\Q\setminus A_i^{\F_p}\bigr|.
    \]
    The rest of the proof consists in bounding the terms $|A_i^\Q\setminus A_i^{\F_p}|$.

    \proofsubparagraph{Bound for $i\leq m$.}

    Let us focus on indices $i \in\{1,\dots,m\}$ for now.

    \begin{claim}
        \label{claim:proof_full_correction_amortized_prefix}
        For every $i \in\{1,\dots,m\}$,
        \[
            |A_i^\Q\setminus A_i^{\F_p}|\leq v_p(\Delta_i),
        \]
        where $v_p$ is the $p$-valuation and $\Delta_i$ is the principal minor of \Cref{eq:principal_minor}.
        Said~differently, 
        \[
            p^{|A_i^\Q\setminus A_i^{\F_p}|} \text{ divides } \Delta_i.
        \]
    Thus, each $\Q$-pivot $\tau_\mathrm{piv}^{\Q}[k]\in A_i^\Q$ that is not an $\F_p$-pivot $\tau_\mathrm{piv}^{\F_p}[\ell]\in A_i^{\F_p}$ contributes to a factor~$p$.
    \end{claim}

    \begin{claimproof}
        To prove it, we consider the greedy procedure described in the proof of Statement~\ref{it:lem_pivot_rank_profile_2}
        of \Cref{lem:pivot_rank_profile}; see \Cref{eq:proof_pivot_rank_profile_det_statement2}.
        Consider the matrix
        \[
            E=\widetilde{D}[:,1 : i] \in \Z^{(n+m)\times i}.
        \]
        We denote by $E^{\F_p}$ its reduction modulo $p$.
        By \Cref{lem:pivot_rank_profile}, the set $A_i^\Q$ (resp.\ $A_i^{\F_p}$) is obtained by greedily selecting independent rows of $E$ (resp.\ $E^{\F_p}$) in canonical priority order (i.e., from bottom to top).
        That is, writing the canonical priority as
        \[
            (n,\dots,1,\bot_m,\dots,\bot_1)=(r_1,\dots,r_{n+m}),
        \]
        we inspect the submatrices
        \[
            E[r_1,:],\quad
            E[(r_1,r_2),:],\quad
            E[(r_1,r_2,r_3),:],\quad\dots
        \]
        and select a new row whenever it increases the rank of the row space over $\Q$ (resp.\ $\F_p$).
        
        Let $q_1\succ \dots \succ q_i$ be the $\Q$-pivot indices selected.
        In particular, their set is $A_i^\Q$.
        Define
        \[
            F=E[(q_1,\dots,q_i),:].
        \]
        Up to a permutation of its rows, it is equal to the matrix in \Cref{eq:principal_minor_det}.
        Consequently,
        \[
            \det F = \pm\Delta_i \neq 0.
        \]
        
        Next, we reduce $F$ to a lower-triangular matrix via invertible column operations over $\Z$.
        More precisely, we scan the rows $1,\dots,i$ from top to bottom.
        At step $k$, we zero out the entries in columns $k+1,k+2,\dots,i$ of row $k$ via Euclidean transformations:
        \begin{equation*}
            F = 
            \setlength{\arraycolsep}{6pt}
            \begin{pNiceArray}{ccccc}[first-col]
                \firstcol{q_1}    &*&*&*&\cdots&* \\
                \firstcol{q_2}    &*&*&*&\cdots&* \\
                \firstcol{q_3}    &*&*&*&\cdots&*  \\
                \firstcol{\vdots~}&\vdots&\vdots&\vdots&\cdots&\vdots
            \end{pNiceArray}
            \to
            \begin{pmatrix}
            *&0&0&\cdots&0\\
            *&*&*&\cdots&*\\
            *&*&*&\cdots&*\\
            \vdots&\vdots&\vdots&\cdots&\vdots
            \end{pmatrix}
            \to
            \begin{pmatrix}
            *&0&0&\cdots&0\\
            *&*&0&\cdots&0\\
            *&*&*&\cdots&*\\
            \vdots&\vdots&\vdots&\cdots&\vdots
            \end{pmatrix}
            \to
            \cdots
        \end{equation*}
        This reduction is possible because $F$ is nonsingular.
        We refer to \Cref{subsubsec:presentation_matrix_tensor_cycles}, and in particular \Cref{eq:euclidean_reduction}, for a definition of Euclidean transformations.
    
        Let $\widehat{F}$ be the result of this process.
        We also apply the same column operations to $E$, yielding the matrix $\widehat{E}$.
        Since the Euclidean transformations are invertible, they preserve $A_i^\Q$, $A_i^{\F_p}$, and $\det(F)$ up to a sign.
        Thus they also preserve $v_p(\det \widehat{F})$.
    
        To continue, consider $k\in\{1,\dots,i\}$ such that the $\Q$-pivot index $q_k\in A_i^\Q$ does not belong to $A_i^{\F_p}$.
        Let $I_k$ denote the set of row indices with strictly higher canonical priority than $q_k$:
        \[
            I_k = \{ r \in \{\bot_1,\dots,\bot_m,1,\dots,n\} \mid r \succ q_k \}.
        \]
        By the greedy procedure over $\Q$, every row of $\widehat{E}$ indexed by $I_k$ lies in the $\Q$-span of the rows
        \[
            \widehat{E}[q_1,:],\quad \widehat{E}[q_2,:],\quad \dots,\quad  \widehat{E}[q_{k-1},:].
        \]
        These are the first $k-1$ rows of $\widehat{F}$.
        Since $\widehat{F}$ is lower triangular, the $k$th coordinates of these rows vanish. We deduce that
        \begin{equation}
            \label{eq:proof_full_correction_amortized_zero}
            \widehat{E}[r,k] = 0
            \qquad\forall r\in I_k.
        \end{equation}
        
        On the other hand, $q_k\notin A_i^{\F_p}$, i.e., the greedy procedure over $\F_p$ skips row $q_k$.
        Therefore, this row belongs to the $\F_p$-span of the previous rows
        \[
            \widehat{E}^{\F_p}[r,:]\qquad \forall r \in I_k.
        \]
        Every row $\widehat{E}^{\F_p}[r,:]$ has zero $k$th coordinate by \Cref{eq:proof_full_correction_amortized_zero}: $\widehat{E}^{\F_p}[r,k]=0$.
        Since $\widehat{E}^{\F_p}[q_k,:]$ is a linear combination of these rows, we deduce that
        \[
            \widehat{E}[q_k,k] = 0 \pmod{p},
        \]
        hence that $\widehat{F}[k,k] = 0 \pmod{p}$.
        
        Thus, every row in $A_i^\Q\setminus A_i^{\F_p}$ forces a diagonal entry of $\widehat{F}$ to be divisible by $p$, hence
        \[
            |A_i^\Q\setminus A_i^{\F_p}|
            \leq \sum_{k=1}^{i}v_p(\widehat{F}[k,k])
            =v_p(\det \widehat{F})
            =v_p(\Delta_i).
        \]
        This is the bound announced in \Cref{claim:proof_full_correction_amortized_prefix}.
    \end{claimproof}

    \proofsubparagraph{Bound for $i> m$.}

    When $i \in\{m+1,\dots,m+n\}$, the principal minor $\Delta_i$ is not defined.
    But we can still obtain a bound similar to \Cref{claim:proof_full_correction_amortized_prefix}.
    
    By \Cref{def:permutation_piv_p}, the last $n$ entries of $\tau_\mathrm{piv}^{\Q\vphantom{\F_p}}$ are inserted in canonical order, and the last $n$ entries of $\tau_\mathrm{piv}^{\F_p}$ are inserted in their order in $\tau_\mathrm{piv}^{\Q\vphantom{\F_p}}$.
    Thus, $\tau_\mathrm{piv}^{\F_p}[i]$ is the $(i-m)$th entry remaining after deleting the $m$ pivot indices $\tau_\mathrm{piv}^{\F_p}[1],\dots,\tau_\mathrm{piv}^{\F_p}[m]$ from $\tau_\mathrm{piv}^{\Q}$.
    Since only $m$ entries have been deleted, $\tau_\mathrm{piv}^{\F_p}[i]$ lies among the first $i=(i-m)+m$ entries of $\tau_\mathrm{piv}^{\Q}$, hence belongs to $A_i^\Q$:
    \[
        \tau_\mathrm{piv}^{\F_p}[i] \in A_i^\Q.
    \]
    Consequently, $A_i^{\F_p}\setminus A_m^{\F_p}\subset A_i^\Q$, and $A_i^{\F_p}\setminus A_i^\Q=A_m^{\F_p}\setminus A_i^\Q$.
    Since $|A_i^\Q|=|A_i^{\F_p}|$, we deduce~that%
    \[
        \bigl|A_i^\Q\setminus A_i^{\F_p}\bigr|
        = \bigl|A_i^{\F_p}\setminus A_i^\Q\bigr|
        = \bigl|A_m^{\F_p}\setminus A_i^\Q\bigr|
        \leq \bigl|A_m^{\F_p}\setminus A_m^\Q\bigr|.
    \]
    Combining this with
    \Cref{claim:proof_full_correction_amortized_prefix} yields
    \begin{equation}
        \label{eq:proof_full_correction_amortized_Tp}
        T_p
        \leq 2\left(\sum_{i=1}^{m}v_p(\Delta_i)+(n-1)v_p(\Delta_m)\right).
    \end{equation}

    \proofsubparagraph{Sum over all exceptional primes.}

    We now conclude with two additional facts.
    First, the minors $\Delta_j$ are determinants of $j\times j$ submatrices of $\widetilde{D}$.
    By Hadamard's inequality, they~satisfy%
    \begin{equation}
        \label{eq:proof_full_correction_hadamard}
        |\Delta_j| \leq \nu^j        
    \end{equation}
    where $\nu=\|D\|_{2,\infty}$.
    Second, if $a \neq 0$ is an integer, then its prime decomposition yields
    \[
        |a| 
        = \prod_{p\mid a} p^{v_p(a)}
        \geq \prod_{p\mid a} 2^{v_p(a)}
        = 2^A
        \qquad\text{where}\qquad
        A = \sum_{p\mid a} v_p(a).
    \]
    By taking logarithms, we obtain
    \begin{equation}
        \label{eq:proof_full_correction_valuation}
        A 
        \leq \frac{\log |a|}{\log 2}.
    \end{equation}
    We can now proceed to the amortized analysis.
    Summing \eqref{eq:proof_full_correction_amortized_Tp} over $p\in\PP$, we obtain
    \begin{align*}
        T
        & \leq \sum_{p\in\PP}2\left(\sum_{i=1}^{m}v_p(\Delta_i)+(n-1)v_p(\Delta_m)\right)
        \tag{by \Cref{eq:proof_full_correction_T,eq:proof_full_correction_amortized_Tp}}\\
        &\leq
        \frac{2}{\log 2}
        \left(\sum_{i=1}^{m}\log|\Delta_i|+(n-1)\log|\Delta_m|\right)
        \tag{by \Cref{eq:proof_full_correction_valuation}}\\
        &\leq
        \frac{2}{\log 2}
        \left(\sum_{i=1}^{m}i+(n-1)m\right)\log\nu
        \tag{by \Cref{eq:proof_full_correction_hadamard}}\\
        &=
        \frac{m(m+2n-1)}{\log 2}\log\nu\\
        &\leq
        \frac{(m+n)^2}{\log 2}\log\nu.
    \end{align*}
    The inequality $1/\log 2<2$ then yields the result.
\end{proof}

\begin{proof}[Proof of \Cref{th:main_algorithm}]
    \phantomsection
    \label{proof:main_algorithm}
    We consider an arbitrary matrix $D\in\Z^{n\times m}$ and put $\nu=\|D\|_{2,\infty}$.

    \proofsubparagraph{Correctness.}

    We analyze the algorithm step by step.

    \begin{itemize}
        \item 
        Step~\ref{it:main_algorithm_step1} computes the rational pivot sequence $\low(\Q,\bullet)$ via the usual persistence algorithm.
        \item
        Step~\ref{it:main_algorithm_step2} computes the principal minors $\Delta_j$ using \Cref{eq:principal_minor}.
        By \Cref{lem:generic_reduction_identify_primes}, the prime divisors of these minors are precisely the exceptional primes $\PP$.
        \item
        Step~\ref{it:main_algorithm_step3} computes an exhaustive reduction with the same canonical pivots by \Cref{lem:exhaustive_reduction}.
        \item
        Step~\ref{it:main_algorithm_step4} computes the exceptional pivots $\low(\F_p,\bullet)$ for each $p\in\PP$ by \Cref{lem:full_correction_modp}.
    \end{itemize}

    \noindent
    This proves the correctness of the computation of $\low(\Q,\bullet)$, $\PP$, and $\low(\F_p,\bullet)$, $p\in\PP$.

    \proofsubparagraph{Complexity of each step.}

    As above, we analyze each step.

    \begin{itemize}
        \item 
        Steps~\ref{it:main_algorithm_step1} and \ref{it:main_algorithm_step3}:
        each of the two rational reductions uses $O(m^2)$ column operations by \Cref{lem:persistence_algorithm_priority}.
        Since a column of $\widetilde{D}$ has $n+m$ entries, the number of arithmetic operations is
        \begin{equation}
            \label{eq:main_algorithm_complexity_1}
            O\bigl((n+m) m^2\bigr).
        \end{equation}
        
        \item
        Step~\ref{it:main_algorithm_step2}:
        computing the minors $\Delta_1,\dots,\Delta_m$ successively requires $O(m)$ multiplications by \Cref{eq:principal_minor}.
        Factorizing each of them is counted as $O(1)$.
        The cost of this step is~thus
        \begin{equation}
            \label{eq:main_algorithm_complexity_2}
            O(m).
        \end{equation}

        \item
        Step~\ref{it:main_algorithm_step4}---Updates:
        by \Cref{lem:full_correction_amortized}, the number of adjacent transpositions performed~is
        \[
            T = O\bigl((n+m)^2\log \nu \bigr).
        \]
        Each transposition requires at most one column operation.
        Therefore this step costs
        \begin{equation}
            \label{eq:main_algorithm_complexity_3}
            O\bigl((n+m)^3\log\nu \bigr).
        \end{equation}

        \item
        Step~\ref{it:main_algorithm_step4}---Finding pivots:
        As explained in \Cref{rem:implementation_details}, for each column we compute
        \[
            g_k=\gcd(a_1,\dots,a_k),
            \qquad k\in\{1,\dots,n+m\},
        \]
        where $a_1,\dots,a_{n+m}$ are the numerators of its entries, from bottom to top.
        This costs $O(n+m)$ operations per column.
        Moreover, we recompute the $g_1,\dots,g_{n+m}$ after each column operation.
        This yields a total cost of
        \[
            O\bigl((n+m)(m+T)\bigr)
        \]
        where $T= O((n+m)^2\log \nu)$ is the number of adjacent transpositions.
        In addition, each exceptional prime requires $m$ pivot queries, each costing $O(\log(n+m))$ divisibility tests (we use binary search on the $g_1,\dots,g_{n+m}$).
        This costs 
        \[
            O\bigl(m|\PP|\log(n+m)\bigr)
        \]
        tests in total.
        To bound this quantity, we first observe that
        \begin{align*}
            \sum_{p\in\PP}\log p
            &\leq \sum_{j=1}^{m}\log|\Delta_j| \tag{every exceptional prime divides a principal minor}\\
            &\leq \sum_{j=1}^{m}\log \nu^j \tag{by \Cref{eq:proof_full_correction_hadamard}}\\
            &= \frac{m(m+1)}{2}\log\nu.
        \end{align*}
        We deduce an upper bound by treating separately the primes $p>n+m$:
        \begin{align*}
            m|\PP|\log(n+m)
            &= \bigg(m \sum_{\substack{p\in\PP\\p\leq n+m}} \log(n+m)\bigg)
                +\bigg(m\sum_{\substack{p\in\PP\\p>n+m}}\log (n+m)\bigg)\\
            &\leq \bigg(m (n+m)\log(n+m)\bigg)
                +\bigg(m\sum_{\substack{p\in\PP\\p>n+m}}\log p\bigg)\\
            &=O\bigl((n+m)^3\max(1,\log\nu)\bigr).
        \end{align*}
        Overall, the cost for maintaining the $g_1,\dots,g_{n+m}$ of each column and finding the pivots~is%
        \begin{equation}
            \label{eq:main_algorithm_complexity_4}
            O\bigl((n+m)(m+T)+(n+m)^3\max(1,\log\nu)\bigr)
            = O\bigl((n+m)^3\max(1,\log\nu)\bigr).
        \end{equation}
    \end{itemize}

    \proofsubparagraph{Total complexity.}

    Combining the bounds in \Cref{eq:main_algorithm_complexity_1,eq:main_algorithm_complexity_2,eq:main_algorithm_complexity_3,eq:main_algorithm_complexity_4} yields the total cost
    \[
        O\bigl( (n+m) m^2 + m + (n+m)^3\log\nu + (n+m)^3\max(1,\log\nu) \bigr)
        = O\bigl( (n+m)^3\max(1,\log\nu) \bigr).
    \]
    This is the first complexity bound stated in the theorem.

    For the second bound, we assume that $D$ is the boundary matrix of a filtration with $N$ simplices of dimension at most $\ell$.
    In this case, $n=m=N$.
    Moreover, as already observed,
    \[
        \|D\|_{2,\infty}\leq\sqrt{\ell+1}.
    \]
    Replacing $\nu$ by this value in the previous bound yields the result.
\end{proof}

\begin{proof}[Proof of \Cref{th:field_independence_problem}]
    \phantomsection
    \label{proof:field_independence_problem}

    Let $(K^t)_{t=0}^T$ be a simplicial filtration, potentially not simplex-wise.%

    \proofsubparagraph{Correctness.}

    For a simplex $\sigma\in K$, we denote by $t(\sigma)\in\R$ its appearance time in the filtration.
    \Cref{alg:field_independence_problem} starts by refining $(K^t)_{t=0}^T$ into a simplex-wise filtration $(\overline{K^t})_{t=0}^N$.
    That is, it orders the simplices as
    \[
        \sigma_1, \dots, \sigma_N
    \]
    so that $i<j$ whenever $\sigma_i \subsetneq \sigma_j$ and $t(\sigma_i) \leq t(\sigma_j)$ whenever $i<j$.
    In the refined filtration $(\overline{K^t})_{t=0}^N$, we can assume that the times are the indices
    \[
        1, \dots, N.
    \]
    \Cref{alg:main_algorithm} is then called on $(\overline{K^t})_{t=0}^N$.
    It computes its exceptional primes $\PP$ and pivot sequences over $\Q$ and $\F_p$, $p\in\PP$.
    For such a field, the barcode is obtained by \Cref{eq:barcode_from_pivots}:
    \[
        \barcode\big(H_d(\overline{K^\bullet};\kk)\big)
        =
        \bigl\{\!\bigl\{
            [i,d_i)
            \mid i\in\{1,\dots,N\},\;
            \dim(\sigma_i)=d,\;
            \low(\kk,i)=\bot_i
        \bigr\}\!\bigr\}
    \]
    where
    \[
        d_i=
        \begin{cases}
        j & \text{if } \low(\kk,j)=i\text{ for some }j\in\{1,\dots,N\},\\
        +\infty & \text{otherwise}.
        \end{cases}
    \]
    The barcodes of the original filtration are obtained by restoring the original filtration values and discarding the empty intervals:
    \[
        \barcode\big(H_d(K^\bullet;\kk)\big)
        =
        \bigl\{\!\bigl\{
            [t(\sigma_i),\widetilde{d_i})
            \mid i\in\{1,\dots,N\},\;
            \dim(\sigma_i)=d,\;
            \low(\kk,i)=\bot_i
        \bigr\}\!\bigr\}
        \setminus \{\emptyset\}
    \]
    where
    \[
        \widetilde{d_i} =
        \begin{cases}
        t(\sigma_j) & \text{if } \low(\kk,j)=i\text{ for some }j\in\{1,\dots,N\},\\
        +\infty & \text{otherwise}.
        \end{cases}
    \]
    This justifies that the algorithm correctly computes the barcodes of $H_d(K^\bullet;\kk)$.
    The last step is to compare all barcodes with one another, and we deduce that the algorithm is correct.

    \proofsubparagraph{Complexity.}

    By \Cref{th:main_algorithm}, computing the exceptional primes and the pivot sequences~costs
    \[
        O\bigl(N^3\log(\ell+1)\bigr)
    \]
    arithmetic operations.
    Moreover, \Cref{lem:number_exceptional_primes} gives
    \[
        |\PP|=O(N^2).
    \]
    Converting a given pivot sequence into a barcode costs linear time.
    Thus, computing the barcode of $H_d(K^\bullet;\kk)$ for every $\kk\in\{\Q\}\cup\{\F_p\mid p\in\PP\}$ costs
    \[
        O\biggl(N^3\log(\ell+1) + N\times N^2\biggr) = O\biggl(N^3\log(\ell+1)\biggr).
    \]
    For each of these $O(N^2)$ barcodes, we sort its bars in lexicographic order.
    This costs
    \[
        O(N\log(N))
    \]
    per barcode.
    Each is then compared to a reference barcode, for instance the rational~barcode
    \[
        \barcode(H_d(K^\bullet;\Q)),
    \]
    in $O(N)$ time.
    Therefore the full procedure costs
    \[
        O\biggl(
        \underbrace{
        N^3\log(\ell+1)
        }_{\text{computing barcodes}}
        + 
        \underbrace{
        N^2 \times N\log(N) 
        }_{\text{sorting barcodes}}
        + 
        \underbrace{
        N^2 \times N \vphantom{N^3\log(\ell+1)}
        }_{\text{comparing barcodes}}
        \biggr) 
        = O\biggl(N^3\log(N)\log(\ell+1)\biggr).
        \qedhere
    \]
\end{proof}

\subsection{Proofs for \texorpdfstring{\Cref*{sec:torsion_on_barcodes}}{Section \ref*{sec:torsion_on_barcodes}}}

\begin{proof}[Proof of \Cref{lem:three_persistence_modules}]
    \phantomsection
    \label{proof:three_persistence_modules}
    
    We prove the two statements successively.
    
    \proofsubparagraph{Epimorphism.}
    
    The first homomorphism comes from the description of $H_d^\mathrm{free}(K^t;\Z)$ as a quotient of $H_d(K^t;\Z)$, as in \Cref{eq:free_part}:
    \[
        H_d^\mathrm{free}(K^t;\Z) = H_d(K^t;\Z) / H_d^\mathrm{tors}(K^t;\Z).        
    \]
    Thus, we have the exact sequence
    \[
        0 \to H_d^\mathrm{tors}(K^t;\Z) \to H_d(K^t;\Z) \to H_d^\mathrm{free}(K^t;\Z) \to 0.
    \]
    Since tensoring with $\kk$ is a right-exact functor, we get that the following sequence is exact:
    \[
        H_d(K^t;\Z)\otimes\kk \to H_d^\mathrm{free}(K^t;\Z)\otimes\kk \to 0.
    \]
    Next, we must show that this sequence is natural.
    First, observe that each structure map 
    \[
        h_{s,t}\colon H_d(K^s;\Z)\to H_d(K^t;\Z)
    \]
    preserves the torsion elements:
    \[
        h_{s,t} \big( H_d^\mathrm{tors}(K^s;\Z) \big) \subset H_d^\mathrm{tors}(K^t;\Z).
    \]
    In particular, it induces a homomorphism between the free parts:
    \begin{equation}
        \label{eq:proof_three_persistence_modules_epimorphism}
        \overline{h}_{s,t}\colon H_d^\mathrm{free}(K^s;\Z) \to H_d^\mathrm{free}(K^t;\Z).
    \end{equation}
    Thus, we have a commutative diagram of exact sequences
    \[
        \begin{tikzcd}[column sep=normal, row sep=normal]
            0 \arrow[r]
            & H_d^\mathrm{tors}(K^s;\Z)
                \arrow[r]
                \arrow[d, "h_{s,t}"']
            & H_d(K^s;\Z)
                \arrow[r]
                \arrow[d, "h_{s,t}"']
            & H_d^\mathrm{free}(K^s;\Z)
                \arrow[r]
                \arrow[d, "\overline{h}_{s,t}"']
            & 0
            \\
            0 \arrow[r]
            & H_d^\mathrm{tors}(K^t;\Z)
                \arrow[r]
            & H_d(K^t;\Z)
                \arrow[r]
            & H_d^\mathrm{free}(K^t;\Z)
                \arrow[r]
            & 0
        \end{tikzcd}
    \]
    As above, tensoring with $\kk$ yields the diagram
    \[
        \begin{tikzcd}[column sep=normal, row sep=normal]
              H_d(K^s;\Z)\otimes\kk
                \arrow[r]
                \arrow[d, "h_{s,t}\otimes\kk"']
            & H_d^\mathrm{free}(K^s;\Z)\otimes\kk
                \arrow[r]
                \arrow[d, "\overline{h}_{s,t}\otimes\kk"']
            & 0
            \\
              H_d(K^t;\Z)\otimes\kk
                \arrow[r]
            & H_d^\mathrm{free}(K^t;\Z)\otimes\kk
                \arrow[r]
            & 0
        \end{tikzcd}
    \]
    This is precisely the required epimorphism of persistence modules:
    \[
        H_d(K^\bullet;\Z)\otimes\kk \to H_d^\mathrm{free}(K^\bullet;\Z)\otimes\kk \to 0.
    \]

    \proofsubparagraph{Monomorphism.}

    For the second homomorphism, we use the fact that the UCT exact sequence in \Cref{eq:uct} is natural.
    Thus, it induces an exact sequence of persistence~modules:%
    \[
        0
        \xrightarrow[]{} H_d(K^\bullet;\Z) \otimes \kk
        \xrightarrow[]{} H_d(K^\bullet;\kk)
        \xrightarrow[]{} \Tor(H_{d-1}(K^\bullet;\Z), \kk)
        \xrightarrow[]{} 0.
    \]
    Its first three terms yield the required monomorphism:
    \[
        0
        \xrightarrow[]{} H_d(K^\bullet;\Z) \otimes \kk
        \xrightarrow[]{} H_d(K^\bullet;\kk).
        \qedhere
    \]
\end{proof}

\begin{proof}[Proof of \Cref{th:birth_death_profiles}]
    \phantomsection
    \label{proof:birth_death_profiles}

    We study, in order, the barcodes of $H_d^\mathrm{free}(K^\bullet;\Z)\otimes\kk$, $H_d(K^\bullet;\Z)\otimes\kk$, and $H_d(K^\bullet;\kk)$.
    We then conclude by identifying the exceptional primes.

    \proofsubparagraph{Free module.}

    We first show that the barcode of $H_d^\mathrm{free}(K^\bullet;\Z)\otimes\kk$ is obtained from the barcode of $H_d(K^\bullet;\Q)$ by successive applications of death time swaps as in \Cref{eq:death_times_swap}.
    The proof has three steps: we characterize the effect of adding a simplex in \Cref{eq:proof_birth_death_profiles_free_dim_d,eq:proof_birth_death_profiles_free_dim_d+1}, then we deduce two relations regarding the birth and death times of the persistence modules in \Cref{eq:proof_birth_death_profiles_multisets_equality,eq:proof_birth_death_profiles_inequality_rank}, and we conclude with a combinatorial argument.
    The last step involves a classical characterization of the strong Bruhat order on permutations, as in \cite[Lemma 11]{Fulton1997}, that we prove in \Cref{lem:bruhat_order} below for completeness.

    First, we observe that multisets of births and deaths in $H_d^\mathrm{free}(K^\bullet;\Z)\otimes\kk$ do not depend on $\kk$.
    To see this, recall that a birth (resp.\ a death) occurs precisely when the dimension of $H_d^\mathrm{free}(K^t;\Z)\otimes\kk$ increases (resp.\ decreases).
    We show that this condition is independent of $\kk$.

    We inspect the effect of adding a simplex.
    Suppose that $K^{t+1} = K^t\cup\{\sigma\}$ where $\sigma$ has dimension $k$.
    The pair $(K^{t+1},K^t)$ has the relative homology of a $k$-sphere:
    \[
        H_{k}(K^{t+1},K^t;\Z) \cong \Z,\qquad
        H_{k-1}(K^{t+1},K^t;\Z) \cong 0.
    \]
    Moreover, the boundary $\partial\sigma$ belongs to $Z_{k-1}(K^t;\Z)$.
    We denote its homology class as
    \begin{equation}
        \label{eq:homology_class_boundary_new_simplex}
        [\partial \sigma]\in H_{k-1}(K^t;\Z).
    \end{equation}
    The long exact sequence for the pair $(K^{t+1},K^t)$ reads
    \[
        0
        \to H_{k}(K^t;\Z)
        \to H_{k}(K^{t+1};\Z)
        \to \underbracket{H_{k}(K^{t+1},K^t;\Z)}_{\cong\Z}
        \xrightarrow[]{\delta} H_{k-1}(K^t;\Z)
        \to H_{k-1}(K^{t+1};\Z)
        \to 0
    \]
    where the map $\delta$ satisfies $\delta(1) = [\partial\sigma]$.
    In particular, the homology group $H_d(K^{t+1};\Z)$ is affected only when the simplex has dimension $k=d$ or $d+1$.
    Let us distinguish these cases.
    
    We start with the case $k=d$. The long exact sequence yields
    \[
        0
        \to H_d(K^t;\Z)
        \to H_d(K^{t+1};\Z)
        \to \ker (\delta)
        \to 0.
    \]
    The $\Z$-module $\ker (\delta)$ is free as it is a subgroup of $\Z$.
    Therefore the sequence splits and we get
    \[
        H_d(K^{t+1};\Z) = H_d(K^t;\Z) \oplus \ker (\delta).
    \]
    Passing to the free parts yields
    \[
        H_d^\mathrm{free}(K^{t+1};\Z) = H_d^\mathrm{free}(K^t;\Z) \oplus \ker (\delta).
    \]
    Thus, the tensor products satisfy
    \begin{equation}
        \label{eq:proof_birth_death_profiles_free_dim_d}
        H_d^\mathrm{free}(K^{t+1};\Z)\otimes\kk \cong \big(H_d^\mathrm{free}(K^t;\Z)\otimes\kk\big) \oplus \big(\ker (\delta)\otimes\kk \big).
    \end{equation}
    In particular, the dimension of $H_d^\mathrm{free}(K^{t+1};\Z)\otimes\kk$ can only increase.
    It occurs precisely when $\ker (\delta)$ is nonzero, since it is free.
    This condition does not depend on $\kk$, as wanted.

    Next, we consider the case $k=d+1$.
    The long exact sequence yields
    \[
        0
        \to \im(\delta)
        \to H_d(K^t;\Z)
        \xrightarrow[]{h_t} H_d(K^{t+1};\Z)
        \to 0.
    \]
    However, unlike in the previous case, this sequence may not split. 
    Instead, we consider
    \[
        H_d^\mathrm{free}(K^t;\Z)
        \xrightarrow[]{\overline{h}_t} H_d^\mathrm{free}(K^{t+1};\Z)
        \to 0.
    \]
    The map $\overline{h}_t$ is a surjection because $h_t$ preserves the torsion part; see \Cref{eq:proof_three_persistence_modules_epimorphism} in the proof of \Cref{lem:three_persistence_modules}.
    Therefore we obtain the short exact sequence
    \[
        0
        \to \ker(\overline{h}_t)
        \to H_d^\mathrm{free}(K^t;\Z)
        \xrightarrow[]{\overline{h}_t} H_d^\mathrm{free}(K^{t+1};\Z)
        \to 0.
    \]
    Now the sequence splits, since $H_d^\mathrm{free}(K^{t+1};\Z)$ is free.
    We deduce that
    \[
        H_d^\mathrm{free}(K^t;\Z) = H_d^\mathrm{free}(K^{t+1};\Z) \oplus \ker(\overline{h}_t).
    \]
    Note that $\ker(\overline{h}_t)$ is also free since it is a submodule of $H_d^\mathrm{free}(K^t;\Z)$.
    Last, tensoring gives
    \begin{equation}
        \label{eq:proof_birth_death_profiles_free_dim_d+1}
        H_d^\mathrm{free}(K^{t};\Z)\otimes\kk \cong \big(H_d^\mathrm{free}(K^{t+1};\Z)\otimes\kk\big) \oplus \big(\ker(\overline{h}_t)\otimes\kk \big).
    \end{equation}
    Thus, the dimension of $H_d^\mathrm{free}(K^{t+1};\Z)\otimes\kk$ can only decrease;
    this occurs when $\ker(\overline{h}_t)$ is nonzero.
    This condition is, again, independent of the field $\kk$.

    We now compare the barcodes of the persistence modules involved.
    Observe that the birth times $b_i(\Q)$ are distinct; this is because the filtration $(K^t)_{t=0}^N$ is simplex-wise.
    Thus we can order the bars in increasing birth order:
    \[
        \barcode\big(H_d(K^\bullet;\Q)\big) 
        = \big( \big[b_i,d_i\big) \big)_{i=1}^r
        \qquad\text{and}\qquad
        \barcode\big(H_d^\mathrm{free}(K^\bullet;\Z)\otimes\kk\big)
        = \big( \big[b_i,d_i'\big) \big)_{i=1}^r
    \]
    with $b_1<\cdots<b_r$.
    Moreover, as we have just proven, the multisets of death times coincide:
    \begin{equation}
        \label{eq:proof_birth_death_profiles_multisets_equality}
        \big\{\!\big\{ d_1,\dots,d_r \big\}\!\big\} = \big\{\!\big\{ d_1',\dots,d_r' \big\}\!\big\}.
    \end{equation}
    Next, consider structure maps of the free persistence module, denoted
    \[
        \overline{h}_{s,t}\colon H_d^\mathrm{free}(K^s;\Z)\to H_d^\mathrm{free}(K^t;\Z),
    \]
    and tensor them with $\Q$ and $\kk$.
    One has the classical inequality
    \[
        \rank\big(\overline{h}_{s,t}\otimes\kk\big)
        \leq 
        \rank\big(\overline{h}_{s,t}\otimes\Q\big).
    \]
    This relation is clear by seeing the map as an integer matrix.
    In terms of bars, it yields
    \begin{equation}
        \label{eq:proof_birth_death_profiles_inequality_rank}
        \big|\big\{ \ell\in \{1,\dots,m\} \mid d'_\ell\geq t \big\}\big|
        \leq 
        \big|\big\{ \ell\in \{1,\dots,m\} \mid d_\ell\geq t \big\}\big|
    \end{equation}
    for all $m \in \{1,\dots,r\}$ and $t \in \{1,\dots,N,+\infty\}$.
    
    \Cref{eq:proof_birth_death_profiles_multisets_equality,eq:proof_birth_death_profiles_inequality_rank} are all that is needed to apply \Cref{lem:bruhat_order}.
    It states that $(d_1,\dots,d_r)$ can be transformed into $(d_1',\dots,d_r')$ by a sequence of swaps as in \Cref{eq:proof_birth_death_profiles_swap_operation}:
    \[
     d_i \longleftrightarrow d_j,
        \quad
        d_i > d_j, 
        \quad
        i<j.
    \]
    Note that this operation corresponds to the death swaps of nested intervals as in \Cref{eq:death_times_swap}:
    \[    
        [b_i,d_i),\; [b_j,d_j) \longrightarrow [b_i,d_j),\; [b_j,d_i),
        \qquad
        b_i < b_j < d_j < d_i.
    \]    
    We conclude that $\barcode(H_d^\mathrm{free}(K^\bullet;\Z)\otimes\kk)$ is obtained from $\barcode(H_d(K^\bullet;\Q))$ by a sequence of such moves, as wanted.
    
    \proofsubparagraph{Tensor module.}

    The fact that the barcode of the tensor module differs from that of the free module by only extending certain intervals to the right is a direct consequence of Bauer and Lesnick's Structure Theorem \cite{BauerLesnick2015} for epimorphisms.
    Recall the epimorphism of \Cref{lem:three_persistence_modules}:
    \[
        H_d(K^\bullet;\Z)\otimes\kk \twoheadrightarrow H_d^\mathrm{free}(K^\bullet;\Z)\otimes\kk.
    \]
    In this case, the Structure Theorem gives an injection
    \[
        \barcode\big(H_d(K^\bullet;\Z)\otimes\kk\big)
        \hookleftarrow 
        \barcode\big(H_d^\mathrm{free}(K^\bullet;\Z)\otimes\kk\big)
    \]
    which sends intervals $[b,d)$ to $[b,d')$ with $d\leq d'$.
    Moreover, this injection is a bijection since the numbers of bars coincide.
    Indeed, a birth occurs each time the rank of $Z_d(K^t;\Z)$~increases.%

    \proofsubparagraph{Field-coefficient module.}

    This case is similar.
    Recall the monomorphism of \Cref{lem:three_persistence_modules}:
    \[
        H_d(K^\bullet;\Z) \otimes \kk \hookrightarrow H_d(K^\bullet;\kk).
    \]
    By the Structure Theorem, there exists an injection
    \[
        \barcode\big( H_d(K^\bullet;\Z) \otimes \kk \big)
        \hookrightarrow 
        \barcode\big(H_d(K^\bullet;\kk)\big)
    \]
    that sends intervals $[b,d)$ to $[b',d)$ with $b'\leq b$.
    As before, the numbers of bars in these barcodes coincide.
    Indeed, the number of births in the second barcode is $\dim( Z_d(K^N;\kk))$, and the number of births in the first one is
    \[
        \dim\big(Z_d(K^N;\Z)\otimes\kk\big),
    \]
    which agrees with $\dim( Z_d(K^N;\kk))$ since $H_{d-1}(K^N;\Z)$ is torsion free by assumption.
    
    \proofsubparagraph{Exceptional primes.}

    Let us study how $\sigma_i(\kk)$, $\epsilon_i^-(\kk)$, and $\epsilon_i^+(\kk)$ depend on $\kk$.
    Recall that a $\kk$-persistence module $(M^t)_{t=0}^N$ is entirely characterized, up to isomorphism, by its rank~invariant%
    \[
        \rho(s,t) = \rank\big(M^s \to M^t\big),
        \qquad
        0\leq s \leq t \leq N.
    \]
    The structure maps of the three persistence modules under consideration are
    \begin{align*}
        f_{s,t}^\kk\colon 
        &H_d^\mathrm{free}(K^s;\Z)\otimes\kk\to H_d^\mathrm{free}(K^t;\Z)\otimes\kk, 
        \\
        g_{s,t}^\kk\colon 
        &H_d(K^s;\Z)\otimes\kk\to H_d(K^t;\Z)\otimes\kk, 
        \\
        h_{s,t}^\kk\colon 
        &H_d(K^s;\kk)\to H_d(K^t;\kk).
    \end{align*}
    An inequality $\rank f_{s,t}^\kk \neq\rank f_{s,t}^\Q$ occurs precisely when the cokernel of the integral map,
    \[
        \operatorname{Coker}(H_d^\mathrm{free}(K^s;\Z)\to H_d^\mathrm{free}(K^t;\Z)),
    \]
    has $p$-torsion, where $p$ is the characteristic of $\kk$.
    Equivalently, this occurs when the Smith normal form of this integral map has an elementary divisor divisible by $p$.
    As $s$ and $t$ vary in $\{0,\dots,N\}$, there is only a finite number of such elementary divisors.
    This shows that the ranks of $f_{s,t}^\kk$ and $f_{s,t}^\Q$ differ only for a finite number of field characteristics.   

    Next, apart from these exceptional primes, the inequality $\rank g_{s,t}^\kk\neq\rank g_{s,t}^\Q$ occurs only when $H_d(K^s;\Z)$ or $H_d(K^t;\Z)$ has $p$-torsion; this happens only a finite number of times.
    Last, excluding all the exceptional primes collected so far, $\rank h_{s,t}^\kk\neq\rank h_{s,t}^\Q$ holds only when $H_{d-1}(K^s;\Z)$ or $H_{d-1}(K^t;\Z)$ have $p$-torsion.
    The same argument applies: the ranks disagree only for a finite number of characteristics.

    Excluding this finite number of exceptional primes $\PP$, all other fields $\kk$ with characteristic $p\notin\PP$ produce persistence modules with the same rank invariant as the generic modules.
    In particular, they admit the same barcodes.
    Thus, the values of $\sigma_i(\kk)$, $\epsilon_i^-(\kk)$ and $\epsilon_i^+(\kk)$ agree with $\sigma_i(\Q)$, $\epsilon_i^-(\Q)$ and $\epsilon_i^+(\Q)$.
    This last statement concludes the proof.
\end{proof}

\begin{lemma}[{based on \cite[Lemma 11]{Fulton1997}}]
    \label{lem:bruhat_order}
    Let $(d_1,\dots,d_r)$ and $(d_1',\dots,d_r')$ be two sequences of elements in $\{1,\dots,N,+\infty\}$ that are equal as multisets:
    \[
        \big\{\!\big\{ d_1,\dots,d_r \big\}\!\big\} = \big\{\!\big\{ d_1',\dots,d_r' \big\}\!\big\}.
    \]
    Moreover, suppose that for every $m \in \{1,\dots,r\}$ and $t \in \{1,\dots,N,+\infty\}$ one has
    \begin{equation}
        \label{eq:lem_bruhat_order}
        \big|\big\{ \ell\in \{1,\dots,m\} \mid d'_\ell\geq t \big\}\big|
        \leq 
        \big|\big\{ \ell\in \{1,\dots,m\} \mid d_\ell\geq t \big\}\big|.
    \end{equation}
    Then we can transform $(d_1,\dots,d_r)$ into $(d_1',\dots,d_r')$ by a sequence of swaps of the form
    \begin{equation}
        \label{eq:proof_birth_death_profiles_swap_operation}
        d_i \longleftrightarrow d_j,
        \quad
        d_i > d_j, 
        \quad
        i<j.
    \end{equation}
\end{lemma}

\begin{proof}[Proof of \Cref{lem:bruhat_order}]
    \phantomsection
    Assume $(d_1,\dots,d_r)\neq(d_1',\dots,d_r')$.
    Let $i$ be the first index such that
    \[
        d_i\neq d_i'.
    \]
    We will swap $d_i$ with another element $d_j$.
    First, note that the hypothesis \Cref{eq:lem_bruhat_order} gives%
    \begin{equation*}
        d_i > d_i'.
    \end{equation*}
    Let $j\in\{i+1,\dots,r\}$ be the first index such that
    \[
        d_i > d_j \geq d_i'.
    \]
    Such a $j$ exists because the multisets are equal by assumption.
    We then perform the swap 
    \[
        d_i \longleftrightarrow d_j
    \]
    of \Cref{eq:proof_birth_death_profiles_swap_operation}.
    After the swap, if $(d_1,\dots,d_r)=(d_1',\dots,d_r')$, then the result is proved.
    Otherwise, we repeat the process.
    We must show that \Cref{eq:lem_bruhat_order} is conserved.

    The only affected indices are the $m$ and $t$ such that $i\leq m < j$ and $d_j < t \leq d_i$.
    By the choice of $j$, none of the $d_i,\dots,d_{j-1}$ lies in $\big[d_i',d_i\big)$.
    Thus,
    \begin{align*}
        \big|\big\{ \ell\in \{i,\dots,m\} \mid d_\ell\geq t \big\}\big|
        &= \big|\big\{ \ell\in \{i,\dots,m\} \mid d_\ell\geq d_i' \big\}\big|\\
        &\geq \big|\big\{ \ell\in \{i,\dots,m\} \mid d_\ell'\geq d_i' \big\}\big|\\
        &\geq \big|\big\{ \ell\in \{i,\dots,m\} \mid d_\ell'\geq t \big\}\big|+1.
    \end{align*}
    The first inequality comes from the original \Cref{eq:lem_bruhat_order}, and the last inequality from $d_i'< t$.
    Thus, \Cref{eq:lem_bruhat_order} is still valid, as required:
    \[
        \big|\big\{ \ell\in \{1,\dots,m\} \mid d_\ell'\geq t \big\}\big| \leq \big|\big\{ \ell\in \{1,\dots,m\} \mid d_\ell\geq t \big\}\big|.
    \]

    After each swap $d_i \longleftrightarrow d_j$ with $d_i > d_j$ and $i < j$, the new sequence $(d_1,\dots,d_r)$ is smaller in lexicographic order.
    Thus, the process terminates.
\end{proof}

\subsection{Proofs for \texorpdfstring{\Cref*{sec:computation_presentation_matrices}}{Section \ref*{sec:computation_presentation_matrices}}}

\begin{proof}[Proof of \Cref{lem:presentation_order}]
    \phantomsection
    \label{proof:presentation_order}
    Denote the last column of $Q$ by
    \[
        \begin{pmatrix}a\\q\end{pmatrix},
        \quad a\in\Z^{s(t)},\quad q\in\Z.
    \]
    Without loss of generality, we can assume that $q\geq 0$.
    On the one hand, \Cref{eq:euclidean_reduction_full_matrices} gives
    \begin{equation}
        \label{eq:proof_presentation_order_Ht}
        H^t a+q\,\partial\tau_{t+1}=\beta.
    \end{equation}
    On the other hand, \Cref{eq:update_full_matrices} gives    
    \[
        \B_d^Y\!(t)\,a+q\,\tau_{t+1} = \gamma.
    \]
    Together with $D_d^t\, U^t = \begin{pmatrix}0 & H^t\end{pmatrix}$ in \Cref{eq:graded_matrix_tensor_degree_d_boundary}, we deduce that
    \[
        \partial\gamma = \beta.
    \]
    Thus $\gamma$ is a cycle when $\beta=0$.

    Suppose now that $\beta=0$.
    To prove that $\gamma$ is independent from the basis $\B_d^Z(t)$ of $Z_d(K^t;\Z)$, it is enough to show that $q\neq 0$.
    Since $\beta=0$, \Cref{eq:proof_presentation_order_Ht} reads
    \begin{equation}
        \label{eq:proof_presentation_order_Ht_2}
        H^t a = - q \partial\tau_{t+1}.
    \end{equation}
    We see that $q$ cannot be zero, otherwise $a$ would have to be zero too by independence of the columns of $H^t$.
    But this is impossible since $Q$ is invertible.
    We deduce that the elements
    \[
        \B_d^Z(t) \cup  \{\gamma\}
    \]
    are independent.

    We turn to the last statement.
    By \Cref{eq:proof_presentation_order_Ht_2}, the class of $\partial\tau_{t+1}$ in $H_{d-1}(K^t;\Z)$~satisfies%
    \[
        q\, [\partial\tau_{t+1}]=0.
    \]
    To show that $q$ is the order of $[\partial\tau_{t+1}]$, we must prove that it is the smallest positive integer satisfying this equality.
    Let $q'>0$ be such that $q'\, [\partial\tau_{t+1}]=0$.
    Thus, $q'\, \partial\tau_{t+1}$ is a boundary.
    Write
    \begin{equation}
        \label{eq:proof_presentation_order_boundary}
        q'\, \partial\tau_{t+1} = - H^ta', \quad a' \in \Z^{s(t)}.
    \end{equation}
    Consider the element defined in coordinates by
    \begin{equation}
        \label{eq:proof_presentation_order_cycle}
        \begin{pmatrix}b\\r\end{pmatrix}
        =
        Q^{-1}\begin{pmatrix}a'\\q'\end{pmatrix},
        \quad 
        b\in\Z^{s(t)}, 
        \quad r\in\Z.
    \end{equation}
    \Cref{eq:proof_presentation_order_boundary} yields
    \[
        \begin{pmatrix}H^t&\partial\tau_{t+1}\end{pmatrix}
        Q\begin{pmatrix}b\\r\end{pmatrix}
        = 
        \begin{pmatrix}H^t&\partial\tau_{t+1}\end{pmatrix}
        Q Q^{-1}\begin{pmatrix}a'\\q'\end{pmatrix}
        = 0.
    \]    
    On the other hand, \Cref{eq:euclidean_reduction_full_matrices} yields
    \[
        \begin{pmatrix}H^t&\partial\tau_{t+1}\end{pmatrix}
        Q\begin{pmatrix}b\\r\end{pmatrix}
        =
        \begin{pmatrix}\widetilde{H}&0\end{pmatrix}
        \begin{pmatrix}b\\r\end{pmatrix}
        =
        \widetilde{H} b.
    \]
    By combining both equalities, we get
    \[
        0 = \widetilde{H} b.
    \]
    Since $H^t$ has linearly independent columns, so does $\widetilde H$.
    Thus, $b=0$, and \Cref{eq:proof_presentation_order_cycle}~yields
    \[
        \begin{pmatrix}a'\\q'\end{pmatrix}
        =
        Q\begin{pmatrix}0\\r\end{pmatrix}
        =
        r \begin{pmatrix}a\\q\end{pmatrix}
    \]
    since the last column of $Q$ is $\begin{pmatrix}a&q\end{pmatrix}^\top$.
    In particular, 
    \[
        q'=rq,
    \]
    hence $q$ divides $q'$.
    We conclude that $q$ is the order of $[\partial\tau_{t+1}]$.
\end{proof}

\begin{proof}[Proof of \Cref{th:presentation_matrix_tensor_module}]
    \phantomsection
    \label{proof:presentation_matrix_tensor_module}

    Let $(K^t)_{t=0}^T$ be a simplicial filtration.
    If not already, we refine it to a simplex-wise filtration by breaking ties.
    The original filtration times are kept as grades.
        
    \proofsubparagraph{Correctness.}

    We introduce simplices successively and verify that the updates in \Cref{subsubsec:presentation_matrix_tensor_full} maintain the presentation.
    After stage $t$, we have a basis of $C_d(K^t;\Z)$ encoded~by
    \[
        U^t=\begin{pmatrix}\B_d^Z(t)&\B_d^Y(t)\end{pmatrix},
        \qquad
        W^t=(U^t)^{-1},
        \qquad
        D_d^t\, U^t=\begin{pmatrix}0&H^t\end{pmatrix},
    \]
    where $\B_d^Z(t)$ is a basis of $Z_d(K^t;\Z)$ and $H^t$ is in column echelon form (its columns are nonzero and have distinct pivot indices).
    Moreover, the maintained matrix $D_{d+1}^{\mathrm{tens}}(t)$ represents~the~map%
    \[
        \partial_{d+1}^t\colon
        C_{d+1}(K^t;\Z)\to Z_d(K^t;\Z)
    \]
    in the simplex basis $\B_{d+1}^C$ of its domain and the basis $\B_d^Z(t)$ of its codomain.

    When we insert a simplex $\sigma$, all cases~\ref{it:algorithm_tensor_case_d+1}, \ref{it:algorithm_tensor_case_d_a}, \ref{it:algorithm_tensor_case_d_b}, and \ref{it:algorithm_tensor_case_d-1} correctly update the variables
    \[
         D_{d+1}^\mathrm{tens}(t),
        \quad H^t,
        \quad W^t.
    \]
    The final matrix is precisely $D_{d+1}^{\mathrm{tens}}$ of \Cref{eq:graded_matrix_tensor}, hence the algorithm is correct.

    \proofsubparagraph{Complexity.}
    
    Every maintained matrix $D_{d+1}^\mathrm{tens}(t),H^t,W^t$ has shape at most $N \times N$.
    The different updates consist of the following operations.
    
    \begin{itemize}
        \item 
        Case~\ref{it:algorithm_tensor_case_d+1}:
        one matrix-vector multiplication $W^t\partial\sigma$.
        It costs 
        \[
            O(N^2)
        \]
        arithmetic operations.

        \item 
        Cases~\ref{it:algorithm_tensor_case_d_a} and \ref{it:algorithm_tensor_case_d_b}:
        an update of the column echelon form of $\left(\begin{matrix} \vphantom{\dfrac{1}{1}} H^t & \partial\sigma\end{matrix}\right)$.
        As justified below \Cref{eq:euclidean_reduction_iterated}, this operation takes $O(N)$ column operations, each costing $O(N)$ arithmetic operations.
        The corresponding inverse row operations are also performed on $W^t$.
        The number of arithmetic operations during those steps is thus
        \[
            O(N^2).
        \]

        \item 
        Case~\ref{it:algorithm_tensor_case_d-1}: 
        only a zero row is appended to $D_d^t$ and $H^t$.
        This requires no arithmetic operations.
    \end{itemize}

    \noindent
    Thus, each simplex insertion costs $O(N^2)$.
    Since the filtration contains $N$ simplices, the total number of arithmetic operations performed by \Cref{alg:presentation_matrix_tensor_module} is
    \[
        O(N^3).
        \qedhere
    \]
\end{proof}

\begin{proof}[Proof of \Cref{lem:free_saturation_primitive_relation}]
    \phantomsection
    \label{proof:free_saturation_primitive_relation}
    
    We have $Qv=ge_1$ by \Cref{eq:free_saturation_euclidean_reduction}.
    Therefore,
    \[
        Q^{-1}e_1=v/g.
    \]
    By \Cref{eq:free_saturation_basis_update}, the $(s(j)+1)$st column of $U_{j+1}$ is
    \[
        u
        = U_{j+1}\big[:\,,s(j)+1\big]
        = U_j\big[:\,,(s(j)+1):n\big]Q^{-1}e_1
        = \frac{1}{g} U_j\big[:\,,(s(j)+1):n\big]v.
    \]
    On the other hand, $w=U_j^{-1}a_{j+1}$ and $v=w[s(j)+1:n]$ in  \Cref{eq:saturation_routine_coordinates} give
    \[
        a_{j+1}
        = U_j w
        = U_j\big[:\,,1:s(j)\big] w\big[1:s(j)\big] + U_j\big[:\,,(s(j)+1):n\big]v.
    \]
    Combining these equalities yields
    \[
        u=\frac{1}{g}
        \bigl(a_{j+1}-U_j[:,1:s(j)]w[1:s(j)]\bigr).
        \qedhere
    \]
\end{proof}

\begin{proof}[Proof of \Cref{th:presentation_matrix_free_module}]
    \phantomsection
    \label{proof:presentation_matrix_free_module}

    Let $D_{d+1}^{\mathrm{tens}}\in\Z^{n\times m}$ be the input presentation matrix.
    We still~identify
    \[
        Z_d(K^N;\Z) \cong \Z^n
    \]
    through the cycle basis $\B_d^Z=(z_1,\dots,z_n)$.
    
    \proofsubparagraph{Correctness.}

    We must show that, at each stage of the algorithm, the matrix $U_j$ is invertible over $\Z$ and its first $s(j)$ columns form an incremental basis of $\Lambda_j^{\mathrm{sat}}$.
    By induction, we suppose that these conditions hold after processing the columns $a_1,\dots,a_j$ of $D_{d+1}^{\mathrm{tens}}$.
    
    We process column $a_{j+1}$.
    Let $w,v$ be the vectors in \Cref{eq:saturation_routine_coordinates}:
    \[
         w=W_j\,a_{j+1},
        \qquad v=w\big[(s(j)+1) : n\big].    
    \]
    Recall that $v$ represents the coordinates of the class of $a_{j+1}$ in the quotient
    \[
        \Z^n / \Lambda^{\mathrm{sat}}_j \cong \Z^{n-s(j)}
    \]
    with respect to the basis formed by the classes of the columns
    \[
        U_j\big[:\,,s(j)+1\big],\quad U_j\big[:\,,s(j)+2\big],\quad \dots, \quad U_j\big[:\,,n\big].
    \]
    If $v=0$, then $a_{j+1}\in\Lambda_j^{\mathrm{sat}}$ and the saturated lattice remains unchanged.
    
    Otherwise, by \Cref{lem:free_saturation_primitive_relation}, the new column of $U_{j+1}$~is%
    \[
        u=\frac{1}{g}
        \bigl(a_{j+1}-U_j[:,1:s(j)]w[1:s(j)]\bigr).
    \]
    Thus, the class of $u$ in $\Z^n/\Lambda_j^{\mathrm{sat}}$ has coordinates $v/g$.
    Moreover, the vector $v/g$ is primitive and spans the same rational line in $\Z^n / \Lambda^{\mathrm{sat}}_j$:
    \[
        \Q\, v \cap \Z^{n-s(j)} = \left(\Q\, v/g\right)\cap \Z^{n-s(j)} = \Z\, v/g.
    \]
    Consequently,
    \[
        \Lambda_{j+1}^{\mathrm{sat}} =\Lambda_j^{\mathrm{sat}}\oplus\Z u.
    \]
    This is encoded by the updates in \Cref{eq:free_saturation_basis_update}.
    This proves the induction.

    \proofsubparagraph{Complexity.}

    For each column $a_{j+1}$, computing $w=W_j a_{j+1}$ of \Cref{eq:saturation_routine_coordinates} requires
    \[
        O(n^2)
    \]
    arithmetic operations.
    Then, if $v\neq0$, its Euclidean reduction requires $O(n)$ elementary transformations.
    As shown in \Cref{eq:euclidean_reduction}, each transformation updates two columns of $U_j$ and two rows of $W_j$, costing $O(n)$ arithmetic operations each.
    In total, reducing $v$ costs
    \[
        O(n^2).
    \]
    Thus each column costs $O(n^2)$ arithmetic operations.
    Since $m$ columns are processed and $n,m\leq N$, the total cost of \Cref{alg:presentation_matrix_free_module} is
    \[
        O\bigl(mn^2\bigr) = O(N^3).
        \qedhere
    \]
\end{proof}

\subsection{Proofs for \texorpdfstring{\Cref*{sec:filtrations_with_large_torsion}}{Section \ref*{sec:filtrations_with_large_torsion}}}

\begin{proof}[Proof of \Cref{lem:exceptional_bars_base}]
    \phantomsection
    \label{proof:exceptional_bars_base}

    Each complex $X_i$ has $O(\log M_i)$ simplices by \Cref{lem:Newman_complex}.
    Using that $M_i=\prod_{\ell=i}^L p_\ell$, we get that the number of simplices in $X$ is
    \begin{align*}    
        O\left(\sum_{i=1}^L\log M_i\right)
        = O\left(\sum_{i=1}^L\log \prod_{j=i}^L p_j\right)
        = O\left(\sum_{j=1}^L j\log p_j\right)
        &= O\left(\left(\sum_{j=1}^L j\right)\log p_L\right) \\
        &= O\left(L^2\log L\right).
        \qedhere
    \end{align*}    
\end{proof}

\begin{proof}[Proof of \Cref{lem:exceptional_bars_pascal}]
    \phantomsection
    \label{proof:exceptional_bars_pascal}

    The matrix of coefficients of $\beta_1,\dots,\beta_L$ with respect to $\alpha_1,\dots,\alpha_L$ is 
    \[
        A =
        \setlength{\arraycolsep}{6pt}
        \begin{pNiceArray}{cccccc}[first-row,first-col]
                                  & \beta_1 & \beta_2 & \beta_3 & \beta_4 & \cdots & \beta_L \\
            \firstcol{\alpha_1}   & 1        & 1        & 1        & 1        & \cdots & 1 \\
            \firstcol{\alpha_2}   & 0        & 1        & 2        & 3        & \cdots & L-1 \\
            \firstcol{\alpha_3}   & 0        & 0        & \displaystyle\binom{2}{2} & \displaystyle\binom{3}{2} & \cdots & \displaystyle\binom{L-1}{2} \\
            \firstcol{\alpha_4}   & 0        & 0        & 0 & \displaystyle\binom{3}{3} & \cdots & \displaystyle\binom{L-1}{3} \\
            \firstcol{\vdots}     & \vdots   & \vdots   & \vdots   & \vdots   & \ddots & \vdots \\
            \firstcol{\alpha_L}   & 0        & 0        & 0        & 0        & \cdots & 1
        \end{pNiceArray}.
    \]
    Now, let $i\in\{1,\dots,L\}$, and choose $i$ indices $1\leq \ell_1<\cdots<\ell_i\leq L$.
    The restricted matrix is
    \[
        A\left[1\colon i,(\ell_1,\dots,\ell_i)\right]
        =
        \setlength{\arraycolsep}{6pt}
        \begin{pNiceArray}{cccc}[first-row,first-col]
                                  & \beta_{\ell_1}       & \beta_{\ell_2}       & \cdots & \beta_{\ell_i} \\
            \firstcol{\alpha_1}   & 1                  & 1                  & \cdots & 1 \\
            \firstcol{\alpha_2}   & \ell_1-1              & \ell_2-1              & \cdots & \ell_i-1 \\
            \firstcol{\vdots}     & \vdots             & \vdots             & \ddots & \vdots \\
            \firstcol{\alpha_i}   & \displaystyle\binom{\ell_1-1}{i-1} & \displaystyle\binom{\ell_2-1}{i-1} & \cdots & \displaystyle\binom{\ell_i-1}{i-1}
        \end{pNiceArray}
    \]
    where, by convention, $\binom{a}{b}=0$ when $a<b$.
    Observe that, for each $k\in\{1,\dots,i\}$, the $k$th row consists of the values of the polynomial
    \[
        x\mapsto\binom{x}{k-1} = \frac{x(x-1)\cdots(x-k+2)}{(k-1)!}
    \]
     at $x=\ell_1-1,\dots,\ell_i-1$.
    This polynomial has degree $k-1$ and leading coefficient $1/(k-1)!$.
    Thus, the formula for the generalized Vandermonde determinant gives
    \[
        \det A\left[1\colon i,(\ell_1,\dots,\ell_i)\right]
        =
        \left(\prod_{k=1}^{i}\frac{1}{(k-1)!}\right)
        \left(\prod_{1\leq\alpha<\beta\leq i} (\ell_\beta-\ell_\alpha)\right).
    \]
    In this product, every difference $\ell_\beta-\ell_\alpha$ is smaller than $L$ and every factorial $(k-1)!$ is a product of integers smaller than $L$.
    Using $p_i\geq L$, we deduce that it is nonzero over~$\F_{p_i}$:
    \[
        \det A\left[1\colon i,(\ell_1,\dots,\ell_i)\right] \not\equiv 0 \pmod{p_i}.
    \]
    In other words, $A\left[1\colon i,(\ell_1,\dots,\ell_i)\right]$ is invertible over $\F_{p_i}$.
    Using that $[\alpha_1],\dots,[\alpha_i]$ are linearly independent over $\F_{p_i}$ by \Cref{eq:exceptional_bars_base_complex_dimension} we get that $[\beta_{\ell_1}],\dots,[\beta_{\ell_i}]$ are also linearly~independent.%

    If we choose $i+1$ or more classes, then they must be linearly dependent, since $[\alpha_1],\dots,[\alpha_L]$ span an $i$-dimensional subspace of $H_1(X;\F_{p_i})$ by \Cref{eq:exceptional_bars_base_complex_dimension}.
\end{proof}

\begin{proof}[Proof of \Cref{lem:exceptional_bars_pascal_construction}]
    \phantomsection
    \label{proof:exceptional_bars_pascal_construction}

    We first recall that, given two triangular loops $x,y$ in a simplicial complex $X$, their sum in $H_1(X;\Z)$ can be represented by a new triangular loop $z$ after adding only a bounded number of simplices.
    For instance, attach a triangulated pair of pants by identifying two of its boundary components with $x$ and $y$.
    With suitable orientations, the third boundary component $z$ satisfies $[z]=[x]+[y]$.
    The resulting complex $\widetilde{X}$ has homology
    \[
        H_1\bigl(\widetilde{X};\Z\bigr) \cong H_1\bigl(X;\Z\bigr) \oplus \Z^m,
    \]
    where $m=1$ if $x,y$ belong to the same connected component and $m=0$ otherwise. 
    Moreover, the inclusion $X\hookrightarrow\widetilde X$ induces an injection $H_1(X;\Z)\hookrightarrow H_1(\widetilde{X};\Z)$ with free cokernel.
    Consequently, it is a split injection, and it induces an injection over any field.

    Now, for the construction, we represent the classes $\beta_1,\dots,\beta_L$ as triangular loops by taking successive sums of $\alpha_1,\dots,\alpha_L$.
    As noted after \Cref{eq:exceptional_bars_pascal}, the coefficients of these new classes correspond to the rows of Pascal's triangle:
    \begin{align*}
        [\beta_1] = [\alpha_1],
        \qquad
        [\beta_2] = [\alpha_1]+[\alpha_2],
        \qquad
        [\beta_3] = [\alpha_1]+2[\alpha_2]+[\alpha_3],
        \qquad \dots
    \end{align*}
    We build intermediate triangular loops
    \begin{align*}
        \omega_1^{(1)}, \qquad \omega_2^{(1)}, \qquad\qquad&\cdots, \qquad\qquad \omega_{L-1}^{(1)}, \qquad \omega_L^{(1)},\\
        \omega_1^{(2)}, \qquad \omega_2^{(2)},\qquad& \cdots, \qquad\qquad\qquad \omega_{L-1}^{(2)},\\
        &~~\vdots\\
        \omega_1^{(L-2)},  \qquad&\omega_2^{(L-2)}, \qquad \omega_3^{(L-2)},\\
        \omega_1^{(L-1)}&, \qquad \omega_2^{(L-1)},\\
        &\omega_1^{(L)}
    \end{align*}
    as follows.
    The initial loops are those already present in $X$: $\omega_i^{(1)}=\alpha_i$ for $i\in\{1,\dots,L\}$.
    The second generation is obtained by summing pairs of consecutive loops: for $i\in\{1,\dots,L-1\}$,
    \[
        \big[\omega_i^{(2)}\big] = \big[\omega_i^{(1)}\big] + \big[\omega_{i+1}^{(1)}\big].
    \]
    We create these loops by inserting $L-1$ triangulated pairs of pants.
    We repeat this pattern at all generations: for $k\in\{2,\dots,L\}$ and $i\in\{1,\dots,L+1-k\}$,
    \[
        \big[\omega_i^{(k)}\big] = 
        \big[\omega_i^{(k-1)}\big] + \big[\omega_{i+1}^{(k-1)}\big].
    \]
    See \Cref{fig:complex_many_bars} for an illustration.
    During the construction, the classes satisfy Pascal's relation:
    \begin{align*}
        \big[\omega_i^{(k)}\big]
        &= \big[\omega_i^{(k-1)}\big] + \big[\omega_{i+1}^{(k-1)}\big]\\
        &= \big(\big[\omega_i^{(k-2)}\big] + \big[\omega_{i+1}^{(k-2)}\big]\big) + \big(\big[\omega_{i+1}^{(k-2)}\big] + \big[\omega_{i+2}^{(k-2)}\big]\big)\\
        &= \cdots\\
        &= \sum_{\ell=1}^{k}\binom{k-1}{\ell-1} \big[\omega_{i+\ell-1}^{(1)} \big].
    \end{align*}
    In particular, the leftmost classes satisfy
    \[
        \big[\omega_1^{(k)}\big] 
        = \sum_{\ell=1}^{k}\binom{k-1}{\ell-1} \big[\omega_{\ell}^{(1)} \big]
        = \sum_{\ell=1}^{k}\binom{k-1}{\ell-1} [\alpha_\ell],
    \]
    which is the definition of $[\beta_k]$ in \Cref{eq:exceptional_bars_pascal}.
    Therefore we can identify $\beta_i = \omega_1^{(i)}$.
    During this construction, we introduced
    \[
         \sum_{k=2}^{L} (L+1-k)=\frac{L(L-1)}{2} = O(L^2)
    \]
    relations.
    Therefore the resulting complex $\widetilde{X}$ contains only $O(L^2)$ additional simplices.
\end{proof}

\begin{proof}[Proof of \Cref{lem:exceptional_bars_filtration}]
    \phantomsection
    \label{proof:exceptional_bars_filtration}

    First, we fix $i\in\{1,\dots,L\}$ and work over the field $\F_{p_i}$.
    As observed above the lemma, introducing the relation $C_{i+1}$ ends the bar that started at $\nu_1$, thus producing
    \[
        \left[b_1, d_{i+1}\right) \in \barcode\bigl(H_1(K^\bullet;\F_{p_i})\bigr).
    \]
    By induction, each new relation $C_{i+t}$ kills the youngest remaining class, producing
    \[
        \left[b_t, d_{i+t}\right) \in \barcode\bigl(H_1(K^\bullet;\F_{p_i})\bigr).
    \]
    This proves the first claim.

    Over $\Q$, all classes $\beta_1,\dots,\beta_L$ vanish, hence $\barcode(H_1(K^\bullet;\Q))$ contains the bars
    \[
        \left[b_1, d_1\right), 
        \quad \left[b_2, d_2\right)
        \quad \dots,
        \quad \left[b_R, d_R\right).
    \]
    This proves the last claim.
    
    Last, observe that each $\F_{p_i}$ contributes at least $R-i$ exceptional bars.
    They sum to
    \[
        \sum_{i=1}^L (R-i)
        =
        LR - L(L+1)/2
        \geq LR - L^2
    \]
    exceptional bars.
    \Cref{eq:exceptional_bars_count} follows.    
\end{proof}

\begin{proof}[Proof of \Cref{prop:number_exceptional_bars}]
    \phantomsection
    \label{proof:number_exceptional_bars}

    As defined in \Cref{eq:exceptional_bars_final_complex}, $K$ is the disjoint union $Y \sqcup \widetilde{X}$ with cylinders $C_1,\dots,C_R$ added.
    The complex $\widetilde{X}$ has $O(L^2\log L + L^2)$ simplices by \Cref{lem:exceptional_bars_base,lem:exceptional_bars_pascal_construction}.
    Moreover, $Y$ has $6R$ simplices since it is the disjoint union of $R$ triangular cycles.
    Last, each cylinder $C_i$, which connects two triangular cycles, requires the introduction of six edges and six triangles (to form three square faces).
    Therefore $K$ has 
    \[
        O\bigl(L^2\log L + L^2 + 6R + 12R\bigr) 
        = O\bigl(L^2\log L + R\bigr)
    \]
    simplices.
    Substituting $L = \lfloor \sqrt{N/\log N}\rfloor$ and $R = \lfloor L^2\log L\rfloor = \Theta(N)$ yields $O(N)$.

    Next, each of the primes $p_1,\dots,p_L$ is exceptional by \Cref{lem:exceptional_bars_filtration}.
    Thus $K^\bullet$ has at least
    \[
        L = \Omega\left(\sqrt{\frac{N}{\log N}}\right)
    \]
    exceptional primes.
    Again by \Cref{lem:exceptional_bars_filtration}, the total number of distinct bars is at least
    \[
        L (R-L) 
        = \Theta\left(\sqrt{\frac{N}{\log N}}\right) \times \Theta\bigl(N\bigr)
        = \Theta\left(\frac{N^{3/2}}{\sqrt{\log N}}\right).
        \qedhere
    \]
\end{proof}

\begin{proof}[Proof of \Cref{prop:number_exceptional_columns_omnifield}]
    \phantomsection
    \label{proof:number_exceptional_columns_omnifield}
    
    After stage \textbf{(5)} of the filtration $K_N^\bullet$, the boundary matrix reduced over $\Q$ has a rational column of the form
    \[
        R^\Q[:,j]=c\gamma_0
    \]
    for some $c\in\Q\setminus\{0\}$.
    Indeed, since $\gamma_0 = \pm E_1\pm E_2 \pm E_3$ is a rational boundary, some reduced column must have pivot $E_2$ ($E_2$ is inserted the latest).
    Moreover, any reduced Newman column whose pivot is a Morozov edge is supported on $\{E_1,E_2,E_3\}$, as all other Newman edges were inserted later.
    Since this column represents a cycle, it must be a multiple of $\gamma_0$.
    
    When the base triangle $T_1$ is inserted, its column reduces to zero over $\Q$ since $\gamma_0=\partial T_1$, whereas its column over $\F_p$ remains $\gamma_0$.
    In particular, it is $p$-exceptional for all these $p$.
    
    Last, we study the insertion of the fin triangles $T'_i$.
    Initially, each $T'_i$ has pivot index $E_2$.
    Over $\Q$, the rational column owning this pivot is $R^\Q[:,j]$.
    However, over every exceptional field $\F_p$, it is the column of $T_1$.
    The omni-field algorithm therefore starts a separate reduction over each of these exceptional fields.
    Consequently, by \Cref{eq:fin_triangle_cost}, each fin triangle~requires%
    \[
        \underbrace{~~\Omega(N^2)~~}_{\substack{
                \text{cost of reducing}\\
                \text{that column}
            }}
        \times
        \underbrace{~~\Omega(L)~~}_{\substack{
                \text{number of}\\
                \text{primes}
            }}
        =
        \Omega\left(\frac{N^3}{\log N}\right)
    \]
    arithmetic operations.
    Since there are $N$ fin triangles, we deduce the result.
\end{proof}

\section{Algorithms}
\label{sec:algorithms}

\begin{algorithm}[H]
    \caption{Persistence algorithm with row priority $\tau$
        (see \Cref{lem:persistence_algorithm_priority})}
    \label{alg:persistence_algorithm_priority}
    \KwData{
        An augmented matrix $\widetilde{D}=\left(\begin{smallmatrix}I_m\\D\end{smallmatrix} \right)$ with $D\in\kk^{n\times m}$ over a field $\kk$;  
        \newline
        a row priority $\tau$ encoded as a permutation of $\{\bot_1,\dots,\bot_m,1,\dots,n\}$
    }

    \BlankLine
    \KwResult{
        The $\tau$-reduction $\widetilde{R}=\Red(\widetilde{D},\tau)$
    }

    \BlankLine
    $\widetilde{R}\gets\widetilde{D}$\;
    
    \BlankLine
    \ForEach{$i\in\{\bot_1,\dots,\bot_m,1,\dots,n\}$}{
        $P[i]\gets 0$
        \tcp*[r]{Initially, no row is owned}
    }

    \BlankLine
    \For(\tcp*[f]{Scan columns from left to right}){$j\gets 1$ \KwTo $m$}{
        $i\gets\low(\widetilde{R},j,\tau)$
        \tcp*[r]{Highest-priority nonzero entry}
        \While(\tcp*[f]{Resolve pivot conflicts}){$P[i]\neq 0$}{
            $k\gets P[i]$
            \tcp*[r]{Owner of the $i$th row}
            
            $\widetilde{R}[:,j]\gets
                \widetilde{R}[:,j]
                -\dfrac{\widetilde{R}[i,j]}
                       {\widetilde{R}[i,k]}
                 \widetilde{R}[:,k]$
                 \tcp*[r]{Cancel the current pivot}

            $i\gets\low(\widetilde{R},j,\tau)$
            \tcp*[r]{Find the next pivot}
        }
        $P[i]\gets j$
        \tcp*[r]{Record the new owned row}
    }

    \BlankLine
    \KwRet{$\widetilde{R}$}\;
\end{algorithm}

\newpage

\begin{algorithm}[H]
    \caption{Update of a $\tau$-reduction under an adjacent transposition
        (see \Cref{subsubsec:efficient_updates_priority})}
    \label{alg:transposition_priority}
    \KwData{
        A row priority $\tau$ and a $\tau$-reduction $\widetilde{R}\in\kk^{(n+m)\times m}$ of $\widetilde{D}$ over $\kk$;
        \newline
        its pivot-owner array $P$: $P[\low(\widetilde{R},j,\tau)]=j$
        and $P[r]=0$ for unowned rows;
        \newline
        an index $i\in\{1,\dots,n+m-1\}$
    }

    \BlankLine
    \KwResult{
        The updated priority $\tau$ obtained by exchanging its $i$th and $(i+1)$st entries;
        \newline
        a $\tau$-reduction $\widetilde{R}$ of $\widetilde{D}$;
        \newline
        its pivot-owner array $P$
    }

    \BlankLine
    $u\gets\tau_i,\quad v\gets\tau_{i+1}$\;
    $a\gets P[u],\quad b\gets P[v]$\;
    $(\tau_i,\tau_{i+1})\gets(v,u)$
        \tcp*[r]{Row $v$ now has higher priority}

    \BlankLine
    \If(\tcp*[f]{Case~\ref{it:update_transposition_2b}}){$a=0$}{
        \KwRet{$(\widetilde{R},\tau,P)$}\;
    }

    \BlankLine
    $x\gets\widetilde{R}[v,a]$\;
    
    \BlankLine
    \If(\tcp*[f]{
            Case~\ref{it:update_transposition_1a} (if $b\neq0$) or
            Case~\ref{it:update_transposition_2b} (if $b=0$)}
        ){$x=0$}{
        
        \KwRet{$(\widetilde{R},\tau,P)$}\;
    }

    \BlankLine
    \eIf(\tcp*[f]{Case~\ref{it:update_transposition_2a}}){$b=0$}{
        $(P[u],P[v])\gets(0,a)$\;
    }{
        $\beta\gets\widetilde{R}[v,b]$\;

        \eIf(\tcp*[f]{Case~\ref{it:update_transposition_1b}})
            {$a<b$}{
            $\widetilde{R}[:,b]\gets
                \widetilde{R}[:,b]
                -\dfrac{\beta}{x}\widetilde{R}[:,a]$\;
            $(P[u],P[v])\gets(b,a)$\;
        }(\tcp*[f]{Case~\ref{it:update_transposition_1c}}){
            $\widetilde{R}[:,a]\gets
                \widetilde{R}[:,a]
                -\dfrac{x}{\beta}\widetilde{R}[:,b]$\;
        }
    }

    \BlankLine
    \KwRet{$(\widetilde{R},\tau,P)$}\;
\end{algorithm}

\begin{algorithm}[H]
    \caption{Update of a $\tau$-reduction to priority $\tau'$
        (see \Cref{lem:full_correction_priority})}
    \label{alg:full_correction_priority}
    \KwData{
        A row priority $\tau$ and a $\tau$-reduction $\widetilde{R}\in\kk^{(n+m)\times m}$ of $\widetilde{D}$ over $\kk$;
        \newline
        its pivot-owner array $P$: $P[\low(\widetilde{R},j,\tau)]=j$ and $P[r]=0$ for unowned rows;
        \newline
        a target row priority $\tau'$ on the same rows
    }

    \BlankLine
    \KwResult{
        The updated priority $\tau=\tau'$;
        \newline
        a $\tau'$-reduction $\widetilde{R}$ of $\widetilde{D}$;
        \newline
        its pivot-owner array $P$
    }

    \BlankLine
    \For(\tcp*[f]{Scan rows from highest to lowest $\tau'$-priority})
        {$i\gets 1$ \KwTo $n+m$}{
        $j\gets i$\;

        \While(\tcp*[f]{Locate $\tau'_i$ in the current order $\tau$})
            {$\tau_j\neq\tau_i'$}{
            $j\gets j+1$\;
        }

        \BlankLine
        \While(\tcp*[f]{Move $\tau_i'$ to position $i$ by adjacent transpositions})
            {$j>i$}{
            Apply \Cref{alg:transposition_priority} to
                $(\widetilde{R},\tau,P)$ with index $j-1$\;
            $j\gets j-1$\;
        }
    }

    \BlankLine
    \KwRet{$(\widetilde{R},\tau,P)$}\;
\end{algorithm}

\begin{algorithm}[H]
    \caption{Correction of an exhaustive rational reduction at a prime $p$
        (see \Cref{lem:full_correction_modp})}
    \label{alg:full_correction_modp}
    \KwData{
        An exhaustive rational reduction
        $\widetilde{R}^{\Q}\in\Q^{(n+m)\times m}$
        of an integral matrix $\widetilde{D}$;
        \newline
        its pivot priority $\tau=\tau_{\mathrm{piv}}^{\Q}$;
        \newline
        its pivot-owner array $P$:
        $P[\low(\widetilde{R}^{\Q},k,\tau)]=k$
        and $P[r]=0$ for unowned rows;
        \newline
        a prime $p$
    }

    \BlankLine
    \KwResult{
        A priority $\tau=\tau_{\mathrm{piv}}^{\F_p}$
        satisfying $(\tau_1,\dots,\tau_m)=\low(\F_p,\bullet)$;
        \newline
        a rational $\tau$-reduction $\widetilde{R}^{\Q}$
        of $\widetilde{D}$;
        \newline
        its pivot-owner array $P$
    }

    \BlankLine
    \For(\tcp*[f]{Scan columns from left to right})
        {$j\gets 1$ \KwTo $m$}{
        $c\gets\widetilde{R}^{\Q}[:,j]\bmod p$
            \tcp*[r]{Well-defined by \Cref{lem:pivot_sequence_prefix}}
        $r\gets\max_{\tau_{\mathrm{can}}}
            \{s\in\{\bot_1,\dots,\bot_m,1,\dots,n\} \mid c[s]\neq0\}$
            \tcp*[r]{$\low(\F_p,j)$}

        \BlankLine
        $k\gets j$\;
        \While(\tcp*[f]{Locate the newly discovered pivot row})
            {$\tau_k\neq r$}{
            $k\gets k+1$\;
        }

        \BlankLine
        \While(\tcp*[f]{Move row $r$ to position $j$})
            {$k>j$}{
            Apply \Cref{alg:transposition_priority} over $\Q$ to $(\widetilde{R}^{\Q},\tau,P)$ with index $k-1$\;
            $k\gets k-1$\;
        }
    }

    \BlankLine
    \KwRet{$(\widetilde{R}^{\Q},\tau,P)$}\;
\end{algorithm}

\begin{algorithm}[H]
    \caption{Main algorithm: reduction over all fields (see \Cref{th:main_algorithm} and \Cref{rem:implementation_details})}
    \label{alg:main_algorithm}
    \KwData{
        An integer matrix $D\in\Z^{n\times m}$
    }

    \BlankLine
    \KwResult{
        The exceptional primes $\PP$;
        \newline
        the rational pivot index sequence $\low(\Q,\bullet)$;
        \newline
        the exceptional pivot index sequences $\low(\F_p,\bullet)$ for all $p\in\PP$
    }

    \BlankLine
    \tcp{Step~\ref{it:main_algorithm_step1}:~Classical reduction with rational coefficients}
    $\widetilde{D}\gets
        \left(\begin{smallmatrix}I_m\\D\end{smallmatrix}\right)$\;
    $\widetilde{R}_{\mathrm{can}}^{\Q}
        \gets\Red(\widetilde{D}^{\Q},\tau_{\mathrm{can}})$
        \tcp*[r]{Via \Cref{alg:persistence_algorithm_priority} with canonical order}
    $\low(\Q,\bullet)\gets
        \low(\widetilde{R}_{\mathrm{can}}^{\Q},\bullet)$
        \tcp*[r]{Generic pivot indices}

    \BlankLine
    \BlankLine
    \tcp{Step~\ref{it:main_algorithm_step2}:~Identification of the exceptional primes}
    $\Delta_0\gets1$\;
    \For{$j\gets1$ \KwTo $m$}{
        $\Delta_j\gets
            \Delta_{j-1}
            \widetilde{R}_{\mathrm{can}}^{\Q}[\low(\Q,\bullet)(j),j]$
            \tcp*[r]{$j$th principal minor}
    }
    $\PP\gets
        \{\text{prime } p \text{ that divides } \Delta_j \text{ for some } j\in\{1,\dots,m\}\}$
        \tcp*[r]{Exceptional primes}

    \BlankLine
    \BlankLine
    \If(\tcp*[f]{If no exceptional primes}){$\PP=\varnothing$}{
        \KwRet{$(\PP,\low(\Q,\bullet),\varnothing)$}\;
    }

    \BlankLine
    \BlankLine
    \tcp{Step~\ref{it:main_algorithm_step3}:~Exhaustive reduction}
    Let $(q_1,\dots,q_n)$ be the rows absent from $\low(\Q,\bullet)$ in canonical order\;
    $\tau_0\gets
        (\low(\Q,1),\dots,\low(\Q,m),q_1,\dots,q_n)$
        \tcp*[r]{$\Q$-pivot priority (Def~\ref{def:permutation_piv_p})}
    $\widetilde{R}_0^{\Q}
        \gets\Red(\widetilde{D}^{\Q},\tau_0)$
        \tcp*[r]{Via \Cref{alg:persistence_algorithm_priority} with order $\tau_0$}

    \BlankLine
    \BlankLine
    \tcp{Step~\ref{it:main_algorithm_step4}:~Correction at exceptional primes}
    Initialize $P_0[\low(\Q,j)]=j$ for $j\in\{1,\dots,m\}$ and $P_0[r]=0$ for the remaining rows\;    
    $\mathcal{G}\gets
        \{(\PP,\widetilde{R}_0^{\Q},\tau_0,P_0)\}$
        \tcp*[r]{Initial state, will be a partition of $\PP$}

    \BlankLine
    \For(\tcp*[f]{Scan columns from left to right}){$j\gets1$ \KwTo $m$}{
        $\mathcal{G}_{\mathrm{new}}\gets\varnothing$\;

        \BlankLine
        \ForEach{$(S,\widetilde{R}^{\Q},\tau,P)\in\mathcal{G}$}{

            \BlankLine
            \tcp{Identify the exceptional pivots (see Rem~\ref{rem:implementation_details} for optimization)}
            \ForEach{$p\in S$}{
                $c\gets\widetilde{R}^{\Q}[:,j]\bmod p$
                    \tcp*[r]{Well-defined by \Cref{lem:pivot_sequence_prefix}}
                $r_p\gets\max_{\tau_{\mathrm{can}}}
                    \{s\in\{\bot_1,\dots,\bot_m,1,\dots,n\} \mid c[s]\neq0\}$
                    \tcp*[r]{$= \low(\F_p,j)$}
            }

            \BlankLine
            \BlankLine
            \tcp{Update the matrix}
            \ForEach{$r\in\{r_p\mid p\in S\}$}{
                $S_r\gets\{p\in S\mid r_p=r\}$
                    \tcp*[r]{Group primes with same $j$th pivot index}
                $(\widetilde{R}_r^{\Q},\tau_r,P_r) \gets\operatorname{copy}(\widetilde{R}^{\Q},\tau,P)$\;

                \BlankLine
                $k\gets j$\;
                \While(\tcp*[f]{Locate the newly discovered pivot row})
                    {$(\tau_r)_k\neq r$}{
                    $k\gets k+1$\;
                }
                \While(\tcp*[f]{Move row $r$ to position $j$})
                    {$k>j$}{
                    Apply \Cref{alg:transposition_priority} over $\Q$ to $(\widetilde{R}_r^{\Q},\tau_r,P_r)$ with index $k-1$\;
                    $k\gets k-1$\;
                }

                \BlankLine
                Add $(S_r,\widetilde{R}_r^{\Q},\tau_r,P_r)$ to $\mathcal{G}_{\mathrm{new}}$\;
            }
        }
        $\mathcal{G}\gets\mathcal{G}_{\mathrm{new}}$\;
    }

    \BlankLine
    \BlankLine
    \tcp{Save the final pivot sequences}
    \ForEach{$(S,\widetilde{R}^{\Q},\tau,P)\in\mathcal{G}$}{
        $\low(\F_p,\bullet)\gets(\tau_1,\dots,\tau_m)$ for every $p\in S$
            \tcp*[r]{Same sequence in each group}
    }
    
    \BlankLine
    \BlankLine
    \KwRet{$(\PP,\low(\Q,\bullet),
        (\low(\F_p,\bullet))_{p\in\PP})$}\;
\end{algorithm}

\begin{algorithm}[H]
    \caption{Field-independence problem (see \Cref{th:field_independence_problem})}
    \label{alg:field_independence_problem}
    \KwData{
        A simplicial filtration $(K^t)_{t=0}^{T}$;
        \newline
        an integer $d\geq0$
    }

    \BlankLine
    \KwResult{
        $\texttt{True}$ if $\barcode(H_d(K^\bullet;\kk))$ is independent of the field $\kk$, $\texttt{False}$ otherwise
    }

    \BlankLine
    \tcp{Refine to a simplex-wise filtration}
    Order the simplices $\sigma_1,\dots,\sigma_N$ by nondecreasing filtration time, breaking ties by increasing dimension\;
    Let $t_1,\dots,t_N \in\{0,\dots,T\}$ be their initial filtration times\;
    Form the integral boundary matrix $D$ in this simplex order\;

    \BlankLine
    \BlankLine
    \tcp{Compute persistence over all fields}
    Apply \Cref{alg:main_algorithm} to $D$, obtaining $\PP$, $\low(\Q,\bullet)$, and $\low(\F_p,\bullet)$ for all $p\in\PP$\;

    \BlankLine
    \BlankLine
    \tcp{Deduce the persistence barcodes}
    \ForEach{$\kk\in\{\Q\}\cup\{\F_p\mid p\in\PP\}$}{
        $e[i]\gets T+1$ for every $i\in\{1,\dots,N\}$
            \tcp*[r]{Encode infinite death by $T+1$}

        \For(\tcp*[f]{Recover finite death times})
            {$j\gets1$ \KwTo $N$}{
            $i\gets\low(\kk,j)$\;
            \If{$i\in\{1,\dots,N\}$}{
                $e[i]\gets t_j$
                    \tcp*[r]{Pair birth $i$ with death $j$}
            }
        }

        \BlankLine
        $\mathcal{B}_{\kk}\gets\varnothing$
            \tcp*[r]{Barcode}
        \ForEach{$i\in\{1,\dots,N\}$ with $\dim\sigma_i=d$ and $\low(\kk,i)=\bot_i$}{
            \If(\tcp*[f]{Discard empty intervals})
                {$t_i<e[i]$}{
                $\mathcal{B}_{\kk}\gets \mathcal{B}_{\kk} \cup \{\!\{(t_i,e[i])\}\!\}$
                    \tcp*[r]{By \Cref{eq:barcode_from_pivots}}
            }
        }
    }

    \BlankLine
    \BlankLine
    \tcp{Compare with the generic barcode}
    \ForEach{$p\in\PP$}{
        \If{$\mathcal{B}_{\Q}\neq\mathcal{B}_{\F_p}$}{
            \KwRet{{\normalfont\texttt{False}}}\;
        }
    }

    \BlankLine
    \KwRet{{\normalfont\texttt{True}}}\;
\end{algorithm}

\begin{algorithm}[H]
    \caption{Presentation matrix for the tensor module (see \Cref{th:presentation_matrix_tensor_module})}
    \label{alg:presentation_matrix_tensor_module}

    \KwData{
        A simplicial filtration $(K^t)_{t=0}^{T}$;
        \newline
        an integer $d\geq0$
    }

    \BlankLine
    \KwResult{
        An integral graded presentation $(D^{\mathrm{tens}},b,c)$ of $H_d(K^\bullet;\Z)\otimes\kk$, valid for every field $\kk$, where $b_i$ is the appearance time of row generator $i$, and $c_j$ is the appearance time of relation $j$.
    }

    \BlankLine
    \tcp{Refine to a simplex-wise filtration}
    Order the simplices $\sigma_1,\dots,\sigma_N$ by nondecreasing filtration time, breaking ties by increasing dimension\;
    Let $t_1,\dots,t_N \in\{0,\dots,T\}$ be their initial filtration times\;

    \BlankLine
    \BlankLine
    \tcp{Main loop}
    $z\gets0,\quad s\gets0$
        \tcp*[r]{Cycle and complement ranks (Equations~(\ref{eq:update_full_matrices}) and~(\ref{eq:euclidean_reduction_iterated_Q}))}
    $W\gets I_0$
        \tcp*[r]{Inverse chain basis $W=(U^t)^{-1}$ (\Cref{eq:full_chain_basis_matrix})}
    $H \gets 0$
        \tcp*[r]{Degree-$(d-1)$ boundaries (\Cref{eq:graded_matrix_tensor_degree_d_boundary})}
    $D^{\mathrm{tens}} \gets 0$
        \tcp*[r]{Degree-$(d+1)$ presentation matrix (\Cref{eq:graded_matrix_tensor})}
    Initialize $b$ and $c$ as empty lists
        \tcp*[r]{Appearance times}

    \BlankLine
    \For{$j\gets1$ \KwTo $N$}{
        \If(\tcp*[f]{Case~\ref{it:algorithm_tensor_case_d+1}})
            {$\dim\sigma_j=d+1$}{
            $w\gets W\,\partial\sigma_j$\;
            $D^{\mathrm{tens}}\gets(D^{\mathrm{tens}}\ w[1:z])$\;
            Append $t_j$ to $c$\;
        }
        
        \BlankLine
        \If(\tcp*[f]{Update the integral cycle basis})
            {$\dim\sigma_j=d$}{
            Compute $Q\in\GL_{s+1}(\Z)$ such that $(H\ \partial\sigma_j)Q=(\widetilde H\ \beta)$ by Euclidean reduction\;

            $H\gets\widetilde H$\;

            $W\gets\operatorname{diag}(I_z,Q^{-1})\operatorname{diag}(W,1)$
                \tcp*[r]{(\Cref{eq:update_chain_basis_inverse})}

            \eIf(\tcp*[f]{Case~\ref{it:algorithm_tensor_case_d_a}})
                {$\beta=0$}{
                Move the last row of $W$ to position $z+1$\;
                $z\gets z+1$\;
                $D^{\mathrm{tens}}\gets\left(\begin{smallmatrix}D^{\mathrm{tens}}\\0\end{smallmatrix}\right)$\;
                Append $t_j$ to $b$\;
            }(\tcp*[f]{Case~\ref{it:algorithm_tensor_case_d_b}}){
                $H\gets(H\ \beta)$\;
                $s\gets s+1$\;
            }
        }

        \BlankLine
        \If(\tcp*[f]{Case~\ref{it:algorithm_tensor_case_d-1}})
            {$\dim\sigma_j=d-1$}{
            $H\gets\left(\begin{smallmatrix}H\\0\end{smallmatrix}\right)$\;
        }
    }
    \BlankLine
    \KwRet{$(D^{\mathrm{tens}},b,c)$}\;
\end{algorithm}

\begin{algorithm}[H]
    \caption{Presentation matrix for the free module (see \Cref{th:presentation_matrix_free_module})}
    \label{alg:presentation_matrix_free_module}

    \KwData{
        The integral graded presentation $(D^{\mathrm{tens}},b,c)$ returned by \Cref{alg:presentation_matrix_tensor_module}, with $D^{\mathrm{tens}}\in\Z^{n\times m}$ and $c_1\leq\cdots\leq c_m$
    }

    \BlankLine
    \KwResult{
        An integral graded presentation $(D^{\mathrm{free}},b,c^\mathrm{free})$ of $H_d^{\mathrm{free}}(K^\bullet;\Z)\otimes\kk$, valid for every field $\kk$, where $b_i$ is the appearance time of row generator $i$, and $c_j^\mathrm{free}$ is the appearance time of relation $j$.
    }

    \BlankLine
    $U\gets I_n,\quad W\gets I_n$
        \tcp*[r]{Change-of-basis matrices, $W=U^{-1}$}
    $s\gets0$
        \tcp*[r]{Rank of the saturated relation lattice}
    Initialize $c^{\mathrm{free}}$ as an empty list\;

    \BlankLine
    \For(\tcp*[f]{Process columns in order})
        {$j\gets1$ \KwTo $m$}{
        $w\gets W\,D^{\mathrm{tens}}[:,j]$
            \tcp*[r]{(\Cref{eq:saturation_routine_coordinates})}
        $v\gets w[s+1:n]$
            \tcp*[r]{Coordinates modulo the saturated relations}

        \If(\tcp*[f]{The rank increases})
            {$v\neq0$}{
            $g\gets\gcd(v_1,\dots,v_{n-s})$\;
            Compute $Q\in\GL_{n-s}(\Z)$ such that $Qv=ge_1$ using the extended Euclidean algorithm\;

            $U\gets U\,\operatorname{diag}(I_s,Q^{-1})$
                \tcp*[r]{(\Cref{eq:free_saturation_basis_update})}
            $W\gets\operatorname{diag}(I_s,Q)\,W$\;

            $s\gets s+1$
                \tcp*[r]{Append one primitive relation}
            Append $c_j$ to $c^{\mathrm{free}}$\;
        }
    }

    \BlankLine
    $D^{\mathrm{free}}\gets U[:,1:s]$\;
    \KwRet{$(D^{\mathrm{free}},b,c^{\mathrm{free}})$}\;
\end{algorithm}

\section{Notation}
\label{sec:notation}

\begingroup
\small
\renewcommand{\arraystretch}{1.15}
\setlength{\tabcolsep}{0.005\linewidth}
\setlength{\LTleft}{0pt}
\setlength{\LTright}{0pt}
\setlength{\emergencystretch}{1em}
\begin{longtable}{p{0.22\linewidth}p{0.76\linewidth}}
    \hline
    \textbf{Symbol} & \textbf{Meaning} \\
    \hline
    \endhead
    \hline
    \endfoot
    \hline
    \endlastfoot

    \multicolumn{2}{l}{\textbf{Standard notation}} \\*

    $\Z$, $\Q$, $\R$, $\F_p$, $\kk$
    & Integers, rationals, reals, the field with $p$ elements
      ($p$ prime), arbitrary field. \\

    $\otimes$, $\oplus$
    & Tensor product over $\Z$ and direct sum of $\Z$-modules. \\

    $\ker f$, $\im f$, $\Coker f$
    & Kernel, image, and cokernel of a homomorphism $f$. \\

    $\rank$, $\dim$
    & Rank of a matrix or free abelian group, dimension of a simplicial complex. \\

    $I_m$, $\GL_m(\Z)$
    & The $m\times m$ identity matrix, the group of invertible integer
      $m\times m$ matrices. \\

    $A[i,j]$, $A[:,j]$, $A[I,J]$
    & Entry, column, and submatrix of $A$ selected by row and column
      indices $I,J$. \\

    $\Matrix(f;\B,\B')$
    & Matrix of $f$ in the domain basis $\B$ and codomain basis $\B'$. \\

    $\{\!\{\cdot\}\!\}$
    & Multiset. \\[.2cm]

    \multicolumn{2}{l}{\textbf{Persistent homology}} \\*

    $\sigma$, $K$, $K^\bullet = (K^t)_{t=0}^T$
    & Simplex, simplicial complex, simplicial filtration. \\

    $T$, $N$
    & Final filtration index of the filtration $K^\bullet$, total number of simplices in $K^T$. For a simplex-wise filtration with $K^0=\emptyset$, one has $T=N$. \\

    $d$, $\ell$
    & Homological degree and dimension $\ell=\dim K^T$ of the final complex. \\

    $C_d(K;\Z)$, $Z_d(K;\Z)$ \newline $B_d(K;\Z)$, $H_d(K;\Z)$
    & Simplicial $d$-chains, $d$-cycles, $d$-boundaries, homology groups in degree $d$. \\

    $H_d(K^t,K^s;\Z)$
    & Relative homology group of the pair $(K^t,K^s)$ for $s\leq t$. \\

    $H_d(K^\bullet;\kk)$
    & Field-coefficient persistence module (\Cref{eq:persistence_module_field_coefficient}). \\

    $H_d(K^\bullet;\Z)$
    & Integral persistence module (\Cref{eq:persistence_module_integer_coefficient}). \\

    $\barcode(M)$, $[b_i,d_i)$
    & Barcode of a field-coefficient persistence module, a bar (\Cref{eq:persistence_barcode}). \\[.2cm]

    \multicolumn{2}{l}{\textbf{\Cref{subsec:reduction_fixed_field}:~Matrix reduction and row priorities}} \\*

    $D\in\kk^{n\times m}$ or $\Z^{n\times m}$
    & $n\times m$ matrix with coefficients in $\kk$ (\Cref{subsec:reduction_fixed_field}) or $\Z$ (\Cref{subsec:reduction_exceptional_primes}). \\

    $R$, $V$
    & Reduction $R=DV$ and its upper-triangular $V$ with diagonal entries $1_\kk$. \\

    $\widetilde D$, $\widetilde R$
    & Augmented matrices $\widetilde D=\binom{I_m}{D}$ and
      $\widetilde R=\widetilde D V=\binom{V}{R}$
      (\Cref{eq:augmented_matrix}). \\

    $\bot_j$
    & Label of the $j$th row in the augmented matrix. \\

    $\low(R,j)$, $\piv(R,j)$
    & Index and value of the lowest nonzero entry in column $j$.
      For a zero column, one has $\low(R,j)=\bot_j$ and $\piv(R,j)=1_\kk$. \\

    $\Red(\widetilde D)$
    & Output of the classical persistence algorithm with sequential reduction
      on~$\widetilde D$. \\

    $\tau=(\tau_1,\dots,\tau_{n+m})$, $\succ$
    & Row priority, encoded from highest to lowest priority:
      $\tau_1\succ\cdots\succ\tau_{n+m}$. \\

    $\tau_{\mathrm{can}}$
    & Canonical priority $(n,\dots,1,\bot_m,\dots,\bot_1)$ (\Cref{eq:canonical_order}). \\

    $\low(\widetilde R,j,\tau)$, \newline $\piv(\widetilde R,j,\tau)$
    & Index of the highest-priority nonzero entry of column $j$ with
      respect to $\tau$, \newline and the corresponding entry. \\

    $\Red(\widetilde D,\tau)$
    & Output of the persistence algorithm with row priority $\tau$
      (\Cref{alg:persistence_algorithm_priority}). \\

    $\tau_{\mathrm{piv}}$
    & Pivot priority $(p_1,\dots,p_m,q_1,\dots,q_n)$: canonical pivot indices in column order, followed by the unused row labels in arbitrary order (\Cref{eq:permutation_piv}). \\

    $\widetilde R_{\mathrm{can}}$, $\widetilde R_{\mathrm{piv}}$
    & Canonical and exhaustive reductions, obtained with priorities
      $\tau_{\mathrm{can}}$ and $\tau_{\mathrm{piv}}$
      (\Cref{eq:canonical_reduction,eq:exhaustive_reduction}). \\

    $K(\tau,\tau')$
    & Kendall tau distance between two permutations (\Cref{subsubsec:efficient_updates_priority}). \\[.2cm]

    \multicolumn{2}{l}{\textbf{\Cref{subsec:reduction_exceptional_primes}:~Exceptional primes and rational correction}} \\*

    $\widetilde D^{\kk}$, $\widetilde R^{\kk}$
    & Integral augmented matrix interpreted over $\kk$, a reduction
      over $\kk$. \\

    $\low(\kk,j)$, $\low(\kk,\bullet)$
    & Canonical pivot index in column $j$ over $\kk$, complete
      pivot index sequence. \\

    $\PP$
    &Primes $p$ for which $\barcode(H_d(K^\bullet;\F_p)) \neq \barcode(H_d(K^\bullet;\Q))$ for some $d$ (\Cref{eq:definition_exceptional_primes}), or for which $\low(\F_p,\bullet)\neq\low(\Q,\bullet)$ for the input matrix~$D$.
    When $D$ is the boundary matrix of a simplex-wise filtration, they coincide. \\

    $\Delta_j$
    & Principal minor $\Delta_j=\prod_{i=1}^{j}\piv(\Q,i)$ (\Cref{eq:principal_minor}). \\

    $\|D\|_{\infty}$, $\|D\|_{2,\infty}$
    & Largest absolute entry of $D$, largest Euclidean norm of a column of $D$. \\

    $\tau_{\mathrm{piv}}^{\Q}$, $\tau_{\mathrm{piv}}^{\F_p}$
    & Pivot priorities of the canonical pivot sequences over
      $\Q$ and $\F_p$ (\Cref{subsubsec:repairing_generic_reduction}). \\

    $\tau_j$, $\widetilde R_j^{\Q}$
    & Intermediate priority and rational reduction (\Cref{eq:intermediate_priority}). \\[.2cm]

    \multicolumn{2}{l}{\textbf{\Cref{sec:torsion_on_barcodes}:~Torsion and hierarchy of persistence modules}} \\*

    $H_d^{\mathrm{tors}}(K^t;\Z)$
    & Torsion subgroup of integral homology (\Cref{eq:free_part}). \\

    $H_d^{\mathrm{free}}(K^t;\Z)$
    & Torsion-free quotient $H_d(K^t;\Z)/H_d^{\mathrm{tors}}(K^t;\Z)$
      (\Cref{eq:free_part}). \\

    $\Tor(H_{d-1}(K^t;\Z),\kk)$
    & Tor term in the universal coefficient theorem
      (\Cref{eq:uct}). \\

    $H_d(K^\bullet;\Z)\otimes\kk$
    & Tensor persistence module (\Cref{eq:persistence_module_tensor}). \\

    $H_d^{\mathrm{free}}(K^\bullet;\Z)\otimes\kk$
    & Free persistence module (\Cref{eq:persistence_module_free}). \\[.2cm]

    \multicolumn{2}{l}{\textbf{\Cref{sec:computation_presentation_matrices}:~Presentation matrices for the tensor and the free module}} \\*

    $D_{d+1}^{\mathrm{tens}}$
    & Tensor presentation matrix (\Cref{eq:graded_matrix_tensor}). \\
    
    $D_d^{\mathrm{free}}$
    & Free presentation matrix (\Cref{eq:graded_matrix_free}). \\
    
    $B_d^{\mathrm{sat}}(K^t;\Z)$
    & Saturation of $B_d(K^t;\Z)$ in $Z_d(K^t;\Z)$ (\Cref{eq:free_homology_via_saturation}). \\
        
    $\B_d^Z$, $\B_d^Z(t)$
    & Incremental integral cycle basis $(z_1,\dots,z_n)$ (\Cref{eq:incremental_basis_cycle}). \\
    
    $\B_d^B$, $\B_d^B(t)$
    & Incremental basis $(b_1,\dots,b_{m'})$ of the saturated
      boundaries (\Cref{eq:incremental_basis_saturated_boundaries}). \\
    
    $n$, $m$, $m'$
    & Rank of $Z_d(K^N;\Z)$, $C_{d+1}(K^N;\Z)$, and $B_d(K^N;\Z)$.
    The presentation matrices $D_{d+1}^{\mathrm{tens}}$ and $D_d^{\mathrm{free}}$ have shapes $n\times m$ and $n\times m'$, respectively. \\
        
    $\Lambda_j$, $\Lambda_j^{\mathrm{sat}}$, $s(j)$
    & Integer span of the first $j$ columns, its saturation (\Cref{eq:saturated_relation_prefix}), its rank. \\
\end{longtable}
\endgroup

\end{document}